\documentclass[fleqn,usenatbib]{mnras}

\usepackage{mathptmx}
\usepackage[T1]{fontenc}

\DeclareRobustCommand{\VAN}[3]{#2}
\let\VANthebibliography\thebibliography
\def\thebibliography{\DeclareRobustCommand{\VAN}[3]{##3}\VANthebibliography}

\usepackage{graphicx}	
\usepackage{amsmath,amssymb}	
\usepackage{amsfonts}
\usepackage{tcolorbox}
\usepackage[colorinlistoftodos,prependcaption,textsize=tiny]{todonotes}
\usepackage{marginnote}
\usepackage{rotating}
\usepackage{multirow}
\usepackage{array}
\usepackage{float}
\usepackage{garamondlibre}
\usepackage{mathtools}
\usepackage{stmaryrd}
\usepackage{subcaption}
\usepackage{tabularx}
\usepackage{lineno}
\usepackage{afterpage}
\usepackage{pdflscape}
\usepackage{rotating}
\usepackage{orcidlink}
\usepackage{hyperref}

\newcommand{\yr}{{\rm yr}}
\definecolor{ppurple}{HTML}{670099}
\newcommand{\mchh}[1]{{\color{black} {#1}}}
\newcommand{\eM}{\text{\emph{e}M}}
\newcommand{\Moyr}{\text{$\mathrm{M}_{\odot}$ yr$^{-1}$}\ }
\newcommand{\Moyrk}{\text{$\mathrm{M}_{\odot}$ yr$^{-1}$kpc$^{-2}$}\ }

\renewcommand{\star}{\dagger}

\title[The Multiscale Structure of U/LIRGS]{The PARADIGM Project I: A Multiscale radio morphological analysis of local U/LIRGS}

\author[G. Lucatelli et al.]{Geferson Lucatelli$^{\orcidlink{0000-0002-2410-1776}\ 1}$\thanks{E-mail: geferson.lucatelli@postgrad.manchester.ac.uk (GL)},
Rob Beswick$^{1}$,
Javier Moldon$^{\orcidlink{0000-0002-8079-7608}\ 1,2}$,
Miguel \'A. P\'erez-Torres$^{\orcidlink{0000-0001-5654-0266}\ 2,3,4}$,
J. E. Conway$^{5}$,
\newauthor
Antxon Alberdi$^{2}$,
Cristina Romero-Ca\~nizales$^{\orcidlink{0000-0001-6301-9073}\ 6}$,
Eskil Varenius$^{\orcidlink{0000-0002-3248-9467}\ 5}$,
Hans-Rainer Klöckner$^{\orcidlink{0000-0002-0648-2704}\ 7}$,
\newauthor
L. Barcos-Mu\~noz$^{8,9}$,
Marco Bondi$^{10}$,
Simon T. Garrington$^{1}$,
Susanne Aalto$^{11}$,
Willem A. Baan$^{12,13}$ and
\newauthor
Ylva M. Pihlstrom$^{\orcidlink{0000-0003-0615-1785}\ 14}$\thanks{YMP is also an Adjunct Astronomer at the National Radio Astronomy Observatory.}
\ \\
$^{1}$Jodrell Bank Centre for Astrophysics, School of Physics and Astronomy, The University of Manchester, Manchester M13 9PL, UK\\
$^{2}$Instituto de Astrof\'isica de Andaluc\'ia (IAA-CSIC), Glorieta de la Astronom\'ia s/n, E-18008 Granada, Spain\\
$^{3}$Facultad de Ciencias, Universidad de Zaragoza, Pedro Cerbuna 12, E-50009 Zaragoza, Spain \\
$^{4}$School of Sciences, European University Cyprus, Diogenes street, Engomi, 1516 Nicosia, Cyprus\\
$^{5}$Department of Space, Earth and Environment, Chalmers University of Technology, Onsala Space Observatory, SE-439 92 Onsala, Sweden\\
$^{6}$Institute of Astronomy and Astrophysics, Academia Sinica, 11F of Astronomy-Mathematics Building, Taiwan, R.O.C.\\
$^{7}$Max-Planck-Institut für Radioastronomie, Auf dem Hügel 69, 53121 Bonn, Germany\\
$^{8}$National Radio Astronomy Observatory, 520 Edgemont Road, Charlottesville, VA  22903, USA \\
$^{9}$Department of Astronomy, University of Virginia, 530 McCormick Road, Charlottesville, VA 22903, USA \\
$^{10}$INAF - Istituto di Radio Astronomia, Via P. Gobetti 101, 40129, Bologna, Italy \\
$^{11}$Department of Space, Earth and Environment, Chalmers University of Technology SE-412 96 Gothenburg, Sweden\\
$^{12}$Netherlands Institute for Radio Astronomy (ASTRON), NL-7991 PD Dwingeloo, the Netherlands\\
$^{13}$Xinjiang Astronomical Observatory, Chinese Academy of Sciences, 150 Science 1-Street, 830011 Urumqi, China\\
$^{14}$Department of Physics and Astronomy, University of New Mexico, Albuquerque, NM 87131, USA\\
}

\date{Accepted XXX. Received YYY; in original form ZZZ}

\pubyear{2015}

\begin{document}
\label{firstpage}
\pagerange{\pageref{firstpage}--\pageref{lastpage}}
\maketitle

\begin{abstract}
Disentangling the radio flux contribution from star formation (SF) and active-galactic-nuclei (AGN) activity is a long-standing problem in extragalactic astronomy, since at frequencies of $\lesssim$ 10 GHz, both processes emit synchrotron radiation. We present in this work the general objectives of the PARADIGM Project, a multi-instrument concept to explore star-formation and mass assembly of galaxies. We introduce two novel general approaches for a detailed multiscale study of the radio emission in local (Ultra) Luminous Infrared Galaxies (U/LIRGs). In this work, we use archival interferometric data from the Very Large Array (VLA) centred at $\sim$ 6 GHz (C band) and present new observations from the \emph{e}-Multi-Element Radio-Linked Interferometer Network (\emph{e}-MERLIN) for UGC\,5101, VV\,705, VV\,250 and UGC\,8696. 
Using our image decomposition methods, we robustly disentangle the radio emission into distinct components by combining information from the two interferometric arrays.
We use \emph{e}-MERLIN as a probe of the core-compact radio emission (AGN or starburst) at $\sim$ 20 pc scales, and as a probe of nuclear diffuse emission, at scales $\sim 100 - 200$ pc. With VLA, we characterise the source morphology and the flux density on scales from $\sim 200$ pc up to and above $1$ kpc.  
As a result, we find deconvolved and convolved sizes for nuclear regions from $\sim 10$ pc to $\sim200$ pc. At larger scales, we find sizes of 1.5$-$2 kpc for diffuse structures (with effective sizes of $\sim$ 300$-$400 pc). We demonstrate that the radio emission from nuclear extended structures ($\sim$ 100 pc) can dominate over core-compact components, providing a significant fraction of the total multiscale SF output. We establish a multiscale radio tracer for star formation by combining information from different instruments. Consequently, this work sets a starting point to potentially correct for overestimations of AGN fractions and underestimates of SF activity. 
\end{abstract}




\section{Introduction}

Luminous and Ultra Luminous galaxies (U/LIRGs) are some of the most energetic extragalactic sources in the local Universe, emitting mostly in infrared (IR) wavelengths, with LIRGs having luminosities of 10$^{11}< L_{\rm IR}[8-1000\mu {\rm m}]< 10^{12}L_{\odot}$ and ULIRGs $L_{\rm IR}[8-1000\mu {\rm m}]> 10^{12}L_{\odot}$ \cite[e.g.][]{sanders1996,Lonsdale2006}. The radio emission from these sources originates from distinct physical processes, such as active galactic nuclei (AGN), jets,  intense star formation at the nuclear regions -- starbursts (SB), and star formation (SF) activity at scales larger than $\sim$ 200 pc. {Currently, there is a challenge on how to decompose the radio emission into each individual mechanism. It is unclear how a multiscale tracer of SF can be constructed in order to comprehend underlying physical processes at all possible scales and frequencies. 
Having such a metric would be ideal, since SF is one of the most fundamental physical processes that can be used as a proxy for galaxy evolution and mass assembly studies \citep[e.g.][]{bauer2013,Ilbert2013,Madau2014}. Star formation is responsible for converting interstellar gas clouds into star clusters, characterised by the efficiency and rate at which the gas is converted \citep{Hattori_2004,Shangguan_2019}. This acts as a major constituent of the dynamics and evolution of a galaxy. Star formation is also related to the interaction between gas, dust and ionised gas caused by supernova explosions which keeps inducing energy to form new stars, and enhancing the magnetic field strength \citep[e.g.][]{Thompson_2006,Thompson2009,Drzazga2011,Schober2016,Vollmer2022}. 
}

{Due to the presence of dust, these nuclear} regions are almost completely optically obscured, especially in the most compact regions at the centre of galaxies \citep[e.g.][]{Ricci2017,Hickox2018,falstad2021}. Infrared and radio observations are thus essential because they are able to see through the dust, which re-emits the ultraviolet radiation from young stars in the far-infrared. Such observations are, therefore, suitable tracers for star formation \citep[e.g][]{Farrah2008,Rieke2009,Tabatabaei2017,Algera_2022,Toro2023} {and provide high-angular resolution ($\lesssim$ 0.1") imaging reconstruction of these regions.}
\mchh{At radio frequencies of $\lesssim 10$ GHz}, the dominating form of radio emission is non-thermal synchrotron radiation, which is produced by both AGN and SF activity \citep{Condon1992}. Non-thermal emission related to SF is produced by core-collapse supernovae from massive stars with $\gtrsim 8\ \mathrm{M}_{\odot}$ \citep{Linden2019}, it is optically thin and not affected by dust. 
At \mchh{intermediate} frequencies, $\sim$ 10$-$100 GHz, the dominant form of radiation is the free-free Bremsstrahlung thermal emission {\citep{burke2019}} produced by ionised gas and photo-ionisation of neutral hydrogen around massive star formation regions (H$_{\textsc{II}}$ regions) {\citep{Murphy2011}}, {and} at frequencies $\gtrsim$ 100 GHz, thermal dust emission becomes dominant. 


Most U/LIRGs in the local 
Universe ($z\lesssim 0.3$) are mergers ($\sim$ 80\%) \citep[e.g.][]{Larson2016} while $\sim$ 20\% are spiral galaxies, 
but these fractions change with redshift \cite[e.g.][]{Hung2014}. Besides these systems being rare in the local Universe, their contribution to the total \mchh{infrared} output and SF budget dominate at higher redshift $z \sim 1$-$2$ \citep{lefloc2005,magnelli2009}. However, there are notable differences between local U/LIRGs and high-$z$ counterparts \citep{HernanCaballero2009,Rujopakarn2011,Whitaker2012,Hung2014} and it is unclear how these systems evolve in time/redshift \citep{Lonsdale2006}. Merger interactions are also relevant in such scenarios since they are a trigger of enhanced SF \citep{Haan2013a,Larson2016,Calabro2019}, resulting in star formation rates (SFR) from dozens to hundreds of \Moyr at the present time and in the 1$-$10 Myr range. As a result, U/LIRGs are great laboratories to study in  detail their structure and evolution \citep{Farrah2003,Petric2011a,Stierwalt2013,Chen2013,HerreroIllana2017}.


Since the emission of SF happens to coexist with AGN on multiple scales, it is relatively significant to understand the structure and the nature of the radio \mchh{emission} at all possible physical scales. \mchh{By mapping the radio emission} from nuclear regions ($\sim$ 10-100 pc) up to galactic-scale structures ($\gtrsim $ 1 kpc), \mchh{we can determine calibrated fractions of the radio power associated to star formation across these scales. Such information is essential for estimating both} the recent and historic rate at which stars are born in a galaxy. For that, \mchh{a non-contaminated and multiscale assessment of the radio emission, by removing all possible contributions from an AGN,} would yield corrected fractions of the total star formation activity. \mchh{By combining information from different instruments, we tackle this gap from previous research, introducing a robust multi-resolution fitting approach to disentangle the radio emissiom.}

\mchh{
Multiple works attempt to separate spatially and spectrally the radio emission coming purely from an AGN and emission coming from star-forming regions, \citep[SED fitting][]{Galvin2018,Dey_2022,Yamada2023}, \citep[using brightness temperatures][]{Morabito2022};  \citep[using data combination and high-resolution observations][]{Biggs2008}, \citep[integral field spectroscopy][]{davies2016}
etc. However, uncertainties are still present, and obtaining unbiased measurements for the AGN/SB-SF contribution is challenging. 

The aims of this work, and subsequent ones, are to establish a panoramic view of the structure of radio emission and its constituents in U/LIRGs, and to investigate their connection with star formation. Understanding the interplay between these processes with the accretion and feedback mechanisms of supermassive black holes at the centres of these sources is crucial to quantify how galaxies evolves over time. 

The technical novelty of this work is based on the combination of information from distinct interferometric arrays. These allow us to investigate the structure of radio emission at different spatial scales simultaneously. The same fitting techniques introduced here can potentially be applied to multi-frequency observations (including both radio and optical wavelengths) and multi-instrument observations, thereby providing resolved spatial and spectral information for each emission region of galaxies.

Using the previous techniques, we can study star formation and physical processes by characterising the multiscale nature of the radio emission from U/LIRGs. By having the properties of each diffuse structure, we can stablish calibration factors for the conversion of radio and infrared emission into SFR estimates. This can be used to explore how the radio-infrared correlation \citep{Bell_2003} behaves through a multiscale characterisation, and evaluate the physical mechanisms responsible for inducing deviations in the correlation.
}

{
In this work, we use \mchh{high-angular resolution observations ($0.05$"$\sim 0.3$")} from the \emph{enhanced}-Multi Element Remotely Linked Interferometer Network (\emph{e}-MERLIN) and from the Karl G. Jansky Very Large Array (VLA). Radio interferometry offers a unique way to reconstruct high angular resolution images, providing unprecedented parsec-scale details of local galaxies, unravelling the mechanisms responsible for producing the associated infrared and radio energies. 
}

To unravel the influence that an AGN/SB have on the kilo-parsec regimes and above, where diffuse star formation is taking place, it is required to understand the contribution that both processes have over the energy generation and how nuclear structures emerge and evolve in such a way that they can be strong enough to shape the global morphologies of these galaxies. Still, it is factually challenging and an open question to discern the contribution of the radio emission between AGN, SB (nuclear star formation) and large scale star formation \citep{P_rez_Torres_2021}.

The key questions we start investigating in this work and future ones are: What is the connection between source and component morphologies? \mchh{How does the location of radio components impact the brightness distribution of a source? How can star-formation processes solely rule or shape a radio source?} What is the behaviour of the radio-infrared correlation when AGN contamination is removed, for both thermal and non-thermal processes? How does the radio-infrared correlation hold when we break down the radio emission into different structural components and scales? We do not expect to answer these questions in this paper, but we post them as a starting point for a series of future works {(see Sec.~\ref{sec:broad_project})}. 

{
By estimating fractions of the radio emission across different physical scales, this work will provide some immediate contributions. We can obtain corrections to the dominant sizes and flux densities on nuclear regions over the influence of an AGN/SB and disentangle that from any existing nuclear emission due to star formation processes. This can pinpoint that the proportions of SF emission can be higher than previously expected, hence probing more massive galaxies constrained on SF activity. As a starting point, we exploit these issues in this work, and our specific steps are:
\begin{enumerate}
    \item Measure the sizes and flux densities of the radio emission at distinct physical scales, from $20\sim 100$ pc covered by \emph{e}-MERLIN to the large scale \mchh{regime} of $\sim 0.5-3$ kpc observed by VLA. 
    \item Disentangle individual sub-regions of the radio emission in order to quantify their structural properties and compute their fractions to the total source flux density.
    \item Obtain a metric for a multiscale star formation tracer: separate the nuclear extended ($\lesssim$ 100 pc) SF from the total SF.  
\end{enumerate}
}

\mchh{
This is the first work of project PARADIGM (see Sec.~\ref{sec:broad_project}), and describes a new analysis technique that can be used to maximize the scientific output of data from radio interferometers with complementary $uv$ spacing. The first phase of PARADIGM focuses on preparatory science for the Square Kilometre Array (SKA), using a combination of instruments that allows us to probe the sky accessible to the SKA in the future. This allows us to efficiently plan SKA observational campaigns with the aim to quantitatively and qualitatively expand the observed galaxies towards higher redshifts, and use larger statistics to measure differences as a function of galaxy mass, morphology, and luminosity.
}

This paper is structured as follows: {In Sec.~\ref{sec:broad_project} we present a general context for a long-term study of U/LIRGs}. In Sec.~\ref{sec:observations}, we present and discuss the observations from \emph{e}-MERLIN and VLA used in this work. We also discuss how the data was analysed, calibrated, combined between \emph{e}-MERLIN and VLA. We end that section presenting the main imaging results and justify the needs of disentangling the radio emission. In Sec.~\ref{sec:image_approach} we introduce two novel methodologies to decompose the radio emission into distinct components at different scales and {show} how integrated fluxes {densities and associated sizes} are estimated. We start by presenting the main results obtained from our combined data in Sec.~\ref{sec:results}, in particular, estimated flux densities and sizes for the nuclear and extended regions, and derived properties such as brightness temperatures and star formation rates. Consequently, we discuss the main findings and limitations of this work in Sec.~\ref{sec:discussion}, and we end the paper with final remarks and future plans in Sec.~ \ref{sec:conclusions}. In the online version of the Appendix, we provide extra material such as notes on individual sources, additional figures and complementary text. 

Regarding the calculation of distances, we adopt the $\Lambda$CMD model with the following constant values: $\Omega_\Lambda$ = 0.692, $\Omega_{m0}$ = 0.308 and $H_0$ = 67.8 km s$^{-1}$ Mpc$^{-1}$.

\section{{Context of a Long-Term Multi-frequency Study of U/LIRGs}}
\label{sec:broad_project}
{In this section we provide a brief overview of the PARADIGM project for a future context, and describe the full LIRGI Sample.}

\subsection{{The PARADIGM project}}

Project PARADIGM (PAnchromatic high-Resolution Analysis of DIstant Galaxy Mergers) is positioned within the overarching framework of exploring the star-formation and mass assembly history of the Universe, focusing on how the combined effects of black hole activity and star formation influence the energetic and chemical evolution of galaxies. Our primary objectives encompass:
\mchh{
\begin{enumerate}
    \item contributing to the calibration of the empirically observed radio-to-infrared correlation across a diverse range of galaxy types and redshifts; 
    \item translating radio and infrared luminosities into star formation rates and efficiency while characterizing the evolution of the Initial Mass Function (IMF) as a function of time and environmental factors; 
    \item disentangle the dominant gas and dust heating mechanism in the nuclear regions of galaxies;
    \item characterizing the clumpiness of the ISM as indicated by H$_{\textsc{II}}$ regions, through the assessment of the free-free emitting medium's smoothness/clumpiness.
\end{enumerate}
}
By pursuing these objectives across a wide spectrum of systems, encompassing various redshifts and a diverse range of phenomenologies, we aim to gain a deeper understanding of the processes that shape galaxy evolution during the most crucial phases of cosmic history.

As an initial step to achieve these goals, we aim to obtain a comprehensive understanding of the evolutionary phases of nuclear starbursts in local interacting galaxies situated within the range of 60 to 250 Mpc, in particular in U/LIRGs. We aim to establish a phenomenological sequence for the evolution of a nuclear starburst combining data spanning FIR, mm, and radio wavelengths. We plan to study a sample U/LIRGs at different merging stages to \mchh{capture the} entire lifecycle of a nuclear starburst from its onset to its decay.

The primary focus of PARADIGM's observations at this stage is centred on using high angular resolution instruments. This enables us to discern various physical phenomena that predominate at different frequencies, while also disentangling the individual contributions of structures from tens to hundreds of pc. For the local systems considered, we need sub-arcsecond resolution for frequencies ranging from MHz to tens of GHz and extends into the millimetre range. Our analysis will cover both morphological and kinematical information from high spatial and spectral resolution data using radio continuum, polarization and spectral line modes. Therefore, the core of the observational data comes from the SKA pathfinders LOFAR-VLBI, \emph{e}-MERLIN (see next section), the EVN, the VLBA and the VLA, together with ALMA and NOEMA at mm wavelengths, each providing resolutions ranging from sub-arcseconds to tens of milliarcseconds at their respective wavelengths. These angular scales align seamlessly with the capabilities of JWST, ELT and Hubble at near-IR and optical wavelengths, underscoring our emphasis on high-resolution panchromatic view of the nuclear regions of U/LIRGs. This will be complemented with the development of data analysis techniques such as the multiscale approach described in this work, as well as other spectral analysis in future works.

Within this scope, this first work of the PARADIGM project describes a new multiscale morphological analysis and applies it to a sample of four U/LIRGs. With this application, we show how the observations from the SKA pathfinders, VLA and \emph{e}-MERLIN, are relevant to disentangle different physical components in the nuclear regions of these galaxies.

\subsection{The LIRGI Sample as a Case Study}

As a part of the \emph{e}-MERLIN Legacy Project, the Luminous Infrared Galaxy Inventory (LIRGI) \citep{Conway2008} studies a northern subset of the Great Observatories All-sky LIRG Survey (GOALS) sample \citep{Armus2009} selected from The Infrared Astronomical Satellite (IRAS) Revised Bright Galaxy Sample \citep{Sanders_2003}. It comprises a representative sample of 33 LIRGs and 9 ULIRGs in the local Universe over a distance range from $\sim 60$ Mpc to $\sim 260$ Mpc. Most of these sources are merging/interacting systems, ranging from a merger stage of 0 (early stage merger) to 6 (final stage merger) \citep[classification of][]{haan2011,Kim2013}, but a few are spiral galaxies undergoing high star formation activity. Star formation rates in these systems span a range from 10 $\sim $ \Moyr up to $\sim$ 400 \Moyr.


Several sources from the LIRGI sample have been studied individually, such as Arp\,299 \cite{torres2009,canizales2011a,Bondi2012,Olivencia2022}, Arp 220 \cite{Varenius2016,loreto2015} and additionally UGC\,8387 \cite{canizales2012b,Modica2012,canizales2017}, IRAS\,23365+3604 \cite{canizales2012a}, UGC\,5101 \citep{Lonsdale_2003},  UGC\,8696 \citep{Carilli_2000,Bondi2005,Klockner2004} among others \cite[see][for a review]{P_rez_Torres_2021}. These studies show how the central and obscured AGN/SB influences the star formation in these galaxies and the possible causes that an AGN is turned on. When it comes to general studies of (massive) star formation activity in galaxies, \cite{Linden2017,Linden2019,Linden2020} explored the multi-frequency nature of the SF in both nuclear and extranuclear regions and \cite{Larson2020} investigate how star-formation clumps are formed in LIRGs.

{The main goals of the Legacy Project focuses on accessing the star formation histories and galaxy assembly of local U/LIRGs using high resolution observations. Equivalent, understand how well the radio continuum emission is converted to star formation rates, without the contamination from an AGN. To that end, specific goals are to (i) quantify the morphology and size of radio emission associated to SF, including diffuse and compact nuclear SF (SB); (ii) detect powerful core-collapse radio supernovae (RSNe), used to constrain the high mass SF; (iii) characterize the existence and morphology of AGNs in these systems; (iv) study the relations between SF, AGN accretion and feedback mechanisms, quantifying how the energy/material flow at the nuclear regions of U/LIRGs.}

\section{Observations, Data Calibration and Imaging}
\label{sec:observations}
In this work, we set up an initial study for a subset of the LIRGI sample (see Tab.~\ref{tab:source_information}) and demonstrate a methodology that will be used later for the entire LIRGI sample, to study the full structure of the radio emission. Our subsample was selected by inspecting existing C band VLA archival data, allowing immediate combination with recent \emph{e}-MERLIN observations at the same frequency (see Tab.~\ref{tab:data_archive}). {The systems are, the two LIRGs VV\,705 (N \& S), VV\,250 (SE \& NW) and the two U/LIRGs UGC\,5101, UGC\,8696 (N \& SE)}. Details of these candidates can be found in Appendix \ref{sec:individual_source_analysis}. We have followed these criteria to select the candidates: 
\mchh{
\begin{itemize}
    \item physical: these sources show evidence of extreme condition environments in radio and infrared, having relatively high star formation activity ($>$ 50 \Moyr) occurring both in the central parts and in their diffuse components.
    \item technical: existing archival data at C band (6 GHz) in the VLA-A configuration ($\sim$ 0.3" angular resolution) to map the major fraction of their extended emission (when existent) in combination with current \emph{e}-MERLIN observations to resolve their nuclear regions (at $\sim$ 0.05").
\end{itemize}
}
These sources contain varied radio structural properties, hence we can experiment with how our approach performs in decomposing their radio emission from  core-compact (AGN or SB) structures, nuclear diffuse emission at $\sim$ 100 pc scales and emission originated from large-scale extended star-forming regions with scales greater than $\sim$ 1 kpc.

\begin{table*}
\caption{Existing source information from literature. Column description:
    (1) System name, with right ascension in (2) and declination in (3). 
    (4) Log of the infrared-luminosity in terms of solar luminosities. 
    (5) Luminosity distance. 
    (6) Merger stage from \citep{haan2011,Kim2013}. 
    (7) Nuclear separation between featured radio sources in each system.
    (8) Radio to infrared $q$ factor \citep{Yun_2001}. 
    (9) The range of total flux densities (for the full system) measured at C band (4 $\sim$ 6 GHz) from the NASA/IPAC Extragalactic Database. The exception is VV\,250, which contains a single measurement.
    (10) The radio morphology.
    }
\label{tab:source_information}
\begin{subtable}[h]{0.99\textwidth}
    \centering
    \begin{tabular}{rllccccccl} 
        \hline
        Galaxy Name    &   RA$^{\dagger}$(J2000.0)    & DEC$^{\dagger}$ (J2000.0)  & $\log_{10}(L_{\rm IR}/L_{\odot})$   & $D_l$[Mpc] & $M_s$        &  $n_c$ [kpc]  & $q_{\rm IR}$   & $S_\nu^{\rm C \ band}$ [mJy]    & Morphology  \\
        (1)            &   (2)                        & (3)                        & (4)                                 & (5)        & (6)          &  (7)          & (8)            & (9)                         & (10)        \\ \hline
        UGC\,5101      &   09h35m51.599s              & +61d21m11.72s              & $11.97$                             & 179.27     & 5            &   0.44        & 1.99           & 61.5 $\sim$ 79.0            & AGN/SB      \\ \hline
        VV\,705 N      &   15h18m06.115s              & +42d44m45.06s              & $11.89$                             & 183.12     & 4            &   6.26        & 2.35           & 19.6 $\sim$ 32.0            & AGN/SB      \\ 
        \quad \dots\quad\ S &   15h18m06.328s         & +42d44m38.11s              &                                     &            &              &               & --             &                             & ?           \\ \hline
        UGC\,8696 N    &   13h44m42.130s              & +55d53m13.50s              & $12.14$                             & 169.78     & 5            &   0.77        & 2.27           & 60.3 $\sim$ 103.0           & SB          \\ 
    \quad \dots\quad SE&   13h44m42.179s              & +55d53m12.79s              &                                     &            &              &               & --             &                             & AGN         \\ \hline
        VV\,250 SE     &   13h15m34.954s              & +62d07m28.80s              & $11.77$                             & 140.58     & 2            &   42.48       & 2.39           & 19.6                            & HII         \\ 
    \quad \dots\quad NW&   13h15m30.676s              & +62d07m45.40s              &                                     &            &              &               & --             &                             & ?           \\ \hline
        \hline
    \end{tabular}
    \caption*{$^{\dagger}$ Coordinates refers to the peak brightness position in the main component of the source (see Figs.~\ref{fig:pre_results} and \ref{fig:results_cont_1}).}
\end{subtable}
\end{table*}

\subsection{\emph{e}-MERLIN and VLA Observations}

\begin{table*}
\centering
\caption{Observational data information for VLA and \emph{e}-MERLIN at C band ($\sim 6$ GHz).}
\label{tab:data_archive}
\begin{subtable}[h]{0.99\textwidth}
    \centering
    \begin{tabular}{llllllll} 
        \hline 
        Galaxy Name   & Project                            & Phase / Flux \&              &  Obs Date                 &   Time on               & Central               & Freq.      & SPWs/ \\ 
                      & Code                               & Bandpass Calibrators         &                           &   Source [hr]           & Freq. [GHz]            & Range [GHz]    & Channels          \\ \hline 
        UGC\,5101     & EVLA 19A-076                       & J0921+6215 / 3C48            &  09.16.19$\sim$10.20.19   &   0.7$^{*}$             & 5.1                   & 4.1 $\sim$ 6.2 & 19 / 128          \\ 
                      & \emph{e}-M LE1014-L14              & ---                          &  10.12.19                 &   8.8                   & 5.1                   & 4.2 $\sim$ 6.0 & 8 / 512          \\  \hline
        VV\,705       & EVLA 19A-277                       & J1549+5038 / 3C286           &  10.08.19                 &   1.0                   & 6.0                   & 4.1 $\sim$ 7.9 & 36 / 64          \\ 
                      & \emph{e}-M LE1014-L26              & ---                          &  03.15.20                 &   8.8                   & 5.2                   & 4.5 $\sim$ 5.8 & 8 / 512          \\  \hline
        UGC\,8696     & EVLA AL746                         & J1349+5341 / 3C286           &  06.21.11$\sim$07.01.11   &   0.4                   & 5.9                   & 4.2 $\sim$ 7.6 & 16 / 128          \\ 
                      & \emph{e}-M LE1014-L24              & J1337+5501 / 3C286           &  10.07.20$\sim$03.13.20   &   26.4                  & 5.2                   & 4.2 $\sim$ 6.2 & 8 / 512          \\  \hline
        VV\,250       & EVLA AL746                         & J1339+6328 / 3C286           &  06.21.11$\sim$07.01.11   &   0.4                   & 5.9                   & 4.2 $\sim$ 7.6 & 16 / 128          \\ 
                      & \emph{e}-M LE1014-L21              & ---                          &  12.27.19                 &   8.8                   & 5.2                   & 4.2 $\sim$ 6.2 & 8 / 512          \\  \hline
    \end{tabular}
        \caption*{$^{*}$Between the 9 epochs, only a single one was used.}
\end{subtable}
\end{table*}

We use VLA and \emph{e}-MERLIN observations at $\sim$ 6 GHz (C band) to conduct a multiscale study of the nuclear and diffuse radio emission of four systems from LIRGI. \mchh{At this frequency, the angular resolutions of the two interferometers are 0.05" and 0.33" for \emph{e}-MERLIN and VLA, respectively. Equivalently, for a source at 150 Mpc, the physical scales of these angular resolutions would be $\sim$ 36 pc and $\sim$ 240 pc, respectively}. A summary of the observational data and their related projects from VLA and \emph{e}-MERLIN archives is presented in Tab.~\ref{tab:data_archive}.

\mchh{
The \emph{e}-MERLIN observations at 6 GHz were observed between 2019 and 2020. They comprise about $\sim$ 370 hours in total, for the 42 sources, giving an average time on source of $\sim 8.8$ hours. The total bandwidth is of about 1.5 $\sim$ 2.0 GHz, and the average sensitivity provided is $\sim $ 10 - 20 $\mu$Jy/beam at C-band.
}


We use VLA archival data on the most extended VLA configuration (A) in order to achieve the highest angular resolution provided by the array \mchh{(which baseline lengths up to $\sim$ 37 km)}. We have acquired the data from NRAO archive for our four sources, with observational dates between 2011 and 2019. The average sensitivity of the data is {\color{black} $10\mu$}Jy for a 30-minute on-source time. However, the project archive list we are using spans an on-source time between 20 minutes to one hour. 

\begin{figure}
\centering
\includegraphics[width=0.82\linewidth]{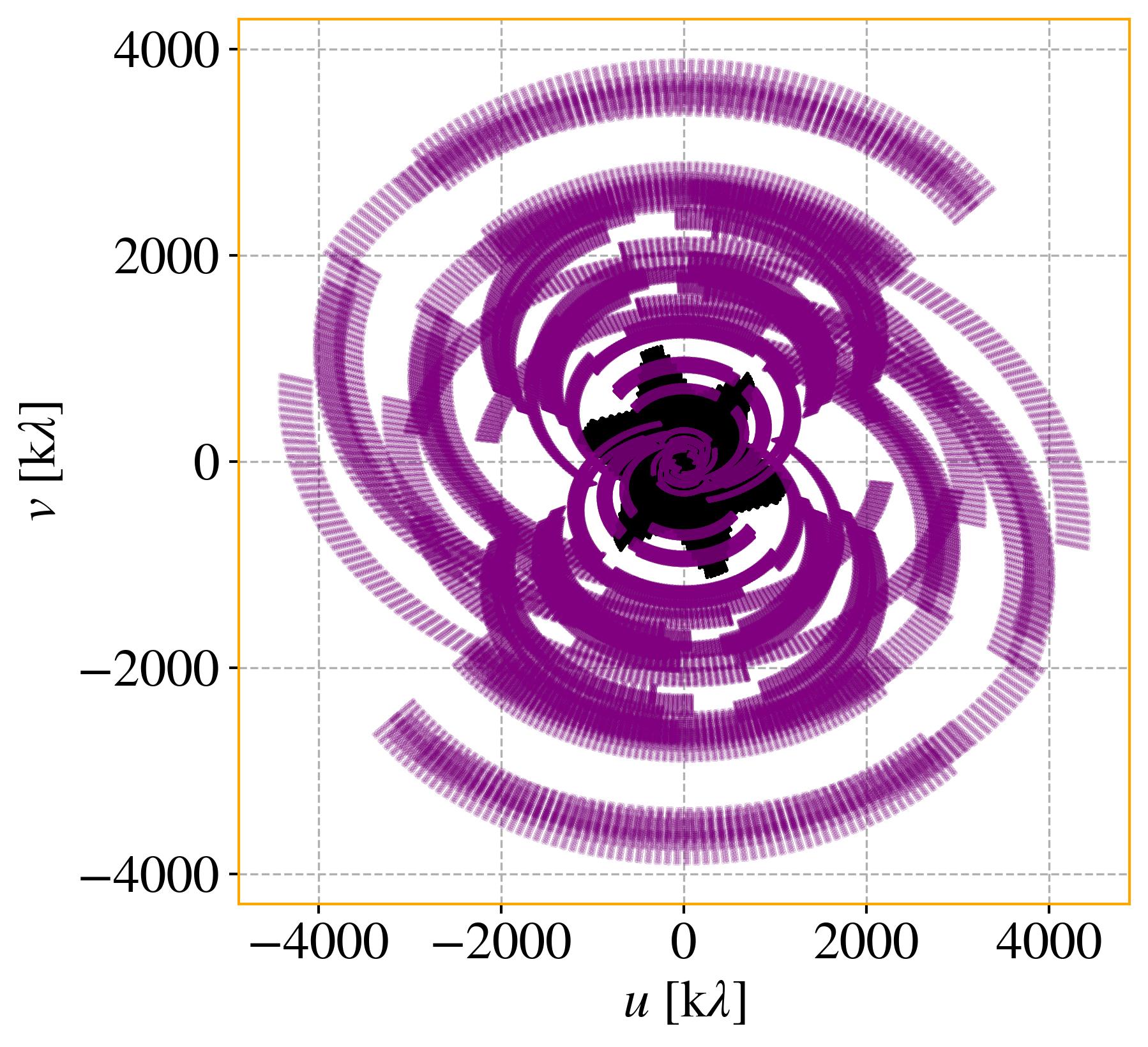}
\caption{
Typical $uv$ coverage (e.g. VV\,705) with \emph{e}-MERLIN ($\sim$ 8.8 hours on source) in purple points and VLA ($\sim$ 1 hour on source) in black points.}
\label{fig:uv_coverage}
\end{figure}

\subsection{Data Calibration}
\label{sec:data_reduction}

\subsubsection{Flagging, Calibration and Averaging}


We have calibrated VLA data using the CASA software (v6.2) \citep{CASAMcMullin2007,TheCASATeam_2022}, adopting standard procedures from the EVLA pipeline (6.2) guidelines. For the \emph{e}-MERLIN data, the \emph{e}-MERLIN CASA Pipeline (v1.1.19) was used \citep{eMCP2021}\footnote{See more information on the \emph{e}-MERLIN CASA Pipeline page at \url{https://github.com/e-merlin/eMERLIN_CASA_pipeline}.} with CASA (v5.7). 

The basic steps of the calibration are summarised as:
\begin{enumerate}
\item Standard pre-flagging of the data (e.g. observational flags, shadowing, clipping, quack). 
\item Visual inspections of each visibility in order to look for bad data, RFIs, antenna issues and others. 
\item Different steps of auto-flagging, using \texttt{tfcrop} prior to calibration and \texttt{rflag} after calibration: 
\begin{enumerate}
    \item A first pass of \texttt{tfcrop} was applied after pre-flagging (before any pre-calibration step) since this method is suitable for uncalibrated data. 
    \item When pre-calibration was established and applied, we performed a first-pass of \texttt{rflag} on the pre-calibrated data.
    \item After achieving the final calibration tables and applied to the data, we performed another pass of \texttt{rflag}, including also the target fields. 
\end{enumerate}
\item If new issues or bad data are identified in the final calibrated data, they are appended to the manual flagging file and the calibration is performed again from start.
\end{enumerate}

Since on average VLA observations contain 128 channels, we consider averaging our \emph{e}-MERLIN data in frequency, from $512$ channels to $128$. We use this averaged data to conduct all phase-referencing calibration through the \emph{e}-MERLIN pipeline. We reserve spectral line analysis for a future work. {For the VLA data, we use the native spectral and temporal resolution of each observation to perform phase-referencing calibration.} During calibration of VLA data, we considered trying three different {shorter} solutions intervals {to correct the more rapid phase variations}, named \texttt{solint\_short = 16s}, \texttt{solint\_mid = 32s} and \texttt{solint\_long = 48s}. The calibration tables were inspected in order to check which one performed better {and flagged fewer data}, given that the signal-to-noise in each {observation} is different. We also combined these solutions with a {longer solution interval of \texttt{60s} and a} final \texttt{solint\_inf = inf} solution interval to calibrate scan-related offsets in phase and amplitude. For project AL746, {the longer shorter} solution interval worked better, while for 19A-076 and 19A-277, {the shorter one was} more suitable. {For reference, see Tab.~\ref{tab:solints}.}

\begin{table}
\centering
\caption{Template of solution intervals used during calibration of VLA data using the phase reference sources.}
\label{tab:solints}
\begin{subtable}[h]{0.49\textwidth}
    \centering
    \begin{tabular}{l|c|cc}
        \hline
        Project    & Shorter Solint (\texttt{p}) & Solint Long (\texttt{p} and \texttt{ap}) \\[0.3ex] \hline \rule{0pt}{1.0em}
        AL746      & 48s                         & 60s,\texttt{inf}                         \\[0.3ex] \hline \rule{0pt}{1.0em}
        19A-277    & 16s                         & 60s,\texttt{inf}                         \\[0.3ex] \hline \rule{0pt}{1.0em}
        19A-076    & 16s                         & 60s,\texttt{inf}                         \\[0.3ex] \hline 
    \end{tabular}
\end{subtable}
\end{table}

\subsubsection{Self-Calibration}
\label{sec:selfcal}
{
After having achieved the desired calibration of each observation in both instruments, we prepared the data for self-calibration. Further time averaging is done up to a factor of two of the integration time for \emph{e}-MERLIN, down to 8s. We do this because there are no concerns related to time smearing and frequency aliasing since all our sources are near the phase pointing centre. We also averaged the VLA data in time up to 8s to match with our \emph{e}-MERLIN observations. A summary of the observation properties is given in Tab.~\ref{tab:data_archive}.
}

We conducted self-calibration individually for VLA, \emph{e}-MERLIN and combined data (see below). Different steps of self-calibration were performed for each source because of the nature of their structures, observational setups and phase-referencing quality. {To image the data, we used the} CASA task \texttt{tclean}. 
{To account for the intrinsic spectral index, we have used the \texttt{mtmfs} deconvolver with three Taylor terms (\texttt{nterms=3}), in combination with the \texttt{multiscale} deconvolver \citep{rau2011}, enabling it by setting different scales, i.e. \texttt{scales=[0,2,4,8,16,32,64]}}. This is required since we consider that, in general, sources have extended emission. The masking during deconvolution was created manually (setting \texttt{iterative=True} and \texttt{usemask=`user'} in \texttt{tclean}).

We have used a standard initial number of steps for self-calibration: \texttt{1:phase} + \texttt{2:phase} + \texttt{3:phase}. When possible, having enough flux density for an amplitude self-calibration, we applied \texttt{1:phase} + \texttt{2:phase} + \texttt{3:amp/phase}. This is the first attempt to improve data quality. However, for sources UGC\,8696 and UGC\,5101 involving VLA data, we had to perform additional steps of phase-self-calibration, decreasing the solution interval each time in order to achieve higher image fidelity. Where required, we also considered combining polarisations (in \texttt{gaincal}, by setting \texttt{gaintype=`T'}) to improve the signal-to-noise of the solutions. In each case, in the first self-calibration iteration, we considered a lower Briggs robust parameter (e.g. \texttt{-0.5} or \texttt{0.0}, depending on how bright the source is), so that only the most compact emission is used in the first loop. In the subsequent iterations, a higher robust parameter was considered (e.g. \texttt{0.5-2.0}), to account for more diffuse structures.

After the self-calibration step, we estimate the average rms-based value of each visibility weight, labelled ${\bf w}^{\rm VLA}$ and ${\bf w}^{\eM}$ for VLA and \emph{e}-MERLIN respectively.  We have used the task \texttt{statwt} in CASA, with the default parameter options. These weights are used for data combination in Section \ref{sec:data_comb}. Finally, after performing all the data calibration steps, the amount of data flagged for sources within project AL746 was $\sim$ 40\%, while for 19A-277 (VV\,705) it was $\sim$ 37\% and 15\% for 19A-076 (UGC\,5101). For \emph{e}-MERLIN, the amount of data flagged was 54\% for VV\,705, {\color{black} 70\% for VV\,250},  55\% for UGC\,5101 and 64\% for UGC\,8696. We report that for the \emph{e}-MERLIN observation of VV\,250, the high fraction of data flagged was also due to a period during the observation that the source was above an elevation of 80$^\circ$, so most of the calibration solutions failed.

\begin{table}
\centering
\caption{{Basic source and observational radio continuum properties derived from the data. 
            This table presents: (1) Source name. (2) From which image the measurement was made. (3) The total continuum integrated flux density $S_{\nu}$ in mJy, here at C band ($\sim$ 5-6 GHz). For simplicity, we have not corrected $S_{\nu}$ for the spectral variation of the flux density from 5 GHz to 6 GHz, given that each observation is centred at a different frequency in that range, see Tab.~\ref{tab:source_information}.  (4) The peak brightness flux density $S_{\rm peak}$, in $\mu$ Jy/beam. (5) The restoring beam size, in arcsec. (7) The median absolute standard deviation noise level ($\sigma_{\rm mad}$), in $\mu$Jy/beam, measured in each image's residual map.
            These values were computed for each source, for four different images (as indicated by column (2)). These images are presented in Figs.~\ref{fig:pre_results} and \ref{fig:results_cont_1}.} }
\label{tab:source_global_properties}
\label{tab:source_properties}
{
\begin{subtable}[h]{0.50\textwidth}
    \centering
    \scalebox{0.82}{%
    \begin{tabular}{l|lcccccc} 
        \hline
        Source                 & Image              & $S_{\nu}$             & $S_{\rm peak}$             & $\theta_a \times \theta_b$              & $\sigma_{\rm mad}$                 \\[0.1ex]             \rule{0pt}{0.3em}
                               &                    & [mJy]                 & [mJy/Beam]                 & [arcsec]                                & [$\mu$Jy/beam]                     \\[0.3ex]        \rule{0pt}{0.3em}
        (1)                    & (2)                & (3)                   & (4)                        & (5)                                     & (6)                                \\[0.3ex] \hline \rule{0pt}{0.3em}
        VV\,705 N              & \emph{e}-MERLIN    & 6.0 $\pm$ 1.3         & 2.4                        & 0.04$\times$0.03                        & 18.6                               \\[0.3ex]        \rule{0pt}{0.3em}
                               & $w_1$              & 7.5 $\pm$ 1.0         & 2.9                        & 0.09$\times$0.06                        & 14.9                               \\[0.3ex]        \rule{0pt}{0.3em}
                               & $w_2$              & 10.2 $\pm$ 0.4        & 4.1                        & 0.19$\times$0.17                        & 8.1                                \\[0.3ex]        \rule{0pt}{0.3em}
                               & VLA                & 12.7 $\pm$ 0.2        & 6.4                        & 0.51$\times$0.40                        & 3.3                                \\[0.3ex] \hline \rule{0pt}{0.3em}
        VV\,705 S              & \emph{e}-MERLIN    & 1.2 $\pm$ 0.3         & 0.5                        & 0.04$\times$0.03                        & 18.4                               \\[0.3ex]        \rule{0pt}{0.3em}
                               & $w_1$              & 1.8 $\pm$ 0.4         & 0.7                        & 0.09$\times$0.06                        & 14.9                               \\[0.3ex]        \rule{0pt}{0.3em}
                               & $w_2$              & 4.5 $\pm$ 0.7         & 1.2                        & 0.19$\times$0.17                        & 8.0                                \\[0.3ex]        \rule{0pt}{0.3em}
                               & VLA                & 6.2 $\pm$ 0.2         & 2.0                        & 0.51$\times$0.40                        & 3.5                                \\[0.3ex] \hline \rule{0pt}{0.3em}
        UGC\,5101              & \emph{e}-MERLIN    & 37.3 $\pm$ 5.5        & 12.6                       & 0.04$\times$0.03                        & 31.6                               \\[0.3ex]        \rule{0pt}{0.3em}
                               & $w_1$              & 54.9 $\pm$ 3.4        & 21.3                       & 0.11$\times$0.08                        & 26.2                               \\[0.3ex]        \rule{0pt}{0.3em}
                               & $w_2$              & 58.5 $\pm$ 1.1        & 29.0                       & 0.29$\times$0.18                        & 14.2                               \\[0.3ex]        \rule{0pt}{0.3em}
                               & VLA                & 60.5 $\pm$ 0.5        & 41.7                       & 0.60$\times$0.49                        & 8.8                                \\[0.3ex] \hline \rule{0pt}{0.3em}
        UGC\,8696              & \emph{e}-MERLIN    & 40.9 $\pm$ 7.1        & 5.2                        & 0.05$\times$0.04                        & 66.0                               \\[0.3ex]        \rule{0pt}{0.3em}
                               & $w_1$              & 47.2 $\pm$ 2.6        & 6.5                        & 0.09$\times$0.06                        & 15.4                               \\[0.3ex]        \rule{0pt}{0.3em}
                               & $w_2$              & 51.4 $\pm$ 1.4        & 17.4                       & 0.22$\times$0.20                        & 10.1                               \\[0.3ex]        \rule{0pt}{0.3em}
                               & VLA                & 53.3 $\pm$ 0.8        & 26.9                       & 0.40$\times$0.34                        & 8.8                                \\[0.3ex] \hline \rule{0pt}{0.3em}
        VV\,250 SE             & \emph{e}-MERLIN    & 5.6 $\pm$ 0.9         & 0.4                        & 0.07$\times$0.03                        & 24.2                               \\[0.3ex]        \rule{0pt}{0.3em}
                               & $w_1$              & 13.3 $\pm$ 1.7        & 1.5                        & 0.14$\times$0.12                        & 29.4                               \\[0.3ex]        \rule{0pt}{0.3em}
                               & $w_2$              & 16.8 $\pm$ 0.9        & 3.2                        & 0.23$\times$0.22                        & 11.0                               \\[0.3ex]        \rule{0pt}{0.3em}
                               & VLA                & 17.9 $\pm$ 0.5        & 7.5                        & 0.60$\times$0.46                        & 7.6                                \\[0.3ex] \hline \rule{0pt}{0.3em}
        VV\,250 NW             & \emph{e}-MERLIN    & 1.4 $\pm$ 0.3         & 0.2                        & 0.12$\times$0.09                        & 24.5                               \\[0.3ex]        \rule{0pt}{0.3em}
                               & $w_1$              & 1.7 $\pm$ 0.4         & 0.4                        & 0.16$\times$0.14                        & 22.2                               \\[0.3ex]        \rule{0pt}{0.3em}
                               & $w_2$              & 2.5 $\pm$ 0.3         & 0.8                        & 0.23$\times$0.22                        & 7.4                                \\[0.3ex]        \rule{0pt}{0.3em}
                               & VLA                & 3.3  $\pm$ 0.4        & 1.2                        & 0.60$\times$0.46                        & 7.1                                \\[0.3ex] \hline 
    \end{tabular}
    }
    \caption*{Notes: 1) For UGC\,5101 and UGC\,8696, since the nuclear separation between the sub-components are small, we display these flux densities for the total emission. 2) Errors for $S_\nu$ are only statistical, computed in the convolved residual map, inside the masked region of the source (see Eq.~\ref{eq:erro_flux}).}
\end{subtable}
}
\end{table}

\subsubsection{Data combination: \emph{e}-MERLIN and VLA}\label{sec:data_comb}

Having \emph{e}-MERLIN and VLA observations, we can achieve a suitable balance between sensitivity and resolution by combining data between both interferometers. VLA observations in A configuration enable us to fill the short spacing for longer-baseline interferometers \citep[e.g.][]{Muxlow2005,Williams2019,Muxlow2020}. An example of the $uv$ coverage for a particular combined visibility is presented in Fig.~\ref{fig:uv_coverage}, with \emph{e}-MERLIN in purple and VLA in black. As can be seen, the shorter baselines in \emph{e}-MERLIN match the longest baselines in VLA and the combination of both arrays provides a smooth transition between scales. Also, both arrays have matching sensitivity, and because of these two factors, it is possible to compute multiple images with different weighting schemes that recover physical structures from small to large scales continuously.

When concatenating the calibrated visibilities with the CASA task \texttt{concat}, weight scaling factors $v_{\rm \bf w}$ must be multiplied to each visibility via the argument \texttt{visweightscale} with the weights ${\bf w}^{\rm VLA}$ and ${\bf w}^{\eM}$ previously calculated. In this approach, if we impose that both visibilities have the same weight, we require that $v_{\bf w}^{\rm VLA} {\bf w}^{\rm VLA} = v_{\bf w}^{\eM} {\bf w}^{\rm \eM}$. We can for simplicity keep the \emph{e}-MERLIN scaling factor to unity, i.e. $v_{\bf w}^{\eM} = 1$, and change the one for VLA accordingly. Therefore, the VLA scaling factor is chosen to satisfy 
\begin{align}
    &v_{\bf w}^{\rm VLA} \times \left(\frac{{\bf w}^{\eM}}{{\bf w}^{\rm VLA}}\right)^{-1} = v_{\bf w}^{\eM} = 1.0 \label{eq:vw2} \qquad (\text{Balanced Visibility})
\end{align}
where $v_{\bf w}^{\rm VLA}$ represents a combined visibility with balanced weights between both arrays. 

\subsubsection{Astrometry Check}

Prior to concatenating the interferometric data, we performed alignment checks and two assumptions were made. We used \emph{e}-MERLIN as the coordinate reference due to its higher resolution and greater positional accuracy (10 $\sim$ 20 mas at C band). Moreover, it is important to note that the peak flux location in \emph{e}-MERLIN may not align perfectly with that of VLA, so we can not use the peak brightness position for alignment when sources are not point-like structures. 

To align the interferometric images, comparing multiple point-like sources is a common approach. However, in our case, there are very few sources in each field, except for VV\,250. Hence, we can utilise the phase reference centre of the \emph{e}-MERLIN observation and shift the VLA centre to match \emph{e}-MERLIN's centre. Furthermore, we can potentially correct significant errors in the phase shift through self-calibration using the combined data by producing a combined model during cleaning.
{
\subsubsection{Relative Flux Density Scaling}
Regarding data combination, additional checks were performed to understand the relative amplitude between \emph{e}-MERLIN and VLA. We have plotted the visibilities between both instruments on common baselines in the $uv-\lambda$ space. The amplitude revealed to be the same. Small corrections were potentially corrected when performing self calibration in the combined data (usually by an infinite amplitude complex gain correction). If large differences are present (not in our four cases), relative scaling factors must be enforced to the data to correct the differences, since self-calibration will not correct for such large differences. 
}

\subsection{Imaging and Gridding Weights}
\label{sec:weights}

Final image products were produced using the \textsc{wsclean} algorithm (v3.1.0), a w-stacking fast imager.
After combining the data, we used two weighting schemes during imaging, by changing the deconvolution gridding weights or by applying $uv$ tapering. In the first case, on-the-fly weighting imaging can be changed with the Briggs \texttt{robust} parameter \citep{briggs1995} in \textsc{CASA} or \textsc{wsclean}. 
For each combined interferometric data (having similar visibility weights, Eq.~\ref{eq:vw2}), we generated a series of images by varying the Briggs robust parameter between \texttt{-1.5} and \texttt{2.0}.  Lower values yielded predominantly \emph{e}-MERLIN characteristics (compact emission), while values above zero trended towards VLA features (unresolved and extended emission). We also individually imaged pure \emph{e}-MERLIN and VLA visibilities with typical robust parameters of \texttt{0.0}, \texttt{0.5} and \texttt{2.0}.

In order to account for flux density systematics due to the PSFs effects during deconvolution of combined interferometric data, we follow the solutions discussed in \cite{Radcliffe2023} (and private communication, Radcliffe). 
With \textsc{wsclean}, we configure the deconvolution by setting the options \texttt{-auto-threshold} and \texttt{-auto-masking} to values  \texttt{0.01} and \texttt{3.0}, respectively. In the particular cases of VV\,250 and UGC\,8696, we also considered the use of $uv$ tapering, which we discuss individually in App.~\ref{sec:imaging_results_sources}.

For standard Stoke I continuum imaging, we have used only \texttt{RR} and \texttt{LL} correlations for imaging the combined data.
With \textsc{wsclean}, we have used the multiscale deconvolver {\citep{rau2011}} with scales \texttt{[0,2,4,8,16,32,64]}, setting both the \texttt{--multiscale-gain} and \texttt{--gain} parameters to \texttt{0.05}. In \textsc{wsclean}, {the Multi-Frequency Synthesis is enabled by cleaning channels jointly, which takes into account the spectral variation of the sky.  The arguments for this mode are \texttt{-join-channels -channels-out nc} where \texttt{nc} is the number of sub-band images} to be used, each one at a different frequency \citep[for more details, see][]{offringa-wsclean-2014,offringa-wsclean-2017}\footnote{See \url{https://wsclean.readthedocs.io/en/latest/index.html}.}. We do not use these spectral images in this work nor perform spectral index analysis, hence we leave this for a future study. However, the final synthesised image, which is a result of an interpolation between these spectral images, contains the spectral information on it. Furthermore, spectral information was also considered during self-calibration with CASA, when the visibilities were imaged with the \texttt{mtmfs} deconvolver {\citep{rau2011}}. {Additionally, since our systems are near the field pointing centre, no direction-dependent corrections were applied to our data, and also, we did not apply primary beam corrections. The only source that contained a bright outlier source far from the pointing centre, that could affect imaging, was VV\,250, about 100". We imaged the entire area for this observation.}

\afterpage{%
\begin{landscape}
\begin{figure}
    \centering
    \includegraphics[width=0.36\textwidth,height=0.30\textwidth]{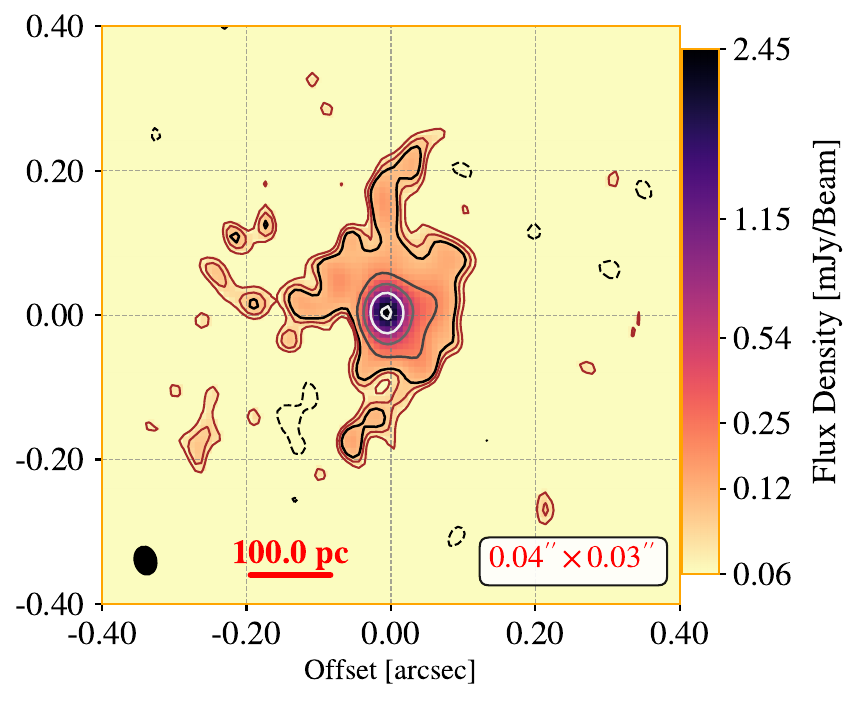}
    \includegraphics[width=0.30\textwidth,height=0.30\textwidth]{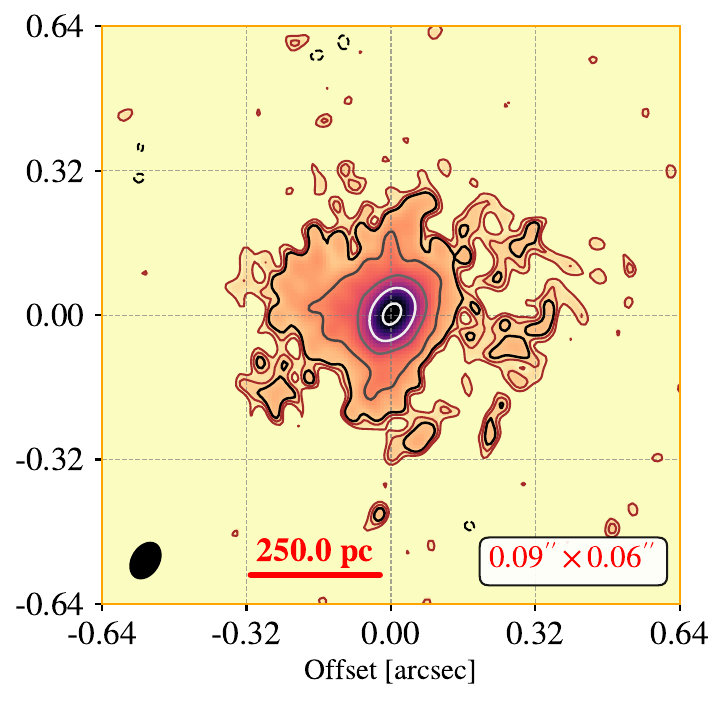}
    \includegraphics[width=0.30\textwidth,height=0.30\textwidth]{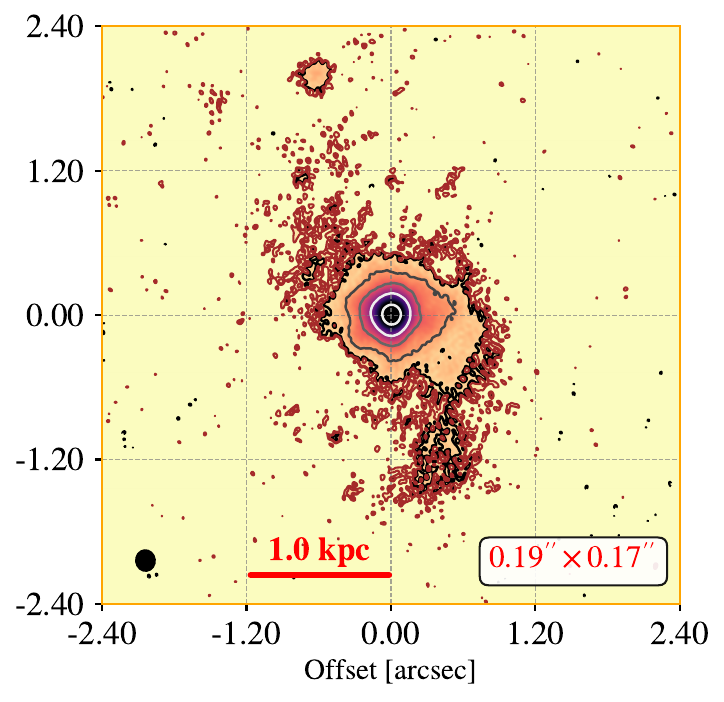}
    \includegraphics[width=0.30\textwidth,height=0.30\textwidth]{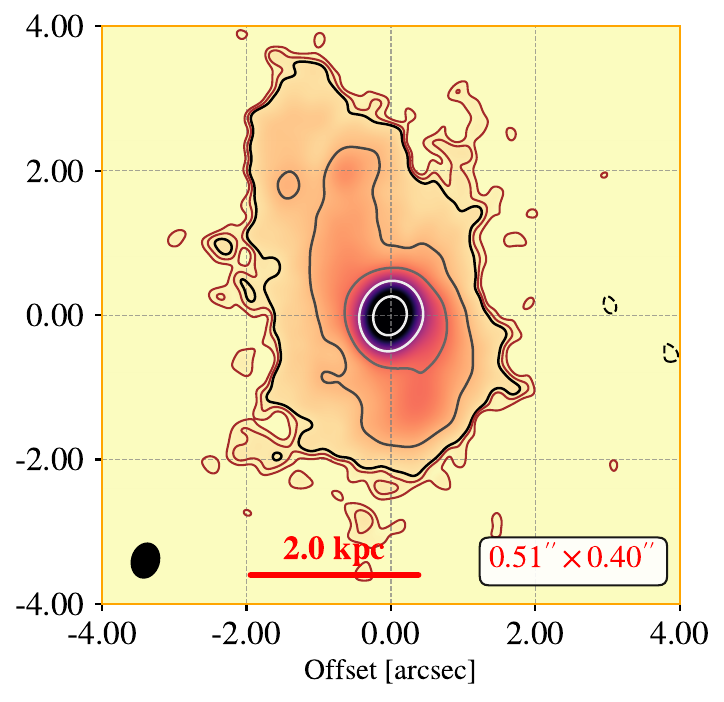}
    \includegraphics[width=0.36\textwidth,height=0.30\textwidth]{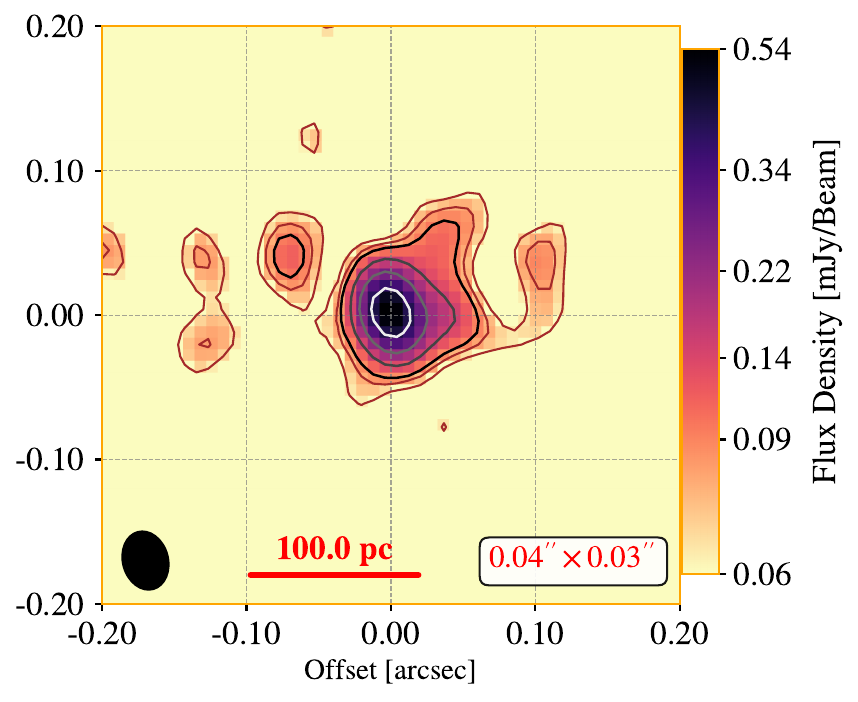}
    \includegraphics[width=0.30\textwidth,height=0.30\textwidth]{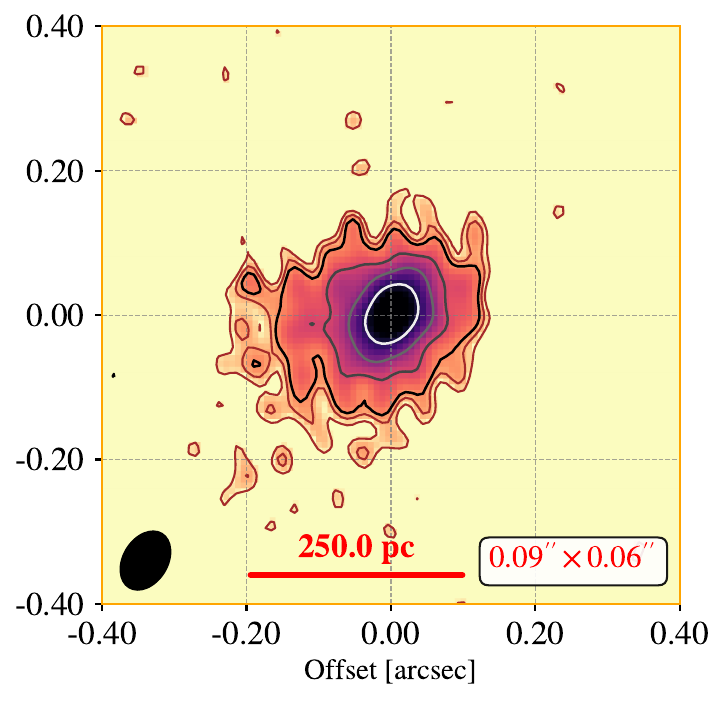}
    \includegraphics[width=0.30\textwidth,height=0.30\textwidth]{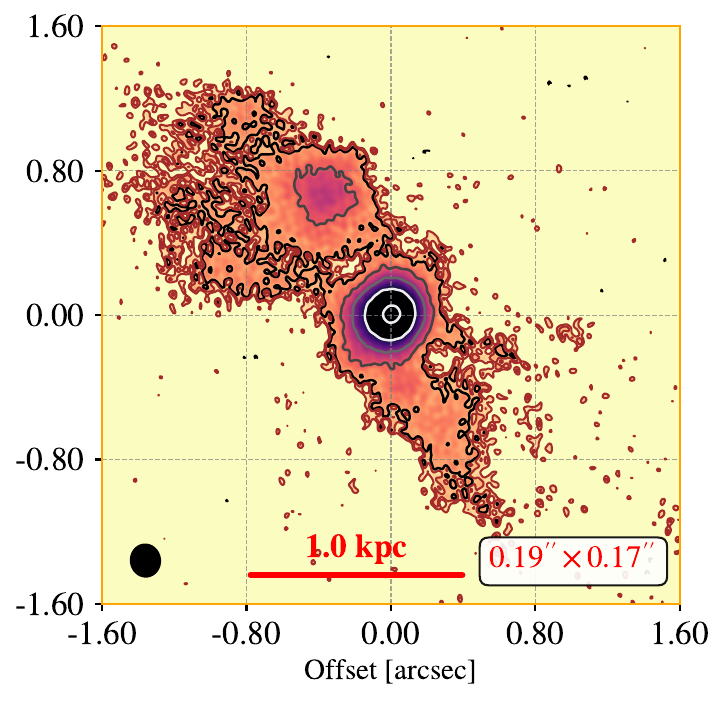}
    \includegraphics[width=0.30\textwidth,height=0.30\textwidth]{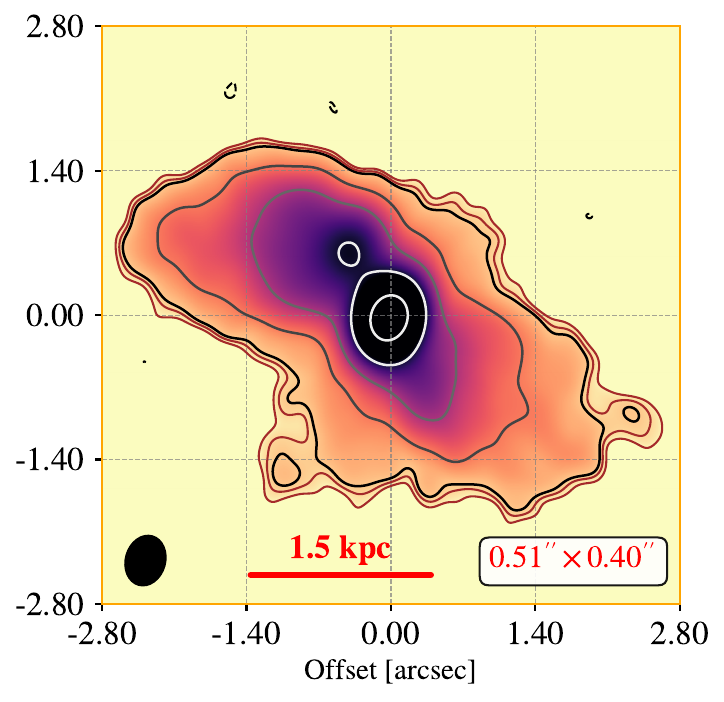}
    \includegraphics[width=0.36\textwidth,height=0.30\textwidth]{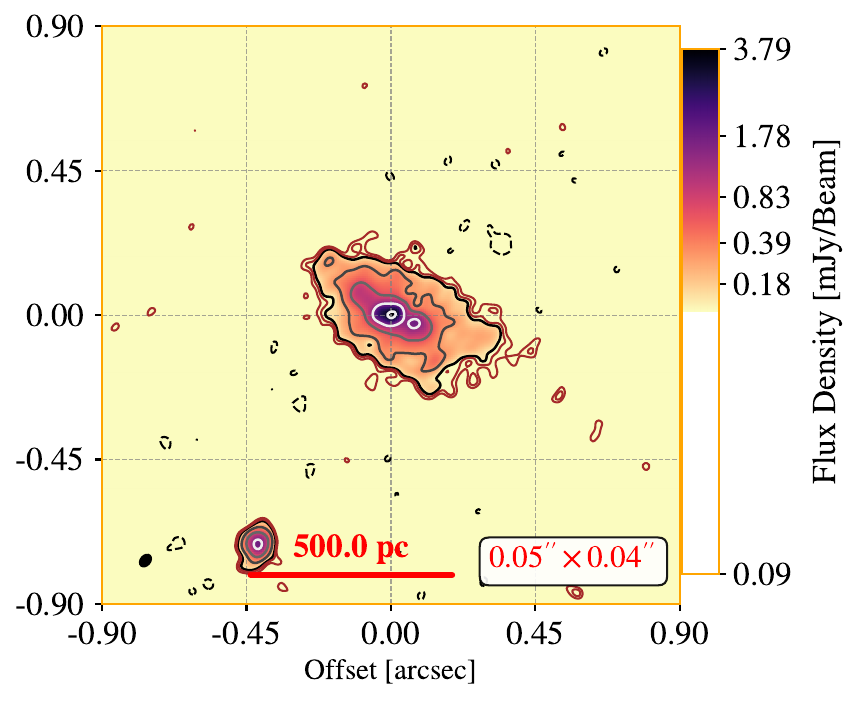}
    \includegraphics[width=0.30\textwidth,height=0.30\textwidth]{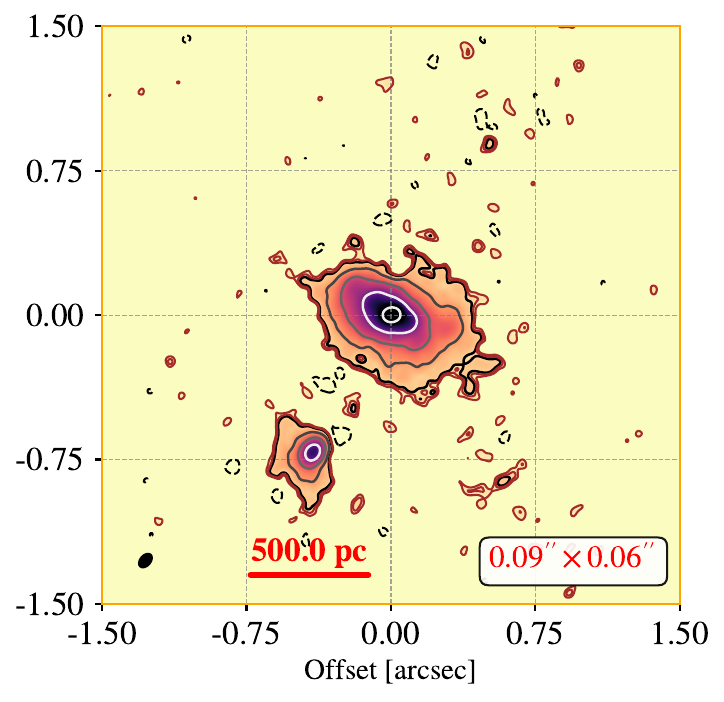}
    \includegraphics[width=0.30\textwidth,height=0.30\textwidth]{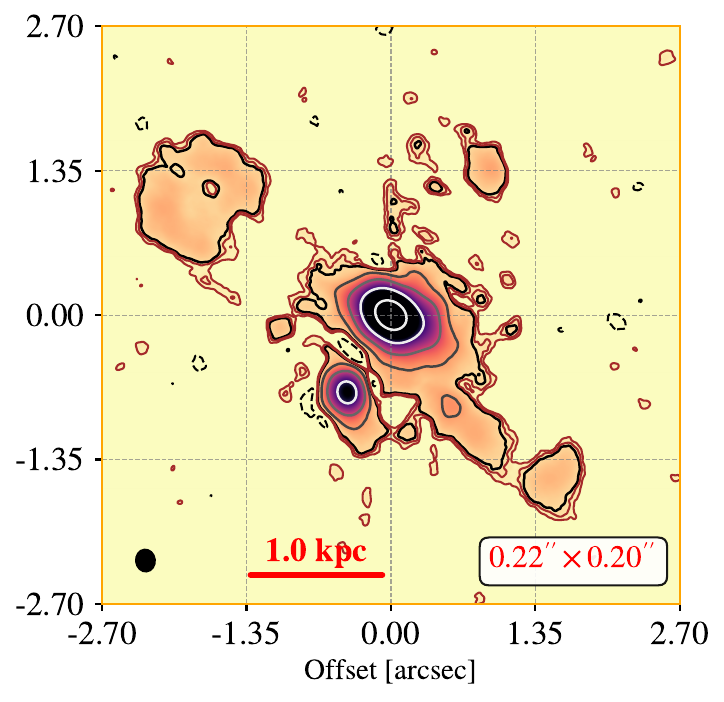}
    \includegraphics[width=0.30\textwidth,height=0.30\textwidth]{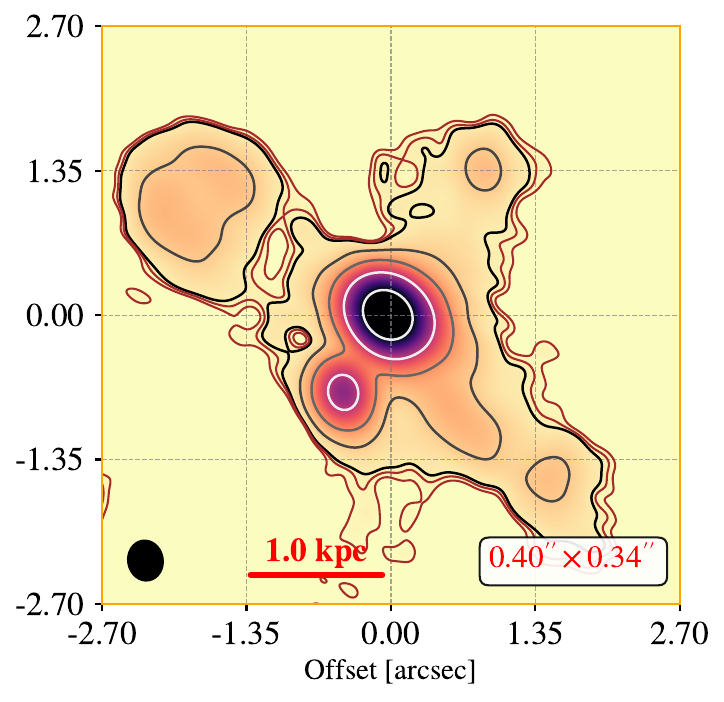}
    \caption{Radio emission from VV705 (VV\,705 N on top row and VV\,705 S on middle row) and UGC\,8696 (third row) obtained with\emph{e}-MERLIN and VLA interferometric data. 
        The first column on the left is a pure \emph{e}-MERLIN cleaned image, while the last on the right is a pure VLA cleaned image. The two images in between were created by using different weights during deconvolution. 
        The solid grey contours (white to black) are created automatically with a geometric progression in log scale, from the peak to $5\times \sigma_{\rm mad}$. The brown continuum contours represent the low level emission, at $4\sigma_{\rm mad}$ and $3\sigma_{\rm mad}$, respectively, and the black dashed lines are the negative contours at $-3\sigma_{\rm mad}$. The beam shape is represented by the black ellipse or additionally, in case of being too small, by the label (as $\theta_{\rm maj}\times \theta_{\rm min}$) in arcsec.
    }
    \label{fig:pre_results}
\end{figure}
\end{landscape}
\begin{landscape}
\begin{figure}
    \centering
    \includegraphics[width=0.36\textwidth,height=0.30\textwidth]{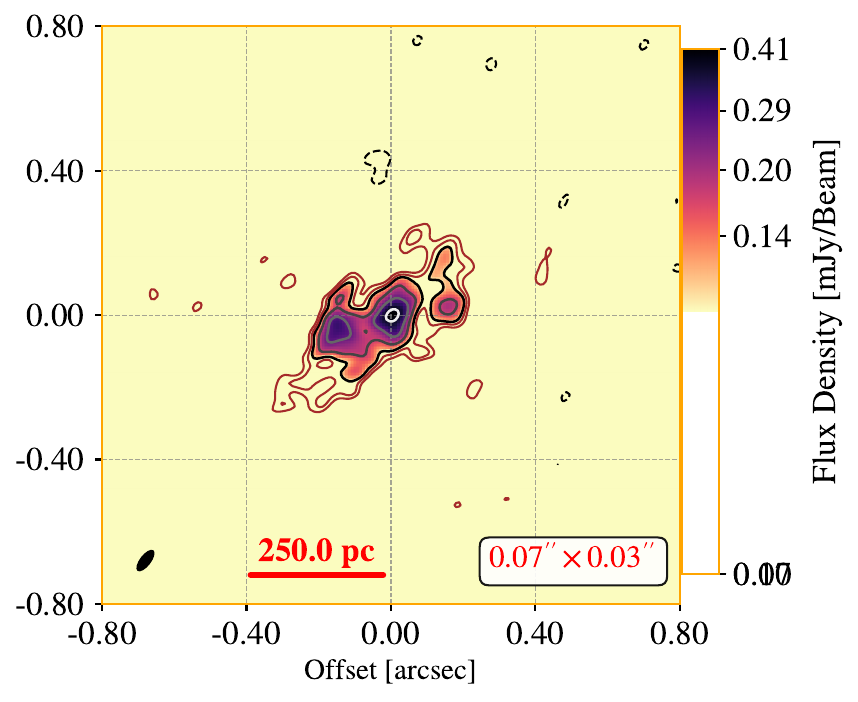}
    \includegraphics[width=0.30\textwidth,height=0.30\textwidth]{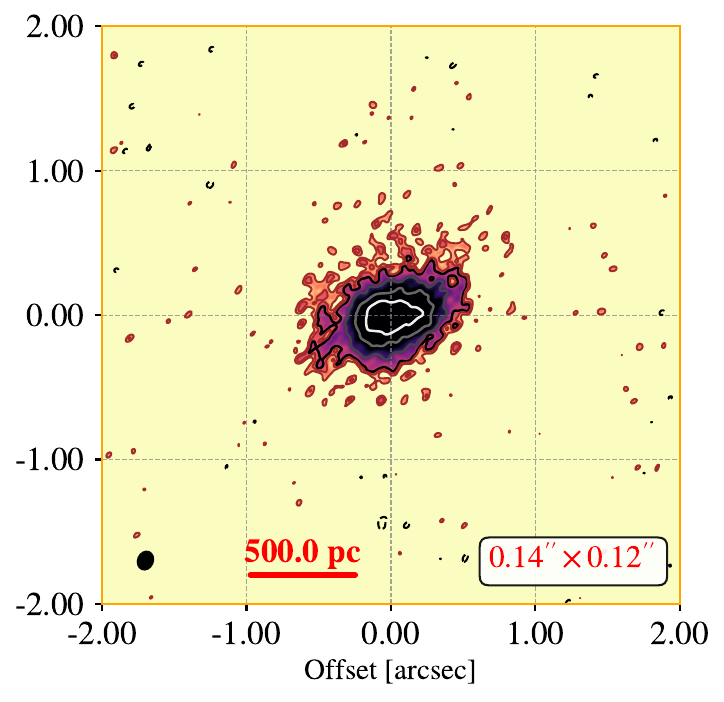}
    \includegraphics[width=0.30\textwidth,height=0.30\textwidth]{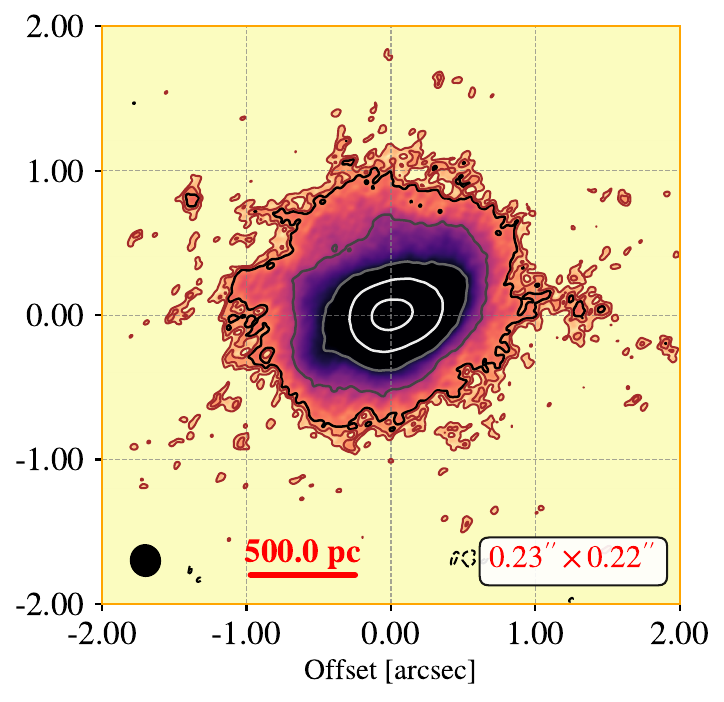}
    \includegraphics[width=0.30\textwidth,height=0.30\textwidth]{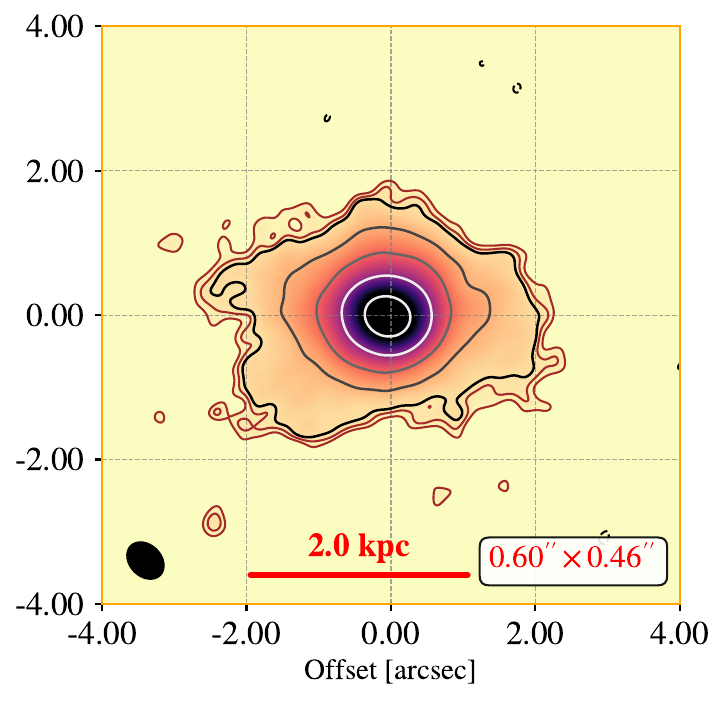}
    \includegraphics[width=0.36\textwidth,height=0.30\textwidth]{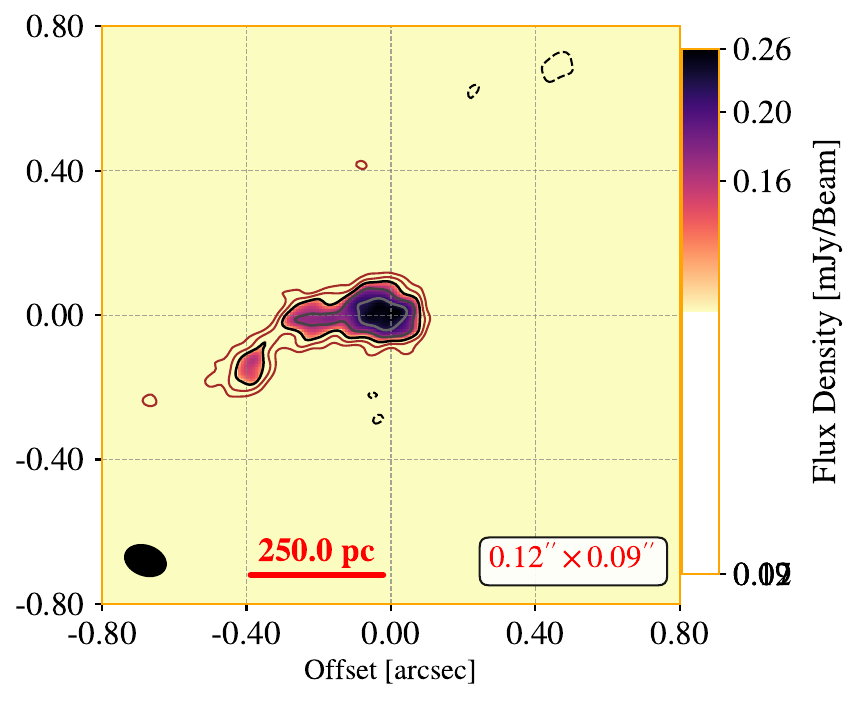}
    \includegraphics[width=0.30\textwidth,height=0.30\textwidth]{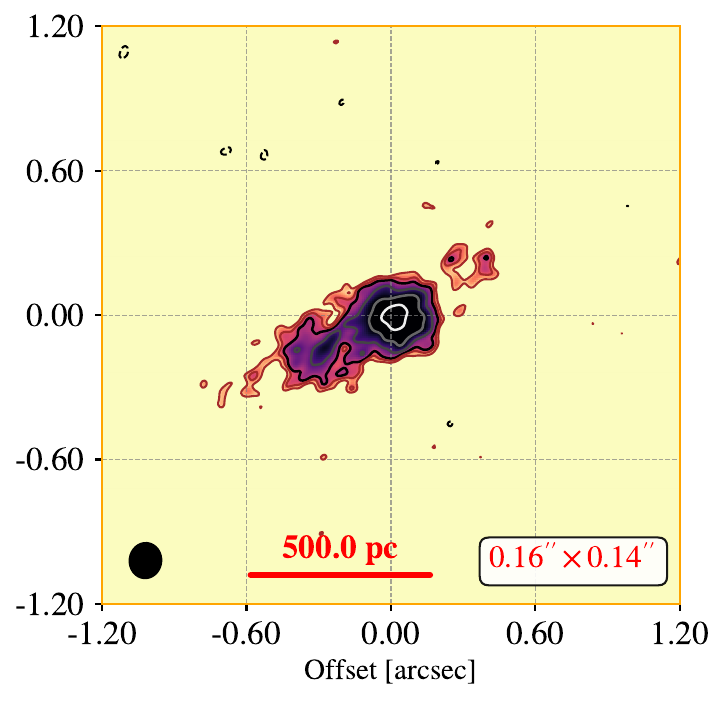}
    \includegraphics[width=0.30\textwidth,height=0.30\textwidth]{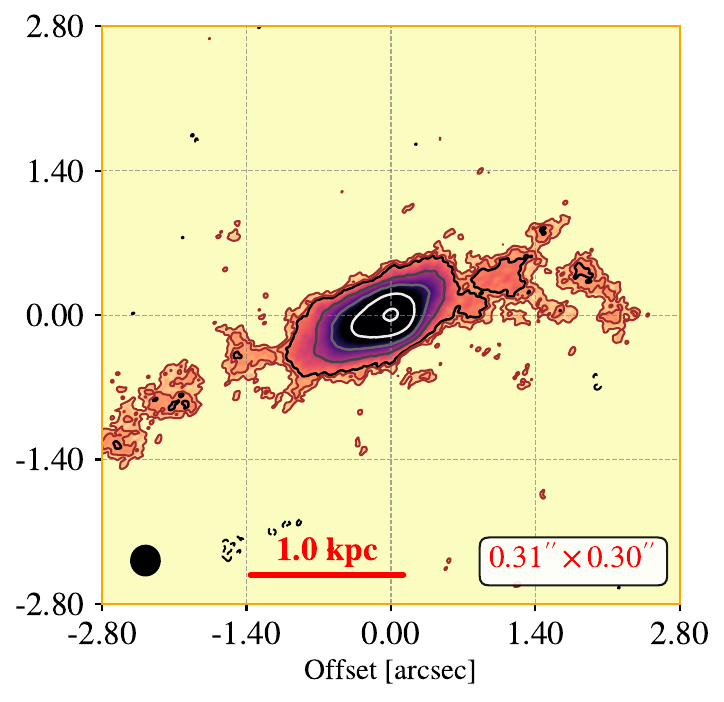}
    \includegraphics[width=0.30\textwidth,height=0.30\textwidth]{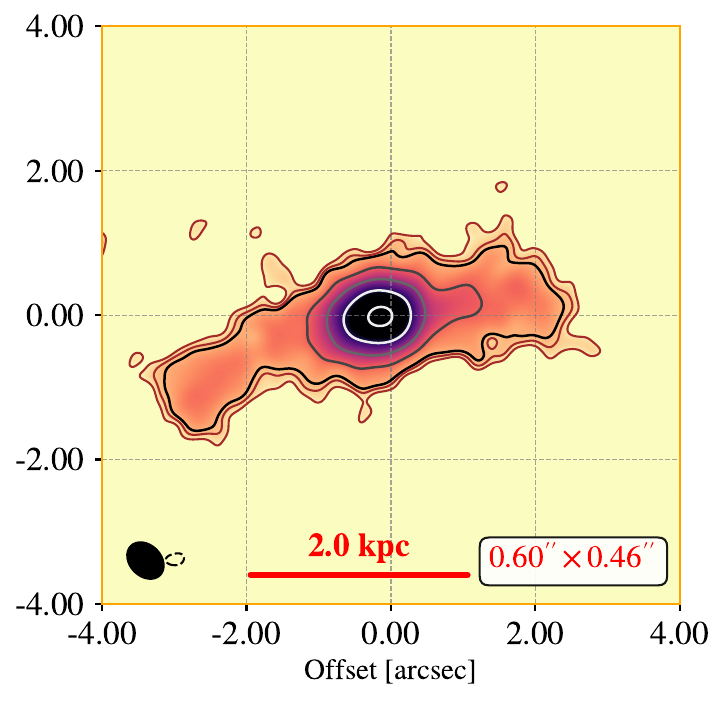}    
    \includegraphics[width=0.36\textwidth,height=0.30\textwidth]{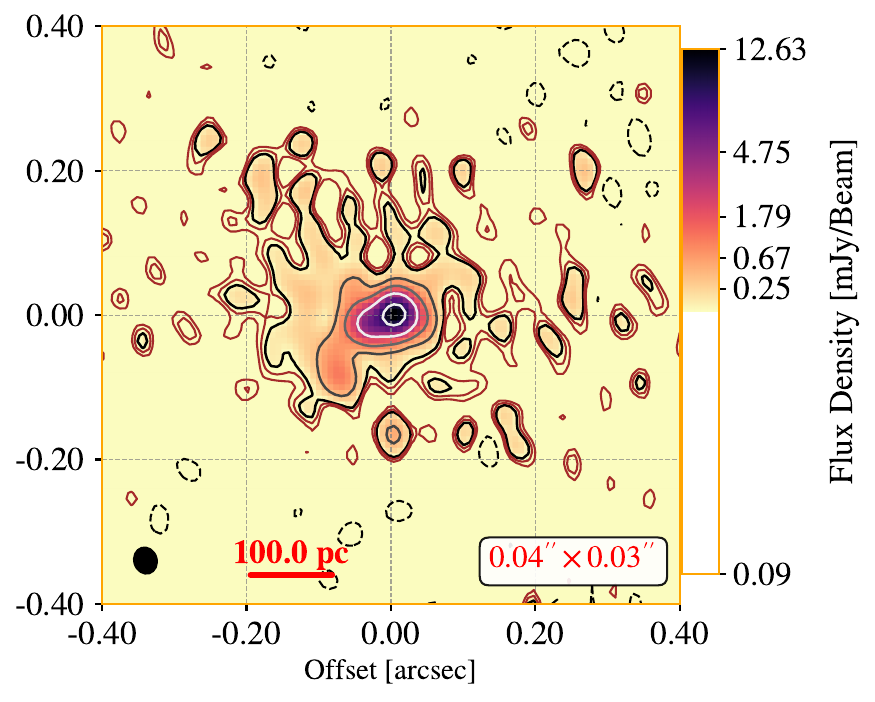}
    \includegraphics[width=0.30\textwidth,height=0.30\textwidth]{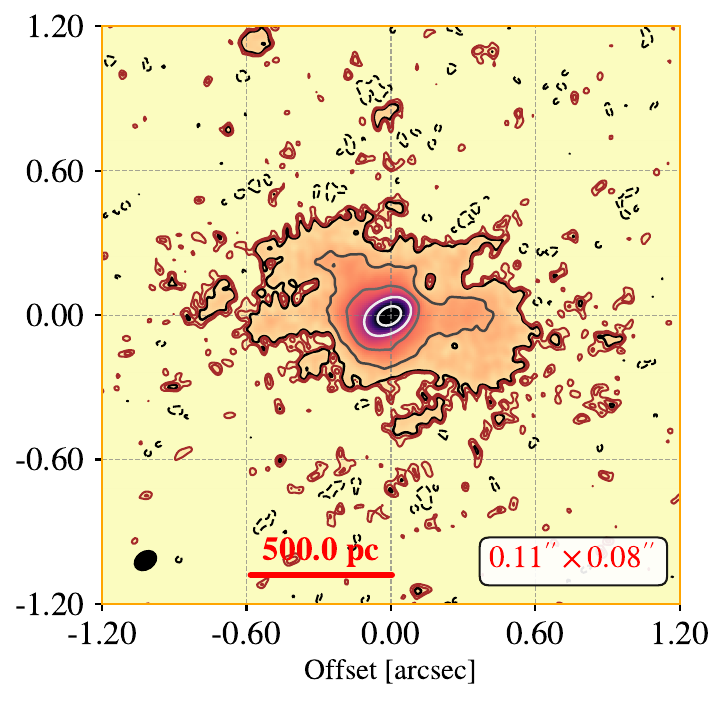}
    \includegraphics[width=0.30\textwidth,height=0.30\textwidth]{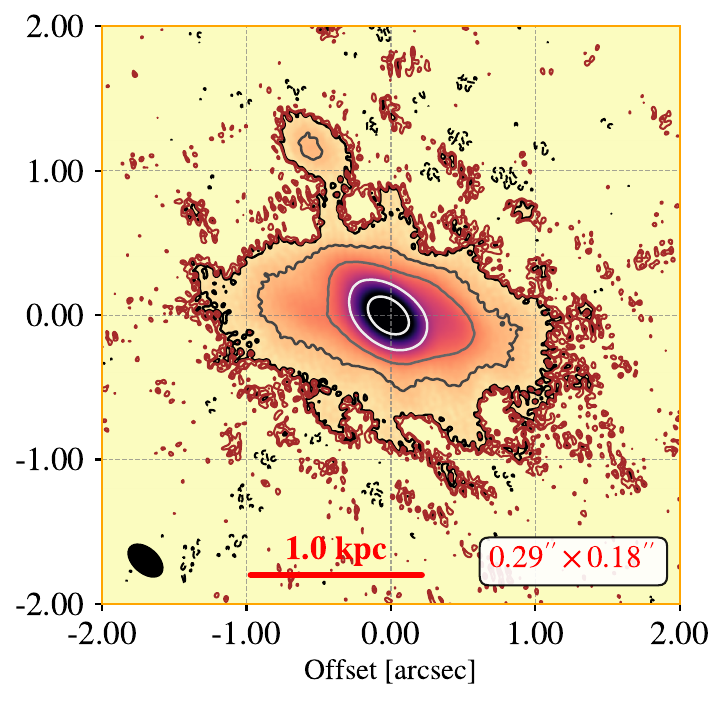}
    \includegraphics[width=0.30\textwidth,height=0.30\textwidth]{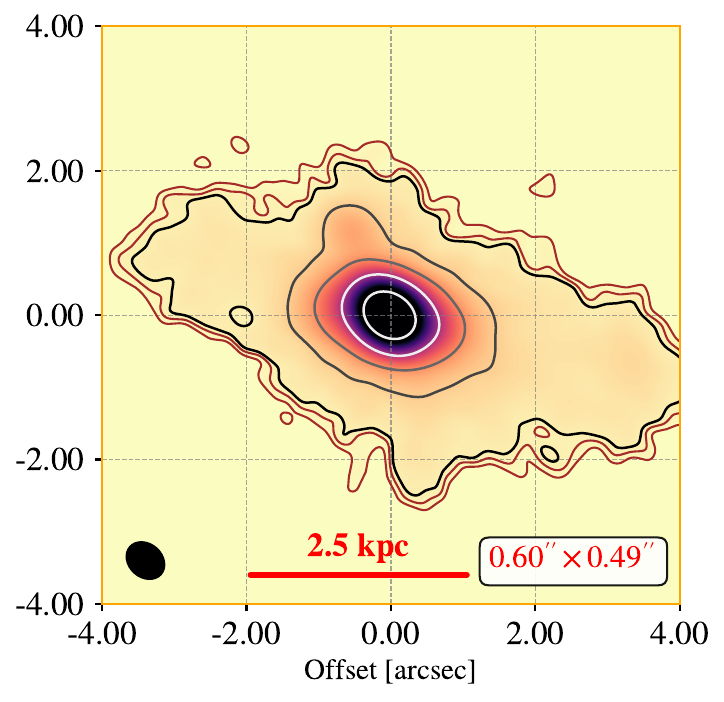}
    \caption{The same as Figure \ref{fig:pre_results} but for the other sources: from top to bottom rows, VV\,250 SE, VV\,250 NW and UGC\,5101. Notes: For VV\,250 NW, the \emph{e}-MERLIN emission was only recovered by using a sky taper of 0.025", with a robustness parameter of \texttt{2.0}.
    }
    \label{fig:results_cont_1}
\end{figure}
\end{landscape}
}

\subsection{Imaging Results: The Multiscale structure of Radio Emission}
\label{sec:pre_results}
In Fig.~\ref{fig:pre_results} we present a grid of multiscale images of VV\,705 (N and S components) and UGC\,8696 (all components), and in Fig.~\ref{fig:results_cont_1} of VV\,250 (SE and NW) and UGC\,5101. These images were produced by using different weighting schemes on combined interferometric data and also individual \emph{e}-MERLIN and VLA data. The first column on the left side displays the radio emission as seen by \emph{e}-MERLIN, by which we use as a probe of core-compact components and nuclear diffuse emission. On the right side, we have pure VLA images, which are the maps we use to quantify the radio emission that it is extended in nature (after removing the contamination from core-compact structures). The two middle images are the combined images (simply labelled by two different weights, $w_1$ and $w_2$) representing the radio emission on intermediate angular/linear scales. They capture the variation of the radio emission when balancing sensitivity against resolution. 

Each image has a different field of view, we decided to keep in this way for better representation of the radio emission on different scales. Therefore, note that we display distinct angular offsets in each panel, according to coordinates of the peak brightness listed in Tab.~\ref{tab:source_information}. The scale bar also informs the {brightness intensity} scale of the maps, distinct for each panel. The restoring beam is also indicated in the bottom of each panel (by a {black} ellipsis {alongside} of a label). Basic properties of these radio maps are listed in Tab.~\ref{tab:source_properties}. In each block property, the quantities are calculated for each image in the same order they appear in Figs.~\ref{fig:pre_results} and~\ref{fig:results_cont_1}: for \emph{e}-MERLIN images, combined images $w_1$, $w_2$, and VLA images.

{
The structure of these systems is distinct at every spatial scale, and valuable information can be obtained by using appropriate instruments that both resolve their nuclear regions and are sensitive enough to map the large-scale extended emission. Therefore, we require a methodology that is able to combine multiscale information and characterize their spatial structures, as we can see in Fig~.\ref{sec:pre_results}. Thus, it allows us to connect their physical and morphological properties such as component positions, sizes , associated flux densities and star-formation rates at each scale. 
}

\section{{Analysing the multiscale structure of radio emission}}
\label{sec:image_approach}
{The structure of the radio emission changes significantly across different linear scales (Figs.~\ref{fig:pre_results} and \ref{fig:results_cont_1}).}
A feasible way to differentiate the total radio power of the source in relation to core-compact components (i.e. AGN/SB nuclei) is to combine interferometric data and compare higher-resolution images (\emph{e}-MERLIN) with lower-resolution, sensitive radio maps (e.g. VLA). We see in these maps clearly that, as the sensitivity of large-scale structures increases, the diffuse emission becomes more evident, shaping the radio morphology of the source. Therefore, with combined data, our objectives are
to
{
\begin{enumerate}
    \item disentangle the core-compact flux from the extended flux in order to account for their contributions to the total flux density and star formation rates;
    \item determine the relative sizes of nuclear regions (core-compact structures, nuclear disks, etc.) and the sizes of large-scale extended components;
    \item and analyse the radio emission for each relevant sub-region/component individually.
\end{enumerate}
}
{
To achieve our goals, we use similar image fitting techniques from optical studies that disentangle structural components from imaging data. However, we recognise that radio emission is complex in nature and presents challenges to be characterised \citep[e.g.][]{Calabro2019b,Panessa2019, P_rez_Torres_2021}.  Decoupling the fraction of luminosity from core-compact structures such as AGN or SB regions in relation to the diffuse surrounding radiation, generated by intense star formation activity or radio jets, is nevertheless challenging \citep[e.g.][]{Mancuso2017,loreto2017,Magliocchetti2022}. However, by successfully doing so, one can characterise the effect and influence that the AGN/SB has over the physics of the source, and hence compute what is the total contribution to the radio power that SF has, from small to large physical scales. Additionally, we reiterate that in this work we do not distinguish AGN from SB components due to data limitation, since these two structures have similar angular sizes when unresolved, typically on scales $\lesssim$ 5-10 pc (unresolved by \emph{e}-MERLIN in our sample). 
}

\mchh{
With the novel fitting method presented here (see below), we can uncover physical information from the radio emission that was previously inaccessible. This process can be further refined by utilising multi-wavelength observations in conjunction with complementary $uv$ coverage. The full scope of our method's applicability extends beyond the objectives of this paper and will be addressed in future publications.
}

\subsection{Quantifying flux densities estimates}
\label{sec:mask_dilation}
\label{sec:flux_estimates}
To robustly estimate flux densities, we use a general approach based on a mask dilation technique \cite[morphological transformation,][]{Serra1984}. {Consider Fig.~\ref{fig:segmentation_decomposition} as an example.} Firstly, only a portion of the emission is selected, above a certain threshold level $\sigma$. Secondly, the mask is dilated to account for more low-level emission in the vicinity of that region. This allows to catch the faint diffuse emission encompassing the source. {To measure $\sigma$, we adopt a robust approach through the standard deviation based on the median absolute deviation} \citep[MAD;][]{astropy_2022}:
\begin{align}
\sigma_{\rm mad} &= \frac{\text{MAD}}{\Phi^{-1}(3/4)} \approx 1.48 \ \text{MAD},\ \ \  
\text{MAD} = \text{med}\left\{|h_i - \text{med}({h})|\right\},
\end{align}
where $\Phi^{-1}(3/4)$ is the normal inverse cumulative distribution with probability 3/4, and $h$ is a generic variable with discrete values $h_i$ and median ${\rm med}(h)$. We made a number of tests and a suitable mask is achieved when considering  emission within $6\times\sigma_{\rm mad}$. The transformation requires a dilation size which is naturally defined to be half of the averaged size of the restoring beam of the image, e.g. $\sqrt{\theta_{\rm maj} \times \theta_{\rm min}}$, in pixels. The total expansion of a region is a unity of the beam size. All contour levels on our images presented in this work are created using multiples of $\sigma_{\rm mad}$.


Flux density errors are estimated over the residual maps generated during cleaning with \textsc{wsclean}. The residual flux density is computed using the same mask that was discussed previously, by computing the sum of all pixels inside the mask. The error based on the root-mean-square of the residual map $\varepsilon_{\rm rms}$ within the area of emission is defined as 
\begin{align}
\label{eq:erro_flux}
\varepsilon_{\rm rms} = \sqrt{\sum_{ij} \left( R_{ij} * \texttt{mask}_{ij} - 
\langle R_{\texttt{mask}}\rangle \right)^2} \times \frac{\sqrt{\texttt{mask}_{\texttt{A}}}}{\mathfrak{B}_{\rm A}} \quad [{\rm Jy}]
\end{align}
with: $\texttt{mask}_{\texttt{A}}$ being the total dilated mask area (in pixels square) where the flux is computed; $R_{ij}$ is the pixel intensity value of the residual map at position $i,j$, in units of Jy/beam; $\langle R_{\texttt{mask}}\rangle$ is the mean of the residual inside the mask, in units of Jy/beam; and $\mathfrak{B}_{\rm A}$ is the restoring beam area, in pixel square, used to convert the sum to Jy. 

Results for flux densities and associated errors for the representative maps presented in Fig.~\ref{fig:pre_results} and Fig.~\ref{fig:results_cont_1} are provided in Tab.~\ref{tab:source_properties}. Results are shown separately for pure \emph{e}-MERLIN and VLA images, as well as combined images (denoted by weights $w_1$ and $w_2$). 


\subsection{Interferometric Decomposition}
\label{sec:interferometric_decomposition}
\begin{figure}
\centering
\includegraphics[width=0.99\linewidth]{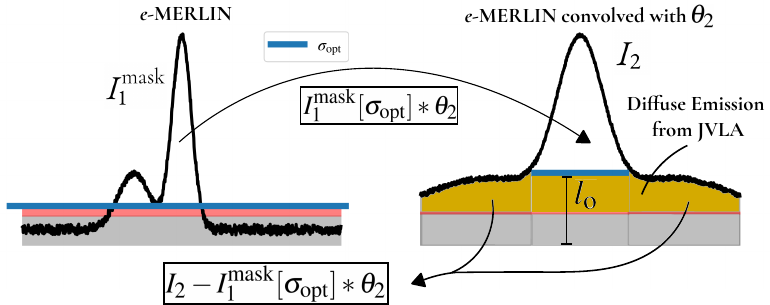}
\caption{A representation of how the interferometric decomposition works. 
    The black line represents the 1D slice of a surface brightness distribution: the left one is a pure \emph{e}-MERLIN and the right one is a combination between \emph{e}-MERLIN and VLA. The grey shaded area represents the background noise level, and the blue horizontal line is the selection of the optimised threshold. On the right, the offset $l_{\mathrm{o}}$ is linked to the amount of \emph{e}-MERLIN flux that should be convolved ($*$) with $\theta_2$ to produce a map having the large-scale diffuse emission only, which is represented by the orange shaded area. Therefore, the optimisation works in the following way: on the left, the blue horizontal line is moved up and down to put the blue horizontal line on the right on the same level (or at least close) to the level of the diffuse emission. The red line on the left highlights the $5\sigma$ minimum limit that we assume during the minimisation so that noise/uncertain flux is not included.}
\label{fig:int_decomposition}
\end{figure}

\begin{figure}
\centering
\includegraphics[width=0.99\linewidth]{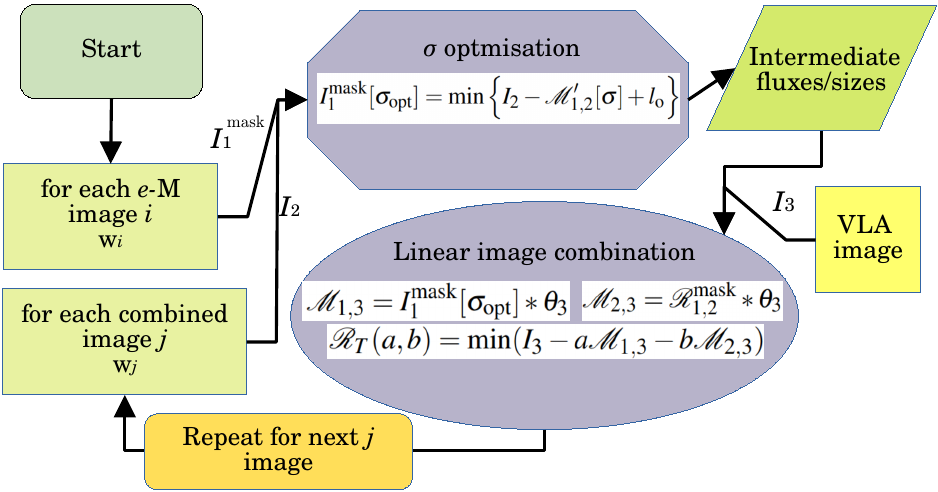}
\caption{{Flowchart showing the iterative process of the interferometric decomposition. See also Fig.~\ref{fig:int_decomposition} and Fig.~\ref{fig:example_sub}.
}}
\label{fig:int_decomposition_flowchart}
\end{figure}

\begin{figure*}
\centering
\includegraphics[width=0.95\linewidth]{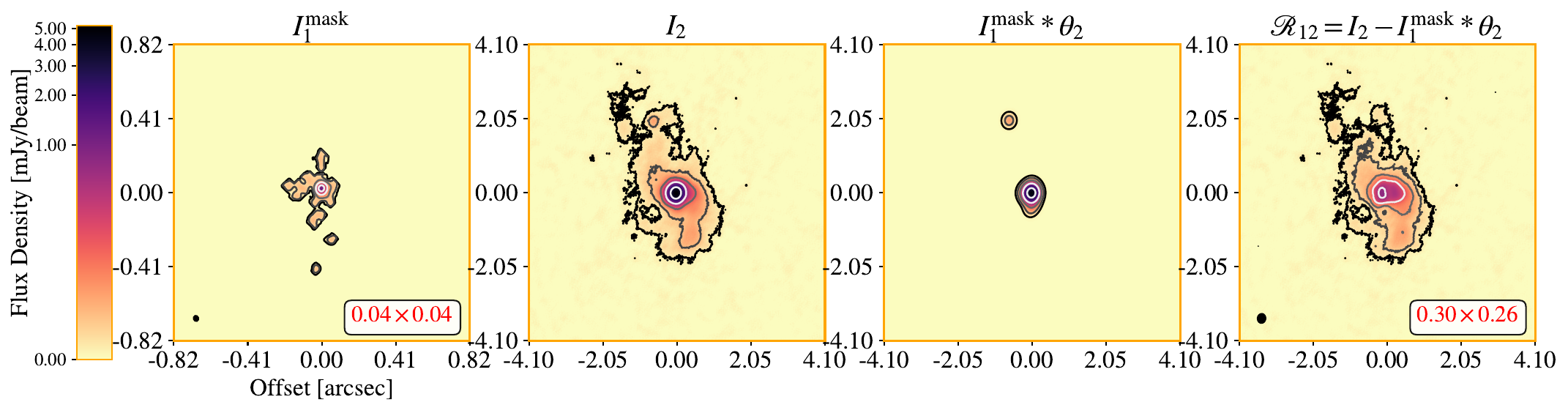}
\includegraphics[width=0.95\linewidth]{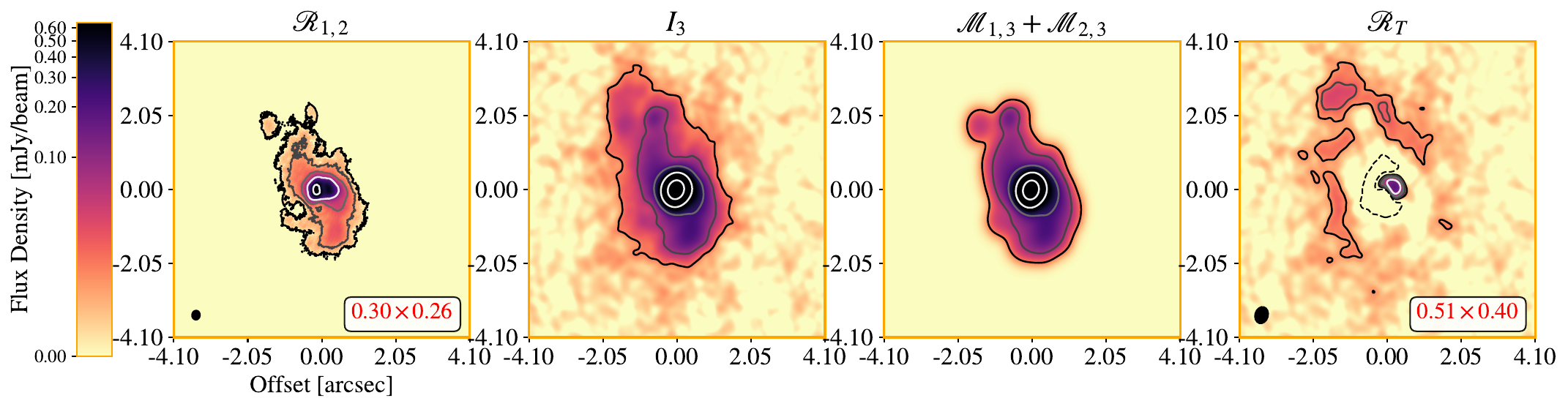}
\caption{Multiscale interferometric decomposition to disentangle the nuclear \emph{e}-MERLIN emission from the VLA extended emission. Top row: We have two main images, $I_1^{\rm mask}$ on the left, which is the \emph{e}-MERLIN image of VV\,705 N (Eq.~\ref{eq:I1_mask_opt}) with the optimised threshold, and $I_2$ which is a combined image between VLA and \emph{e}-MERLIN. Hence, the quantity $\mathcal{M}_{1,2} = I_{1}^{\rm mask} * \theta_2$ (Eq.~\ref{eq:M12}) is removed from $I_2$ in order to obtain the extended residual emission $\mathcal{R}_{1,2}$ (Eq.~\ref{eq:R12} on the combined image. See Fig.~\ref{fig:int_decomposition}. Bottom row: {Last step} of the method (Eq.~\ref{eq:RT}), where $\mathcal{R}_{1,2}$ is taken to be the emission to be convolved with a pure VLA beam $\theta_3$ from $I_3$, which results in the map in the third lower panel. Then, $I_3$ is subtracted from that map, providing the total residual $\mathcal{R}_T$. For this example, the values of the linear combination parameters in Eq.~\ref{eq:RT} are: $a = $ 0.73, $b = $ 1.02.}
\label{fig:example_sub}
\end{figure*}

Consider Figs.~\ref{fig:int_decomposition},~\ref{fig:int_decomposition_flowchart} and~\ref{fig:example_sub}, and let $I_1$ be a pure \emph{e}-MERLIN image,  $I_2$ a combined cleaned image between VLA and \emph{e}-MERLIN, and $I_3$ a pure VLA image. Also, consider that the restoring beams for each image are $\theta_1$, $\theta_2$, and $\theta_3$, respectively, where $\theta_1<\theta_2<\theta_3$.
{
Whilst the emission is being observed by a larger beam ($\theta_2$ or $\theta_3$), we can still simulate how the higher resolution emission ($\theta_1$) downscales by convolving it with $\theta_2$ (or $\theta_3$).}
{To estimate the fraction of extended emission in a lower resolution map in relation to \emph{e}-MERLIN, one can convolve $I_1$ with $\theta_2$ and remove the result from $I_2$ (see Fig.~\ref{fig:int_decomposition}). Mathematically, this is expressed as
\begin{align}
\label{eq:M12}
\mathcal{M}_{1,2} &= I_{1}^{\text{mask}}[\sigma_{\rm opt}] * \theta_2,\\
\label{eq:R12}
\mathcal{R}_{1,2} &= I_2 - \mathcal{M}_{1,2}.
\end{align}
In the above equation, ``$*$'' stands for a convolution, and $\mathcal{R}_{1,2}$ is a residual map containing the extended emission on intermediate scales. The quantity $\mathcal{M}_{1,2}$ is the convolution result of a masked \emph{e}-MERLIN emission with $\theta_2$. This involves an optimisation process to select only trusted emission, so that noisy information is not propagated over the convolution:
\begin{align}
\label{eq:I1_mask_opt}
I_1^{\text{mask}}[\sigma_{\rm opt}]= \min \left\{ I_2 - \mathcal{M}_{1,2}[\sigma]+l_{\rm o} \right\}.
\end{align}
In the above equation, $\sigma_{\rm opt}$ is the optimised standard deviation level of the \emph{e}-MERLIN emission that is selected to be convolved with $\theta_2$ (third upper panel in Fig.~\ref{fig:example_sub}) and removed from the lower resolution map, minimising negative residuals (fourth upper panel in Fig.~\ref{fig:example_sub}) through the offset parameter $l_{\rm o}$. The emission level after subtraction is controlled with $l_o$ (see Fig.~\ref{fig:int_decomposition}), and a good compromise for convergence is $l_{\rm o} = \frac{1}{2}\times \text{std}(I_2^{\rm mask})$. For this case, we must use the traditional standard deviation, which takes into account the relative amplitude of the signal. We also impose a minimum limit to $\sigma$, $\sigma_{\min} = 5.0$, to avoid including significant low-level emission that is uncertain or noisy. A chart of these steps is presented in Fig.~\ref{fig:int_decomposition_flowchart}. It is an iterative procedure that can be performed, in principle, by any pair of interferometric images with distinct resolutions, matching baselines and sensitivities. 
}

Now, consider the pure VLA image $I_3$ (where $\theta_3>\theta_2$) and again the flowchart in Fig.~\ref{fig:int_decomposition_flowchart}. Since $I_2$ (and $\mathcal{R}_{1,2}$) are at an intermediate resolution, containing partial information of the extended emission, they are not capturing the full size of the emission. We can apply the same procedure as previously but taking $\mathcal{R}_{1,2}$ as the high-resolution emission to be convolved with the restoring beam of $I_3$. This example is given in the lower left panel of Fig.~\ref{fig:example_sub}.

The simplest case of a composite radio emission is defined by a compact/unresolved and an extended component (in a generic scale). When modelling these, one can assume that the total model is a linear combination between the extended part with the compact/unresolved parts\footnote{The same principle is applied when modelling a complex radio structure with multi-Gaussian or multi-S\'ersic fitting. Because the total model is simply the linear combination of all individual model subcomponents, see Sec.~\ref{sec:image_fitting}.}. {In this way, we can write
\begin{align}
\label{eq:RT}
\mathcal{R}_T(a,b) = \min(I_3 - a \mathcal{M}_{1,3} - b \mathcal{M}_{2,3}).
\end{align}
where $a$ and $b$ are the coefficients of the linear combination used to minimize the residual on the pure VLA emission. $\mathcal{M}_{1,3}$ and $\mathcal{M}_{2,3}$ have similar definitions as $\mathcal{M}_{1,2}$:
\begin{align}
\label{eq:M13_M23}
\mathcal{M}_{1,3} = I_{1}^{\text{mask}}[\sigma_{\rm opt}] * \theta_3, \quad
\mathcal{M}_{2,3} = \mathcal{R}_{1,2}^{\rm mask} * \theta_3.
\end{align}
The quantity $\mathcal{R}_T$} represents the total VLA residual map of the {emission}, which can still contain significant emission, depending on which previous combined image ($I_2$) was used to perform the decomposition. Still, we can use this extended residual emission to correct for the total flux density which was not printed in $\mathcal{R}_{1,2}$, therefore getting the full morphology of the most extended emission (see $\mathcal{R}_T$ in the lower right panel of Fig.~\ref{fig:example_sub}). Further, {we note} that $\mathcal{R}_T(a,b) \sim $ 0 when $\theta_2 \sim \theta_3$. 

To deal with any bias caused by the choice of which image we use to do this decomposition, we use the continuous set of images between \emph{e}-MERLIN and VLA, balancing the trade-off between angular resolution and sensitivity to diffuse emission.

We do this for each \emph{e}-MERLIN image $I_1$, iterating over the combined images $I_2$ mapping intermediate scales in the diffuse and core-compact parts in comparison with a pure VLA image $I_3$ (see again Fig.~\ref{fig:int_decomposition_flowchart}). 
{Complementing to Fig.~\ref{fig:example_sub}, we display in Fig.}~\ref{fig:int_decomposition_radial_profiles} the associated azimuthally averaged intensity profiles of the previously discussed images. We discuss the results of this method in Sec.~\ref{sec:multi_scale_properties}.

\subsection{General S\'ersic Fitting}
\label{sec:image_fitting}
Modelling the brightness distribution in galaxy astronomical images in order to characterise individual components is a widely adopted technique: in optical studies \citep{Simard_2002,de_Souza_2004,Peng_2010,DeJong2004,Gao_2017,Ferrari_2015,lucatelli2019}, infrared \citep{Gadotti2007}, X-ray \citep{Iwasawa2011} and radio \citep{loreto2015,Hodge2019,Song2021}.  The number of mathematical models is diversified but usually confined to a group of three main functions: Gaussian \citep{Condon_1997,Condon_1998} and multi-Gaussian fitting \citep[e.g. PyBDSF][]{pybdsf_ref,Calabro2019}; exponential functions (describing the brightness distribution of disks of spiral and lenticular galaxies) \citep{Freeman1970}; and the S\'ersic law \citep{sersic1963,caon1993,Ciotti1999A}, a generalisation that can reproduce the first two by introducing the S\'ersic index $n$. 

In the majority of radio sources, the surface brightness distribution of unresolved structures (e.g. nuclear region, blobs of star formation, etc) follow a Gaussian shape, resembling the Gaussian beam shape. Hence, it is common to perform image decomposition with this set of Gaussian basis functions. Additionally, extended emission can be approximated by exponential disk-like distributions \citep[e.g.][]{Murphy_2017}. {All these cases can be expressed in terms of a combination of S\'ersic functions.}

\subsubsection{Implementation}
For simplicity, for a one-dimensional function, the alternative S\'ersic law $\mathfrak{S}$ is given by \citep[e.g.][]{caon1993}
\begin{align}
\label{eq:sersic_law}
\mathfrak{S}(n,R) = I_n \exp\left\{-b_n\left[\left(\frac{R}{R_n}\right)^{1/n} \right]-1\right\},\qquad  b_n \approx 2n - \frac{1}{3}
\end{align}
where $R$ is the projected radius on the image plane; $R_n$ is known as the effective radius, the radial distance that contains half of the total integrated luminosity or flux density $S_\nu$: $\mathfrak{S}(n,R = R_n) = S_\nu/2$; and $I_n$ is the intensity at $R = R_n$, i.e. $I_n = \mathfrak{S}(n,R=R_n)$.  

{
By using the S\'ersic profile, we can robustly model the radio emission with distinct components. We have implemented the minimisation in a general way, however we do not force arbitrary S\'ersic indexes $n$ during the fitting. Most radio sources are well described by elliptical Gaussian functions \citep{Condon_1998}. With Eq.~\ref{eq:sersic_law}, a Gaussian distribution is recovered when $n=0.5$ and a disk distribution when $n=1$. }

For the radial distance in a 2D space, a generalised ellipse is described by a radial geometrical grid of the form \citep[e.g.][]{Peng_2010}
\begin{align}
\label{eq:general_ell}
R_G = \left( 
|x - x_0|^{C + 2} + \left| \frac{y - y_0}{q}\right|^{C+2}
\right)^{\frac{1}{C+2}}
\end{align}
where: $C$ is a parameter that controls how round or boxy the ellipse is. For a usual ellipse, $C=0$, and we use this value in the current work; $q$ is the factor that deforms the ellipse in relation to its semi-major $R_a$ and minor $R_b$ axes (e.g. $q = R_b/R_a$). This parameter is controlled by the aspect ratio of the radial grid; $x_0$ and $y_0$ are the origin of the radial grid {(which will correspond to the centre of each component). To obtain the position angle, rotations are performed on the radial grid $R_G$ for each model component. Hence, in Eq.~\ref{eq:sersic_law}}, the one-dimensional radius $R$ is simply replaced by $R_G$. 

In Fig.~\ref{fig:cartoon_ulirg_structure}, we show a cartoon that clarifies the scales we aim to perform the image decomposition. We use that cartoon as a guide to our ideas for the next sections. More details of how the modelling of the radio emission is performed are given in the {App.~\ref{app:multi_sersic} and in the GitHub repository \textsc{Morphen}\footnote{\url{https://github.com/lucatelli/morphen}.}. We also show a description of how we perform source extraction from the data, using \texttt{SEP} \citep{Bertin1996,barbary2016} and \textsc{PetroFit} \citep{Geda_2022}, so that the initial conditions for each model component during the fitting are constrained with the data itself.} An example of the application of this technique is presented in Fig.~\ref{fig:example_fitting_example_VV705N}, {where we demonstrate} the multi-component detection and image decomposition of the radio emission.

\subsection{Image Shape Analysis}
\label{sec:image_shape_analysis}

Our primary objectives are to disentangle the properties of the nuclear region in terms of the emission that may be linked to an AGN/SB and the emission that is due to the nuclear diffuse and large-scale extended emission. {The latter is interpreted to be related to star formation, but we do not distinguish between AGN and SB emission, where SB is also a SF process. To quantify the radio emission,} we perform a series of global morphometric measurements on all continuum images, and also quantify properties of each individual model components fitted to our continuum images. The strategy is summarised below:
\begin{enumerate}
\item From the flux growth curve (Fig.~\ref{fig:segmentation_decomposition}), we estimate the effective area/region enclosing half of the total flux density, $A_{50}$. This is a conservative size \citep[as it is done in][]{loreto2017} but provides a “safe” characterisation of the source dimension in the most energetic region. However, for sources that have both a diffuse and a core-compact or unresolved components, the $A_{50}$ region has more intrinsic information about the core-compact part itself. That is why we require an alternative description for the diffuse size (see below).
\item We calculate the total area/region of the radio map, which we assume to be the one that encloses $95\%$ of the total integrated flux density, $A_{95}$. Since we are using the masking dilation, the $95\%$ area represents a good indicator of the total emission area.
\item From these previous areas, we infer the equivalent averaged circular radii using $A = \pi R^2$ $\to$ $R_{50}=\sqrt{A_{50}/\pi}$ and $R_{95}=\sqrt{A_{95}/\pi}$ (see Sec.~\ref{sec:limitations} for limitations). 
\item When separating the diffuse emission (nuclear and large-scale) from the core-compact emission using the image fitting approach (see Fig.~\ref{fig:example_fitting_example_VV705N}), we recover the convolved and deconvolved model components, see Eqs.~\ref{eq:deconvolved_mini} and~\ref{eq:convolved_mini}. From that, we compute $A_{50}$ and $A_{95}$ (as well the radii $R_{50}$ and $R_{95}$) for the deconvolved model components, respectively $A_{50,\rm d}$, $A_{95,\rm d}$, $R_{50,\rm d}$ and $R_{95,\rm d}$. 
\item Using the deconvolved quantities, brightness temperatures are calculated for the main core-compact structure (see Sec.~\ref{sec:brightness_temperatures}).
\item Computation of star formation rates (SFR) and surface density star formation rates $\Sigma_{\rm SFR}$. We measure these quantities in two regions: i) in the nuclear region (within the extended total nuclear area $A_{95,\rm d}^{\text{ext-nuc}}$) after the AGN/SB contribution is subtracted; ii) the total multiscale SFR and $\Sigma_{\rm SFR}$, which we take as the emission enclosed within the half-flux area $A_{50}$. Details of these calculations are given in Sec.~\ref{sec:SFR}.
\end{enumerate}

\section{Results}
\label{sec:results}

Throughout the results and discussion sections, we will frequently refer to distances in the image plane between radio components, which are actually projected distances. But for simplicity, we just refer to them as distances, unless said otherwise. We also do not apply corrections factors for estimated sizes due to the inclination of our sources \citep[e.g.][]{loreto2015}. {For simplicity,} we assume that the image-projected distances/sizes and areas are a good representation of the source structure.

\subsection{Global Sizes of the Radio Emission}
\label{sec:sizes_of_emission}
Before applying any decomposition to the radio emission, {we summarise} in Tab.~\ref{tab:source_global_properties} the main global properties of our sources derived from all continuum images. 
{
To capture the variation of the emission morphology across scales, we measure image properties for three sets of images: combined images, pure \emph{e}-MERLIN and pure VLA images. To accommodate the value of a generic quantity $Q$ measured in these set of images with a single notation, we adopt the following format to display the measurements:}
\begin{align}
\label{eq:measured_quantities}
\text{measured quantity } Q = \langle Q_{\rm VLA + \emph{e}M}\rangle^{\langle Q_{\rm VLA}\rangle}_{\langle Q_{\rm \emph{e}M}\rangle}
\end{align}
where:  $\langle Q_{\rm VLA}\rangle$ represents a quantity measured in a set of pure VLA images and $\langle Q_{\rm \emph{e}M}\rangle$ in a set of pure \emph{e}-MERLIN images; $\langle Q_{\rm VLA + \emph{e}M}\rangle$ is the intermediate (or nominal) value of a quantity measured in a set of combined interferometric images. In the following discussions, we refer to this mixed case as the ``intermediate'' measurement. 
{By quantifying properties in this way,} we avoid image selection bias and also minimise the effects of errors associated with low signal-to-noise images, capturing the variance between common linear scales. We use this notation to display measurements in Tabs.~\ref{tab:source_global_properties} $\sim$  \ref{tab:sizes_image_fitting}. 

\begin{table}
\centering
\caption{Derived global source properties from our imaging data. For a given quantity $Q$, we express the results as $\langle Q_{\rm VLA + \emph{e}M}\rangle^{\langle Q_{\rm VLA}\rangle}_{\langle Q_{\rm \emph{e}M}\rangle}$, where upper indices refer to pure VLA images while lower indices to pure \emph{e}-MERLIN images. 
The {intermediate (nominal)} value was determined in images from combined interferometric data.}
\label{tab:source_global_properties}
\begin{subtable}[h]{0.49\textwidth}
    \centering
    \begin{tabular}{l@{\hspace{5pt}}c@{\hspace{5pt}}c@{\hspace{5pt}}c@{\hspace{5pt}}c@{\hspace{5pt}}} 
        \hline
        Galaxy Name    & $\langle S_{\rm peak} \rangle$ [mJy/beam]             
        &  $\langle R_{50}\rangle$ [pc]          
        & $\langle R_{95}\rangle$ [pc]           
        & $R_{\max}^{\rm VLA}$ [pc]         \\
        {\hspace*{0.5cm} (1)}            & (2)                           &  (3)                    & (4)                       &  (5)       \\[0.3ex] \hline \rule{0pt}{1.0em}
        VV\,705 N      &  3.7$_{\ 2.4}^{\ 5.6}$        &  109$_{\ 27}^{\ 235}$   & 450$_{\ 76}^{\ 1147}$     &  1805                       \\[0.3ex] \hline \rule{0pt}{1.0em} 
        VV\,705 S      &  1.0$_{\ 0.5}^{\ 1.7}$        &  148$_{\ 19}^{\ 384}$   & 404$_{\ 39}^{\ 1073}$     &  1686                       \\[0.3ex] \hline \rule{0pt}{1.0em} 
        UGC\,5101      &  28.2$_{\ 14.8}^{\ 37.6}$     &  132$_{\ 41}^{\ 214}$   & 548$_{\ 175}^{\ 801}$     &  1775                       \\[0.3ex] \hline \rule{0pt}{1.0em} 
        UGC\,8696      &  17.4$_{\ 5.9}^{\ 22.4}$      &  156$_{\ 87}^{\ 180}$   & 622$_{\ 206}^{\ 758}$     &  1464                       \\[0.3ex] \hline \rule{0pt}{1.0em}
        UGC\,8696 N    &  17.4$_{\ 5.9}^{\ 22.4}$      &  136$_{\ 81}^{\ 155}$   & 422$_{\ 192}^{\ 475}$     &  905                        \\[0.3ex] \hline \rule{0pt}{1.0em} 
        UGC\,8696 SE   &  3.9$_{\ 2.5}^{\ 4.4}$        &  97$_{\  32}^{\ 119}$   & 233$_{\ 74}^{\ 281}$      &  518                        \\[0.3ex] \hline \rule{0pt}{1.0em} 
        VV\,250 SE     &  2.9$_{\ 0.6}^{\ 4.9}$        &  162$_{\ 58}^{\ 215}$   & 475$_{\ 103}^{\ 688}$     &  1393                        \\[0.3ex] \hline \rule{0pt}{1.0em} 
        VV\,250 NW     &  0.6$_{\ 0.2}^{\ 0.9}$        &  156$_{\ 50}^{\ 197}$   & 401$_{\ 79}^{\ 507}$      &  1286                       \\[0.3ex] \hline \rule{0pt}{1.0em} 
    \end{tabular}
    \caption*{{Notes on columns -- } 
        {(1): source name.}
        (2): the peak brightness of the source.
        (3): circular aperture containing half of the total flux (i.e. using $A_{50} = \pi R_{50}^2$ (convolved quantity).
        (4): Total estimated size of the source (convolved quantity) for the set of combined images at the discussed threshold level. Note that lower values for $R_{95}$ are an estimate of the total size of the emission in \emph{e}-MERLIN images. 
        (5): The circular size of the source using a pure natural weighted VLA image at $\sim$ 6 GHz. Since this image is the lowest resolution possible and improved sensitivity to large-scale structures, this is the maximum recoverable size of the emission from the available data (see footnote \ref{foot:sources_sizes}).
    }
\end{subtable}
\end{table}
{
From Tab.~\ref{tab:source_global_properties}, our estimated global averaged half-flux sizes among all sources is $\langle R_{50}^{\eM} \rangle\sim$ 47 pc for \emph{e}-MERLIN maps and $\langle R_{50}^{\rm VLA}\rangle \sim $ 251 pc for VLA. At intermediate resolutions, we obtained $\sim$ 142 pc. Comparing the intermediate values with \emph{e}-MERLIN values,} these results are consistent with previous studies showing evidence that the radio emission originating in nuclear disks have typical sizes of $\sim$ 100-200 pc or less \citep[e.g.][]{Kawakatu2008,Herrero2012,Song2021}, and the sizes of very compact sources have the typical extent of $\lesssim$ 50 pc. In Sections~\ref{sec:multi_sersic} and~\ref{sec:multi_scale_properties} we compare these estimates with the sizes of disentangled core-compact components as well as with the disentangled diffuse emission. 

Another critical comparison is the approximated total size of the emission (which we adopt $R_{95}$), particularly in \emph{e}-MERLIN images. It is established that if there is ``missing flux'' between two instruments with distinct resolving power, such as \emph{e}-MERLIN and VLA, that is an indication that diffuse emission exists on the intermediate scales \citep{Orienti2010}. In higher resolution maps, the average full emission size found is $\langle R_{95}^{\eM}\rangle\sim$ 125 pc, indicating that the larger contribution of radio flux density is enclosed in a region smaller than $\sim $ 100 pc. With respect to VLA images, the full extent obtained is $\langle R_{95}^{\rm VLA}\rangle \sim$ 1.5 kpc. However, since we are using the A-configuration, these sizes can only represent a lower limit. As an illustration, the extent of UGC\,8696 at C band using the VLA-D configuration exceeds 50 kpc \citep[e.g.][]{Kukreti2022}. 
In App.~\ref{sec:individual_source_analysis}, we provide a detailed discussion with information for every source with our derived results from Tabs.~\ref{tab:source_global_properties}$\sim$\ref{tab:star_formation_rates}.

\begin{table}
\caption{Nuclear and extended sizes estimation using the interferometric decomposition.  
The two columns on the left side of the table (under ``Nuclear'') refers to the sizes of the optimised \emph{e}-MERLIN emission ($I_{1}^{\mathrm{mask}}$, as exemplified in Fig.~\ref{fig:example_sub}) while the right side (under ``Extended'') refers to the size of the large-scale emission characterized by $\mathcal{R}_{1,2}$ (as exemplified in Fig.~\ref{fig:example_sub}).
{Quantities for the extended emission from \emph{e}-MERLIN images (lower indices) are not shown because of the limitation of the method.}
}
\label{tab:sizes_interferometric_decomposition}
\begin{subtable}[h]{0.49\textwidth}
\centering
\begin{tabular}{l|cc|cc} 
\hline
Interferometric \\
Decomposition	  	    &                                 & {\hspace*{-1.8cm} Nuclear}   &   		                      & {\hspace*{-1.7cm} Extended}       \\ \hline
Source	        	    & $\langle R_{50}\rangle$ [pc]    & $\langle R_{95}\rangle$ [pc] & $\langle R_{50}\rangle$ [pc]   & $\langle R_{95}\rangle$ [pc]      \\ \rule{0pt}{1.0em}
(1)	        	        & (2)                             & (3)                          & (4)                            & (5)                               \\ \hline \rule{0pt}{1.0em}
VV705 (N) 	     		& $107_{\ 34}^{\ 196}$            & $273_{\ 109}^{\ 471}$        & $329_{\ --}^{\ 563}$           & $748_{\ --}^{\ 1669}$             \\[0.3ex] \hline \rule{0pt}{1.0em}
VV705 (S) 	     		& $95_{ \ 27}^{\ 186}$            & $204_{\ 58}^{\ 408}$         & $337_{\  --}^{\ 553}$          & $698_{\ --}^{\ 1371}$             \\[0.3ex] \hline \rule{0pt}{1.0em} 
UGC\,5101 	     		& $138_{\ 37}^{\ 235}$            & $428_{\ 168}^{\ 623}$        & $255_{\ --}^{\ 396}$           & $769_{\ --}^{\ 1696}$             \\[0.3ex] \hline \rule{0pt}{1.0em} 
UGC\,8696 	     		& $115_{\ 60}^{\ 170}$            & $285_{\ 123}^{\ 442}$        & $207_{\ --}^{\ 315}$           & $658_{\ --}^{\ 1160}$             \\[0.3ex] \hline \rule{0pt}{1.0em} 
VV250 (SE)    		    & $118_{\ 74}^{\ 189}$            & $250_{\ 130}^{\ 420}$        & $273_{\ --}^{\ 383}$           & $615_{\ --}^{\ 1091}$             \\[0.3ex] \hline \rule{0pt}{1.0em}
VV250 (NW)    		    & $108_{\ 50}^{\ 180}$            & $210_{\ 79 }^{\ 371}$        & $220_{\  --}^{\ 465}$          & $457_{\  --}^{\ 1029}$            \\[0.3ex] \hline 
\end{tabular}
\caption*{{Notes on columns -- 
          (1): Source name.
          (2): Effective radius of the nuclear components and
          (3) total estimated radius.
          (4): Effective radius of the diffuse emission and
          (5) total estimated radius.
          }}
\end{subtable}
\end{table}

\subsection{S\'ersic Fitting Decomposition Results}
\label{sec:multi_sersic}
We denote \emph{e}-MERLIN flux densities by $S_{\nu}^{\eM}$ and the VLA flux densities by $S_{\nu}^{\rm VLA}$. Our aim is to identify relevant regions of radio emission in both instruments, using a source detection algorithm, which it is explained in App.~\ref{sec:sub_region_analysis}. Then, we proceed to split the flux density contribution into different components and scales. At the smallest scales, we have the core-compact flux density $S_{\nu}^{\text{core-comp}}$ and the extended nuclear flux density $S_{\nu}^{\text{ext-nuc}}$, both probed by \emph{e}-MERLIN. At the largest scales, {we have} the core-unresolved flux density $S_{\nu}^{\text{core-unre}}$ (which is basically the ``compact'' unresolved component by VLA) and the large-scale diffuse flux density $S_{\nu}^{\text{ext}}$.

\begin{figure}
\centering
\includegraphics[width=0.95\linewidth]{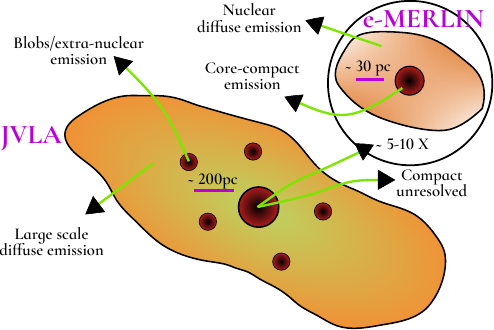}
\caption{Representation of the structure of a U/LIRG, composed of distinct components on different spatial scales. The objectives are to disentangle the radio emission properties in each individual component. In Tabs.~\ref{tab:sizes_image_fitting},~\ref{tab:image_decomposition_b} and ~\ref{tab:star_formation_rates} we present the decomposition results based on this cartoon. }
\label{fig:cartoon_ulirg_structure}
\end{figure}

{
For simplicity, Fig.~\ref{fig:example_fitting_example_VV705N} highlights the S\'ersic fitting decomposition for a pure VLA image of VV\,705 N. Source detected structures are labelled with ``ID'' and model components with ``COMP\_''. For this source, we have used five model components, three were automatically added from the source detection, {COMP\_1}, {COMP\_2} and {COMP\_3}, and two were manually added under visual inspection (see App.~\ref{sec:sub_region_analysis}). To characterise the diffuse emission around the most compact structure ({COMP\_1}), we used model component {COMP\_4}. To model the larger scale (>1 kpc) of the diffuse emission, we used component {COMP\_5}.  Hence, ID1 is the parent structure, having COMP\_4 and COMP\_5 as subcomponents of COMP\_1.}

The top panel of Fig.~\ref{fig:example_fitting_example_VV705N} shows the maps related to the data (left), model (centre) and residual (right). The lower panel, display the multi-component detection (left) (see Appendix~\ref{app:multi_sersic}), the diffuse emission after removing the compact components {COMP\_1} and {COMP\_3} (centre) and the 1D slices of their azimuthally averaged surface brightness profiles (right). {These compare individual model components (dashed lines) with data (purple dotted line), residual (black dotted line) and total model (thick dashed lime-green line). }

The fitting in pure \emph{e}-MERLIN images is more complicated since the signal-to-noise is lower and there are fewer sampling points to provide a statistical good fit. The starting point was an attempt to model the core-compact part with a S\'ersic $n=0.5$ (Gaussian) -- generally good \citep[e.g.][]{Condon_1997,Condon_1998} -- plus a second component ($n=0.5$ or $n=1.0$) to model the nuclear-diffuse emission. However, the fit was not successful for VV\,705 N \& S. For these cases, a single function ($n=0.5$) was used \mchh{to fit the most compact or blob region. Since we have not modelled the diffuse structure, we have considered that the total extended emission is simply represented by the residual image (subtracting the compact/blob model from the data).} In these two cases, deconvolved areas for the nuclear diffuse emission were not established. Nevertheless, for UGC\,5101, UGC\,8696 (N \& SE) and VV\,250, a composite fit \mchh{(compact+extended)} was successful and the total nuclear diffuse flux density was considered to be the flux density from the nuclear-extended model component plus the remaining residual \mchh{emission}.

\subsection{Nuclear vs Large Scale Emission Properties}
\label{sec:multi_scale_properties}
Using the concepts of Sec.~\ref{sec:interferometric_decomposition} and Sec.~\ref{sec:image_fitting} we describe the disentangled properties from the nuclear radio emission probed by \emph{e}-MERLIN (both core-compact and nuclear diffuse structures) and VLA (unresolved compact and large scale diffuse emission).

\begin{figure*}
\centering
\includegraphics[width=0.85\linewidth]{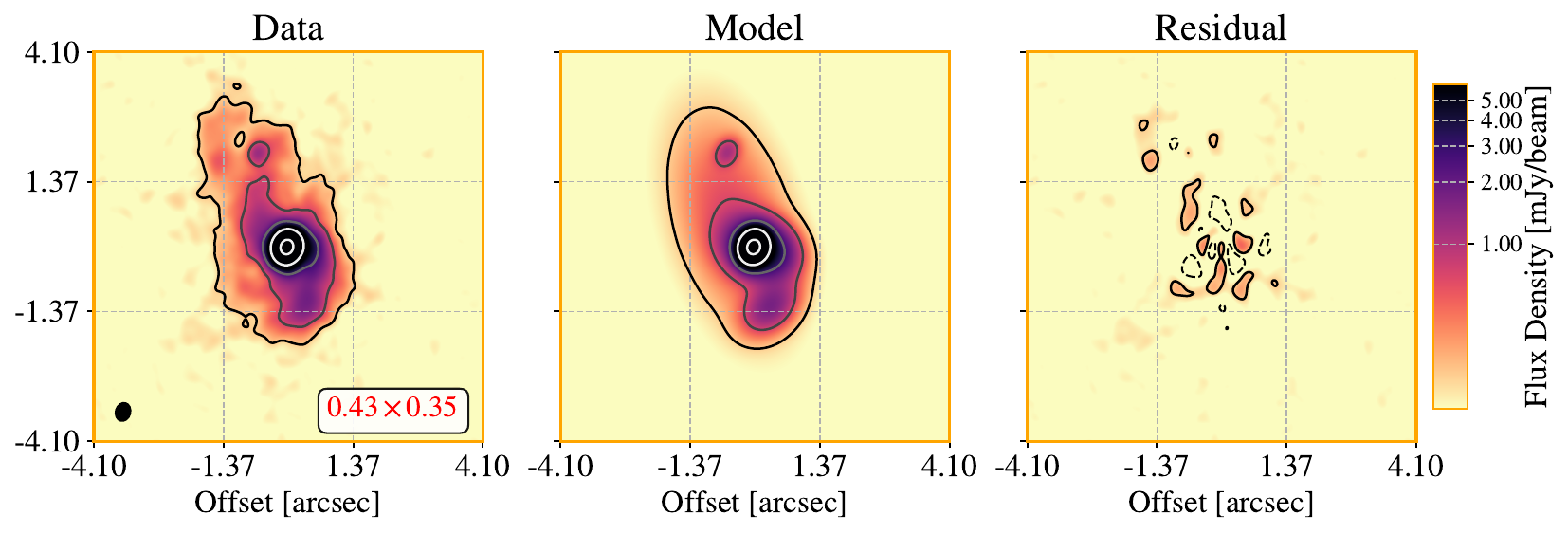}
\includegraphics[width=0.27\linewidth]{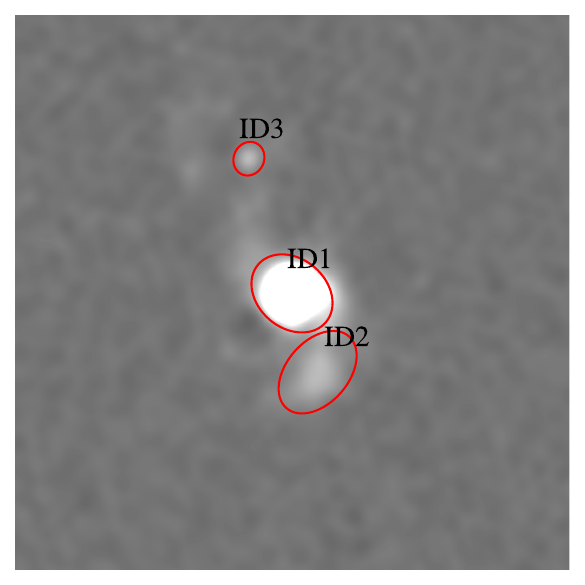}
\includegraphics[width=0.268\linewidth]{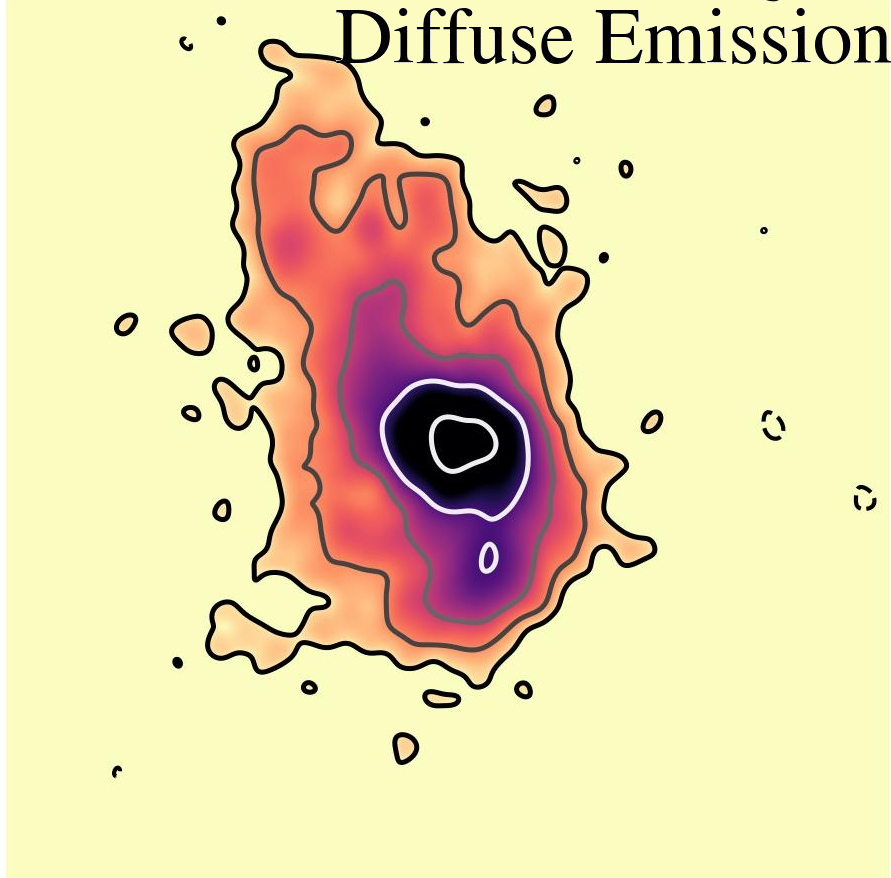}
\includegraphics[width=0.33\linewidth]{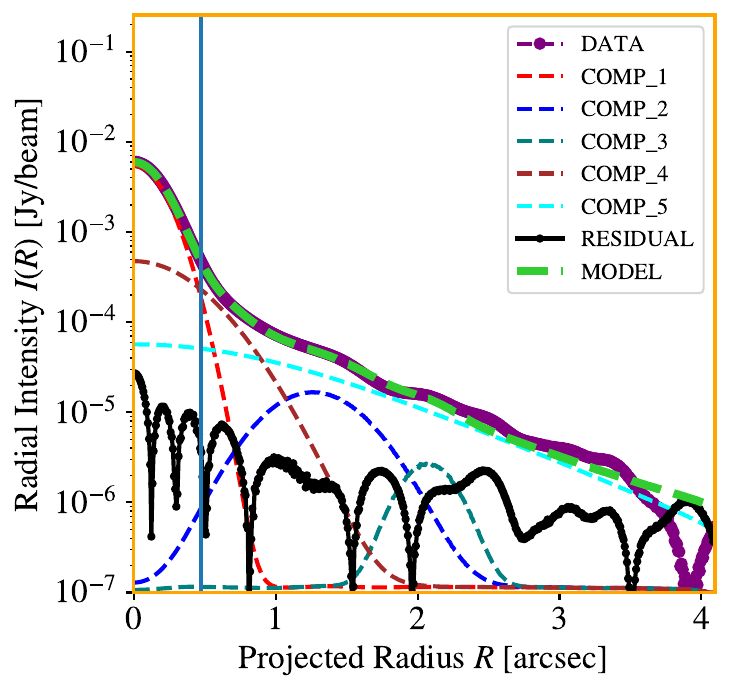}
\vfill
\caption{Results on fitting multiple components to VV\,705 N. Top row (from left to right, respectively) shows the original pure VLA radio map, the model (sum of all components) and the residual. Lower left: it is demonstrated how individual regions of radio emission are identified with \texttt{SEP} in order to compute basic photometry so that they are used to constrain the non-linear least-square minimisation. Lower centre: Diffuse emission with the VLA unresolved component removed (\text{COMP\_1}). Lower right: The one-dimensional azimuthally averaged surface brightness profile of the source jointly with all decomposed model components, five in total. }
\label{fig:example_fitting_example_VV705N}
\end{figure*}

With the image fitting decomposition (see Section.~\ref{sec:multi_sersic}), we can reconstruct the total flux density\footnote{\label{foot:sources_sizes}Note that this refers to the structure probed by VLA in A-configuration. For some sources, there is a significant fraction of emission coming from scales larger than 10 kpc, as it is the case of UGC\,8696 when using the VLA-D configuration \citep[e.g.][]{Kukreti2022}.} $S_{\nu}^{\text{ext-tot}}$ that originates from any structure that is not powered by (unresolved) core-compact regions by taking 
\begin{align}
\label{eq:s_ext_tot}
S_{\nu}^{\text{ext-tot}} = S_{\nu}^{\rm VLA} - S_{\nu}^{\text{core-comp}}, \qquad 
S_{\nu}^{\text{ext-nuc}} = S_{\nu}^{\eM} - S_{\nu}^{\text{core-comp}}. 
\end{align}
With this approach, we probe any significant resolved nuclear diffuse emission recovered by \emph{e}-MERLIN, providing a correction for the total multiscale flux density that is not generated by core-compact components, such as nuclear emission from nuclear star-forming regions as well as large-scale SF. Also, {we estimate} the ratio of core-compact components to the total radio flux density as $S_{\nu}^{\text{core-comp}}\ /S_{\nu}^{\rm VLA}$. This is the fraction that we compare in App.~\ref{sec:imaging_results_sources}, individually for each source with values from the literature, when it comes to AGN fractions {determinations}. For completeness, we also compute the integrated flux density in the unresolved VLA components, $S_{\nu}^{\text{core-unre}}$, described by the Gaussian/S\'ersic fit at the unresolved parts (e.g. {COMP\_1} in Fig.~\ref{fig:example_fitting_example_VV705N}). 

\begin{table*}
\centering
\caption{Decomposed properties of the radio emission using S\'ersic image fitting approach. 
The column block in the left ``Core-compact and unresolved components'' represent the sizes of the modelled core-compact or unresolved structures, while the block column in the right ``Nuclear and large scale diffuse components'' represent the sizes of the nuclear diffuse (\emph{e}-MERLIN) and large-scale extended (VLA) emission. The sizes are calculated on the resulting deconvolved and convolved images. 
The notation is the same as discussed in Eq.~\ref{eq:measured_quantities}: the main value is the nominal quantity measured on combined images, lower-indices refers to pure \emph{e}-MERLIN images and upper indices to pure VLA images. We refer the reader to Figure \ref{fig:cartoon_ulirg_structure} for clarification. 
Additionally, we also differentiate between the deconvolved quantities as well the convolved quantities with the lower index ``d''. 
The brightness temperature is measured on the {brightest} \emph{e}-MERLIN core-compact deconvolved components. 
}
\label{tab:sizes_image_fitting}
\begin{subtable}[h]{0.99\textwidth}
\qquad \qquad\qquad \qquad \qquad  \qquad \quad Core-compact and unresolved components \qquad \qquad Nuclear and large scale diffuse components \\
\centering
\begin{tabular}{l|ccccc|cccc}
\hline      
Source              &  $\langle R_{50,\rm d}\rangle$  & $\langle R_{50}\rangle$  & $\langle R_{95, \rm d}\rangle$ &  $\langle R_{95}\rangle$  & $\langle T_b \rangle$ & $\langle R_{50,\rm d}\rangle$  & $\langle R_{50}\rangle$    & $\langle R_{95, \rm d}\rangle$  & $\langle R_{95}\rangle$    \\[0.1ex] 
                    &    [pc]                     &     [pc]                 & [pc]                       & [pc]                      & $\times 10^{5}$ [K]           & [pc]                           & [pc]                       &   [pc]                      &  [pc]                       \\[0.1ex]
 (1)                &    (2)                      &     (3)                  & (4)                        & (5)                       & (6)                           & (7)                            & (8)                        &   (9)                       &  (10)                       \\[0.3ex] \hline \rule{0pt}{1.0em}
VV\,705 N$^{\rm a}$ & $44_{\ 12}^{\  56}$         & $106_{\ 18}^{\ 153}$     & $114_{\ 26}^{\ 140}$       & $269_{\ 41}^{\ 516}$      & $3.3 \pm 0.9$                 & $306_{\ -- }^{\ 448}$          & $371_{\ 48}^{\ 516}$       & $893_{\ -- }^{\ 1336}$      & $1195_{\ 77}^{\ 1590}$      \\[0.3ex] \hline \rule{0pt}{1.0em} 
VV\,705 S$^{\rm b}$ & $62_{\ 15}^{\  58}$         & $115_{\ 21}^{\ 151}$     & $138_{\ 29}^{\ 131}$       & $258_{\ 42}^{\ 341}$      & $0.4 \pm 0.1$                 & $429_{\ -- }^{\ 513}$          & $504_{\ -- }^{\ 583}$      & $781_{\ --}^{\  1126}$      & $995_{\ --}^{\  1346}$      \\[0.3ex] \hline \rule{0pt}{1.0em} 
UGC\,8696 N         & $81_{\ 22}^{\  76} $        & $119_{\ 27}^{\ 138}$     & $162_{\ 44}^{\ 153 } $     & $ 239_{\ 53}^{\ 278}$     & $7.7 \pm 0.8$                 & $196_{\ 139 }^{\ 172} $        & $215_{\ 140}^{\ 219}$      & $392_{\ 279}^{\ 674}$       & $430_{\ 281}^{\ 803}$         \\[0.3ex] \hline \rule{0pt}{1.0em} 
UGC\,8696 SE        & $53_{\ 21}^{\  53} $        & $103_{\ 26}^{\ 127}$     & $106_{\ 42}^{\ 105 } $     & $ 205_{\ 52}^{\ 254}$     & $ 5.1 \pm 0.2 $               & --                             &    --                      & --                          &       --                    \\[0.3ex] \hline \rule{0pt}{1.0em} 
UGC\,5101           & $40_{\ 14}^{\  38} $        & $100_{\ 29}^{\   165}$   & $101_{\ 60}^{\  96}  $     & $247_{\ 82}^{\ 412}$      & $ 19.1 \pm 3.0 $              & $235_{\ 101}^{\ 267}  $        & $267_{\ 130}^{\ 324}$      & $691_{\ 177}^{\ 1143}$      & $911_{\ 240}^{\ 1303}$      \\[0.3ex] \hline \rule{0pt}{1.0em} 
VV\,250 SE          & $86_{\ 54 }^{\ 91}$         & $117_{\ 61 }^{\ 145}$    & $179_{\ 96}^{\ 204}  $     & $249_{\ 115}^{\ 325}$     & $ 0.2 \pm 0.1 $               & $198_{\ 56}^{\ 270}   $        & $207_{\ 68}^{\ 290}$       & $383_{\  89}^{\ 645}$       & $388_{\ 113}^{\ 657}$       \\[0.3ex] \hline \rule{0pt}{1.0em} 
VV\,250 NW          & $61_{\ 26}^{\  86}$         & $ 109_{\ 45}^{\ 155}$    & $131_{\ 40}^{\ 187}  $     & $233_{\ 75}^{\ 337}$      & $ < 0.1 $                     & $281_{\ 46}^{\ 380}   $        & $298_{\ 58}^{\ 407}$       & $490_{\ 73}^{\ 638}$        & $ 526_{\ 98}^{\ 697}$      \\[0.3ex] \hline 
\end{tabular}
\caption*{{Notes on columns -- 
(1): Source name.
(2): Deconvolved effective radius of the core-compact and unresolved components.
(3): Similarly, but the convolved effective radius.
(4): Total deconvolved radius of the core-compact and unresolved components.
(5): Similarly, but the total convolved radius.
(6): Brightness temperature measured using the core-compact \emph{e}-MERLIN deconvolved sizes.
(7): Deconvolved effective radius of diffuse components (nuclear diffuse emission and VLA diffuse emission). 
(8): Similarly, but the convolved effective radius.
(9): Total deconvolved radius of diffuse components (nuclear diffuse emission and VLA diffuse emission).
(10): Similarly, but the total convolved radius.}\\
$^{\rm a}$For VV\,705 N, the nuclear diffuse emission in \emph{e}-MERLIN was not modelled, hence only its size (convolved) is provided. 
$^{\rm b}$For VV\,705 S, the nuclear diffuse emission in \emph{e}-MERLIN is almost absent, therefore associated sizes are uncertain and we have not shown the lower limits {extracted from \emph{e}-MERLIN images}.}
\end{subtable}
\end{table*}

In Tab.~\ref{tab:sizes_image_fitting}, we present the size results for the cases previously discussed. On average, across all sources, the value of the half-light radii for the core-compact components is $\sim$ 30 pc (convolved) and $\sim$ 20 pc (deconvolved). The full sizes are $\sim$ 60 pc (convolved) and $\sim$ 40 pc (deconvolved). The averaged convolved half-light radii for the nuclear diffuse emission among the successful measurements (UGC\,8696 N, UGC\,5101, VV\,250 SE \& NW) is $\sim$ 100 pc up to $\sim$ 200 pc in total extent. For these decomposed structures, we present the fractional flux densities computed with the S\'ersic image fitting method in Tab.~\ref{tab:image_decomposition_b}.

Regarding the total flux densities and fractions calculated with the interferometric decompositions from \emph{e}-MERLIN and VLA maps, they are presented in Tab.~\ref{tab:image_decomposition_a}. We quantify the extended flux density in VLA, $S_{\nu}^{\text{ext}}$ as the sum of the emission from two maps: $\mathcal{R}_{1,2}$ (Eq.~\ref{eq:R12}) and $\mathcal{R}_T$ (Eq.~\ref{eq:RT}), that is 
\begin{align}
    S_{\nu}^{\text{ext}} = \frac{1}{\mathfrak{B}_{\rm A}^{1,2}}\sum \left(\mathcal{R}_{1,2} \times {\texttt{mask}}_{1,2}\right)
    +
    \frac{1}{\mathfrak{B}_{\rm A}^{T}}\sum \left(\mathcal{R}_{T} \times {\texttt{mask}}_{T}\right)
\end{align}
where $\mathfrak{B}_{\rm A}$ is the beam area in each iteration image (see Fig.~\ref{fig:int_decomposition_flowchart}).

{When comparing fitting results in VLA and \emph{e}-MERLIN images, }Tabs.~\ref{tab:image_decomposition_b} and~\ref{tab:image_decomposition_a}, we note a significant difference between fractions of \emph{e}-MERLIN to total VLA flux densities $S_{\nu}^{\eM}/S_{\nu}^{\rm VLA}$ in comparison with $S_{\nu}^{\text{core-comp}}/S_{\nu}^{\rm VLA}$. \mchh{
These differences arise from the contribution of the nuclear extended emission, which reduces and imposes a limit to the value of $S_{\nu}^{\text{core-comp}}$.
}
For the two brightest sources, UGC\,5101 and UGC\,8696, the nuclear emission dominates significantly over the VLA-A large-scale flux density, and for the other sources, the latter dominates over the nuclear flux density probed with \emph{e}-MERLIN, in exception to VV\,705 N, which shows equivalent contribution. It is relevant to mention that since the total \emph{e}-MERLIN flux density is not pure core-compact, it is necessary to disentangle that component from any nuclear diffuse portion (see Fig.~\ref{fig:cartoon_ulirg_structure}), which can be \mchh{reported} as compact/unresolved in a lower resolution instrument. Thus, we see that the multiscale diffuse flux density, interpreted as a by-product of star formation, plays a significant role in the total radio emission. This idea is being discussed in recent studies, mainly in radio-quiet AGNs \citep{Panessa2019} and radio-quiet quasars \citep[e.g.][]{Condon_2013,Wang2023,Wang2023b}. {Accurately measuring these differences is critical when studying high-redshift sources, since there is an instrumental limitation regarding angular resolution to separate such physical structures.}

\begin{table*}
\centering
\caption{Separation of nuclear or compact fluxes probed by \emph{e}-MERLIN against diffuse emission, comparing two different approaches:  (a) image S\'ersic fitting. (b) interferometric decomposition}
\label{tab:image_decomposition}
\begin{subtable}[h]{0.99\textwidth}
\centering
\begin{tabular}{lccccccc} 
\hline
S\'ersic Fitting \\ 
Decomposition	  	     &                     		           &                                   &						             &  				           &                                  &				                                                &  \\ \hline
Source	        	     &  $S_{\nu}^{\text{core-comp}}$ [mJy] & $S_{\nu}^{\text{ext-nuc}}$ [mJy]  & $S_{\nu}^{\text{core-unre}}$ [mJy]  & $S_{\nu}^{\text{ext}}$ [mJy]& $S_{\nu}^{\text{ext-tot}}$ [mJy] & $\dfrac{S_\nu^{\text{ext-tot}}}{S_{\nu}^{\text{core-comp}}}$ & $\dfrac{S_{\nu}^{\text{core-comp}}}{S_{\nu}^{\rm VLA}}$   \\
                         &  \emph{e}-MERLIN 		              &	\emph{e}-MERLIN       	          &	VLA  	 				            & VLA                        & Diffuse Total                         &                                                             &                 \\ \hline
VV\,705 (N) 			 &  3.7 $\pm$ 0.2                      &	2.3 $\pm$ 1.1  		           & 6.6 $\pm$ 1.0    			         & 6.1 $\pm$ 0.9               & 9.3 $\pm$ 0.3                    & 2.52 $\pm$ 0.16                                             & 0.28 $\pm$ 0.02 \\ \hline
VV\,705 (S) 			 &  0.9 $\pm$ 0.1                      &	0.2 $\pm$ 0.1		           & 2.4 $\pm$ 0.6				         & 3.6 $\pm$ 0.5               & 5.4 $\pm$ 0.2                    & 6.00 $\pm$ 0.71                                             & 0.14 $\pm$ 0.02 \\ \hline
UGC\,5101    			 &  23.8 $\pm$ 0.9                     &	17.7 $\pm$ 4.0			       & 33.0 $\pm$ 2.2   			         & 26.5 $\pm$ 1.3              & 37.5 $\pm$ 1.0                   & 1.57 $\pm$ 0.07                                             & 0.38 $\pm$ 0.02 \\ \hline
UGC\,8696 N    			 &  9.5 $\pm$ 0.1                      &	24.5 $\pm$ 6.6			       & 22.9 $\pm$ 3.0 				     & 23.4 $\pm$ 1.9              & 39.4 $\pm$ 0.9                   & 4.11 $\pm$ 0.13                                             & 0.19 $\pm$ 0.01 \\ \hline
UGC\,8696 SE    	     &  4.4 $\pm$ 0.4                      &	$\sim$ 0	     		       & 5.3 $\pm$ 0.2 				         & $\sim$ 0                    &  $\sim$ 0                        & $\sim$ 0                                                    &> 0.83 $\pm$ 0.08 \\ \hline
VV\,250 SE    	         &  $\lesssim$ 6.2 $\pm$ 2.2           &	$\gtrsim$ 2.2 $\pm$ 1.8           & 7.9 $\pm$ 0.9			             & 10.0 $\pm$ 0.8              & 11.7$\pm$2.3                     & 1.89 $\pm$ 0.76                                            &< 0.34 $\pm$0.12 \\ \hline
VV\,250 NW               & $\sim$ 0.5 $\pm$ 0.6                &	$\sim$ 0.9 $\pm$ 0.5           & 1.6 $\pm$ 0.1	    		         & 1.7 $\pm$ 0.4               & $\gtrsim$ 2.8                    & $\gtrsim$ 5.6                                               & $\lesssim$ 0.15  \\ \hline

\end{tabular}
\caption{Flux densities estimates for the nuclear region{, core-compact and nuclear diffuse components, and the VLA} extended emission, using the image fitting decomposition. \\ 
Notes on columns --
(1): Source name.
(2): $S_{\nu}^{\text{core-comp}}$ - Recovered core-compact component flux density from \emph{e}-MERLIN. 
(3): $S_{\nu}^{\text{ext-nuc}}$ - Estimated nuclear diffuse flux density in \emph{e}-MERLIN. 
(4): $S_{\nu}^{\text{core-unre}}$ - Unresolved ``compact'' flux density  in VLA. 
(5): $S_{\nu}^{\text{ext}}$ - Large scale diffuse flux density  in VLA. 
(6): $S_{\nu}^{\text{ext-tot}}$ - Total estimated diffuse flux density (small scale + large scale), where we assume $S_{\nu}^{\text{ext-tot}} = S_{\nu}^{\rm VLA} - S_{\nu}^{\text{core-comp}}$.}
\label{tab:image_decomposition_b}
\end{subtable}
\ \\
\begin{subtable}[h]{0.99\textwidth}
\centering
\begin{tabular}{lcccccc} 
\hline
Interferometric \\
Decomposition	&                                      &							   &                                      &                                                  &  				                 &                                   \\ \hline
Source	  	    & $S_{\nu}^{e\text{-M}}$ [mJy]         & $S_{\nu}^{\rm ext}$ [mJy]     & $S_\nu^{\rm ext}/S_\nu^{e\text{-M}}$ & $S_\nu^{e\text{-M}}/S_{\nu}^{\rm VLA}$          & $S_\nu$ on $\mathcal{R}_T$ [mJy]        & $S_{\nu}^{\rm VLA}$ [mJy]        \\ \hline
VV705 (N) 		& 6.0  $\pm$ 1.0  		               & 6.1  $\pm$ 0.8                & 1.02 $\pm$ 0.21        		      & 0.46 $\pm$ 0.08                                  & 1.9 $\pm$ 1.0	                 & 13.0                              \\ \hline
VV705 (S) 	    & 1.2  $\pm$ 0.3  		               & 4.5  $\pm$ 0.7                & 3.75 $\pm$ 1.10        		      & 0.19 $\pm$ 0.04                                  & 1.2 $\pm$ 0.9	                 & 6.3                               \\ \hline
UGC\,5101	    & 39.4 $\pm$ 3.0  		               & 20.6 $\pm$ 0.5                & 0.52 $\pm$ 0.04        		      & 0.64 $\pm$ 0.05                                  & 1.5 $\pm$ 0.8	                 & 61.3                              \\ \hline
UGC\,8696	    & 31.6 $\pm$ 2.7  		               & 20.6 $\pm$ 1.3                & 0.65 $\pm$ 0.07        		      & 0.58 $\pm$ 0.05                                  & 1.5 $\pm$ 3.0	                 & 54.0                              \\ \hline
VV\,250 (SE) 	& 7.3  $\pm$ 1.0  		               & 10.2 $\pm$ 0.9               & 1.40 $\pm$ 0.12        		      & 0.42 $\pm$ 0.06                                  & 1.4 $\pm$ 1.3	                 & 17.9                              \\ \hline
VV\,250 (NW)    & 1.0  $\pm$ 0.4  		               & 2.2 $\pm$ 0.3                 & 2.29 $\pm$ 0.36        		      & 0.31 $\pm$ 0.14                                  & 0.7 $\pm$ 0.6	                 & 3.3                               \\ \hline
\hline
\end{tabular}
\caption{Flux densities estimates for the extended and compact emission, using the interferometric decomposition method. In this table, there is no distinction between compact or non-compact components. The values only refers to \emph{e}-MERLIN emission that is removed from VLA maps. In the general case, this emission can be core-compact (e.g. VV\,705) and/or nuclear diffuse emission around those (e.g. UGC\,8696) and/or blobs of emission. \\
Notes on columns --
(1): Source name.
(2): The flux density originated from nuclear regions or from blobs probed by \emph{e}-MERLIN. 
(3): The estimated flux density of the diffuse emission after removing the \emph{e}-MERLIN contribution (see Fig.~\ref{fig:int_decomposition}). 
(4): Fraction of VLA extended flux density against \emph{e}-MERLIN flux density. 
(5): residual flux density on the resulting image after subtraction. 
(6): Total VLA flux density recovered at $\sim$ 6 GHz.}
\label{tab:image_decomposition_a}
\caption*{\hrule \ \\ Notes: i) In these two tables, error estimates are statistical, represented by the statistical significance of the variance of measurements that are computed between multiple images. ii) When computing the fractions, for example, $\frac{S_\nu^{\text{ext-tot}}}{S_{\nu}^{\text{core-comp}}}$, we have added in quadrature the errors associated to $S_{\nu}^{\text{ext-tot}}$ and $S_{\nu}^{\rm VLA}$ in Tab.~\ref{tab:source_properties} so that the errors attributed to the diffuse emission are properly propagated. }
\end{subtable}
\end{table*}

\begin{table*}
\centering
\caption{Representative values of core-compact, nuclear and large-scale diffuse emission areas, alongside star formation and surface density star formation rates, SFR and $\Sigma$, respectively. Star formation rates are calculated using the extended emission only, both nuclear and large scale.}
\label{tab:star_formation_rates}
\begin{subtable}[h]{0.99\textwidth}
\centering
\begin{tabular}{lccccccccc} 
\hline
Source	        & $A_{95,\rm d}^{\rm core-comp}$   & $A_{95,\rm d}^{\text{ext-nuc}}$  & $A_{50,\rm d}^{\text{ext-tot}}$ & $A_{95}^{\text{ext-tot}}$      & SFR$^{\text{ext-nuc}}$  & SFR$^{\text{ext-tot}}$          &  $\Sigma_{\text{ext-nuc}}^{95}$  & $\Sigma_{\text{ext-tot}}^{50,\rm d}$          \\
                & [kpc$^{2}$ $\times 10^{3}$]    & [kpc$^{2}$ $\times 10^{3}$]      & [kpc$^{2}$]                     & [kpc$^{2}$]                     & $[\mathrm{M}_{\odot}\yr^{-1}]$   & $[\mathrm{M}_{\odot}\yr^{-1}]$            &  [$\mathrm{M}_{\odot}\yr^{-1}$ kpc$^{-2}$] & [$\mathrm{M}_{\odot}\yr^{-1}$ kpc$^{-2}$] \\
                & (2)                              & (3)                              & (4)                             & (5)                             & (6)                     & (7)                              &  (8)                              & (9)                              \\\hline \rule{0pt}{1.0em} 
VV\,705 N$^a$ 	&  2 $\pm$ 1                       & 22$\pm$ 14                       & 0.982                           & 9.099                           & 17.6 $\pm$ 8.6          & 63.2 $\pm$ 3.2                   &  454 $\pm$ 221                    & 32 $\pm$ 3                     \\[0.3ex] \hline \rule{0pt}{1.0em}
VV\,705 S 	    &  3 $\pm$ 1                       & --                               & 1.264                           & 6.029                           & --                      & 37.0 $\pm$ 2.7                   &  --                               & 15 $\pm$ 2                     \\[0.3ex] \hline \rule{0pt}{1.0em}
UGC\,5101 	    &  13 $\pm$ 13                     & 177 $\pm$ 105                    & 0.302                           & 10.295                          & 95.6 $\pm$ 31.3         & 254.6 $\pm$ 10.8                 &  227 $\pm$ 74                     & 420 $\pm$ 12                    \\[0.3ex] \hline \rule{0pt}{1.0em}
UGC\,8696	    &  15 $\pm$ 4                      & 83 $\pm$ 29                      & 0.232                           & 4.595                           & 157.7 $\pm$ 32.9        & 251.4 $\pm $ 8.4                 &  1046 $\pm$ 218                   & 542 $\pm$ 23                    \\[0.3ex] \hline \rule{0pt}{1.0em}
VV\,250 SE 	    &  35 $\pm$ 23                     & 27 $\pm$ 49                      & 0.211                           & 3.895                           & 6.8 $\pm$ 4.0          & 57.5 $\pm$ 10.7                   & 154 $\pm$ 127                   & 135 $\pm$ 10                    \\[0.3ex] \hline \rule{0pt}{1.0em}
VV\,250 NW    	&  $\sim$ 7 $\pm$ 6                 & $\sim$ 22 $\pm$ 18              & 1.129                           & 3.390                          & $\sim$ 1.7 $\pm$ 1.4   & 11.0 $\pm$ 2.1                   &  $\sim$ 51 $\pm$ 58              & 5 $\pm$ 1                        \\[0.3ex] \hline \rule{0pt}{1.0em}

\end{tabular}
\caption*{Notes on columns --
(1): Source name.
(2): Deconvolved total area of the core-compact components. 
(3): Total deconvolved area of the nuclear diffuse emission after removing the core-compact structures. 
(4): Deconvolved half-light area of the large-scale diffuse emission. This is used to compute $\Sigma_{\mathrm{ext-tot}}^{50,\rm d}$.
(5): Total convolved area of the large-scale diffuse emission. 
(6): Estimated star formation rate for the nuclear diffuse emission, after removing the core-compat contribution and assuming that the remaining emission is pure synchrotron (see Sec.~\ref{sec:SFR} for details). 
(7): Estimated total star formation rate for the entire diffuse emission (nuclear and large-scale) assuming that it is pure synchrotron.
(8): Surface density star formation rate at the nuclear region without core-compact components.
(9): Overall surface density star formation rate for the total diffuse emission at the half-light area.
$^a$For VV\,705 N, the convolved diffuse nuclear area was used instead, see text for details.
}
\end{subtable}
\end{table*}

\subsection{Brightness Temperatures}
\label{sec:brightness_temperatures}
The brightness temperature $T_b$ is defined with reference to a black body object \citep{burke2019}. If we consider that the black body has a physical temperature $T_b$, then we can relate a source with brightness temperature $T_b$ that will have the same brightness intensity as the black body \citep{Morabito2022}. Since most of the sources deviates from a black body, $T_b$ is distinct for different physical processes existent in these sources, such as AGN, SB and star forming regions. For star-forming regions, $T_b$ is caped at a maximum limited value imposed by the physics of the environment, \citep[e.g.][]{Walter2009,Hopkins2010,Crocker2018} and also it depends on the frequency of observation. Therefore, $T_b$ is a good discriminant to distinguish radio components such as AGN, SB and diffuse star-formation mechanisms \citep{Morabito2022}. Brightness temperatures that exceed a value of $\sim$ 1 $\times 10^{5}$ K at $\sim$ 6 GHz can provide an indication of the existence of extreme compact sources. We use $T_b$ to compare our measurements of the nuclear regions with literature and check AGN classification. 

The brightness temperature $T_b$ is defined by \citep{Ulvestad_2005,Kovalev_2005}
\begin{align}
\label{eq:Tb}
T_b = \frac{2 \ln 2}{\pi k_B} \frac{c^{2} (1+z)}{\phi_{\rm maj} \phi_{\rm min}} \frac{S_\nu}{\nu^2}
\end{align}
where $k_B$ is the Boltzmann constant, $z$ the redshift and $\phi_{\rm maj}$, $\phi_{\rm min}$ are the deconvolved semi-major and minor axis of the source (Eqs. \ref{eq:phi_maj} and \ref{eq:phi_min}). In our image fitting implementation, recall that radii quantities are converted from half-light radius to FWHM Gaussian quantities through $\phi = 2 R_n \approx 2 R_{50,{\rm d}}$ (see Eq.~\ref{eq:Rn05_fwhm})\footnote{Therefore, $\phi_{\rm maj} = 2R_n$ and $\phi_{\rm min} = 2 q R_n$ where $q$ is the fitted axis ratio, $q=R_b/R_a$.}. 

Results for $T_b$ are just calculated for \emph{e}-MERLIN images, on the brightest core-compact component. Because of that, the results in Tab.~\ref{tab:sizes_image_fitting} do not follow the formatting introduced in Eq.~\ref{eq:measured_quantities}. The errors presented are statistical, in terms of the variance of $T_b$ calculated over multiple \emph{e}-MERLIN images recovered with different weights.

\subsection{Morphology of the diffuse radio emission}
After minimising the flux contribution of the nuclear region due to core-compact components, we can explore the properties of the diffuse structures. These are primarily associated with synchrotron emission resulting from star formation activity, which extends from the immediate vicinity of the nuclear regions (a few parsecs) to larger scales ($>$1kpc). For each map of the diffuse emission, quantities such as $R_{50}$ and $R_{95}$ are re-calculated, since now we do not have the contamination from core-compact structures. The results are presented in Table \ref{tab:sizes_interferometric_decomposition} and \ref{tab:sizes_image_fitting} under the label ``Extended emission''. It should be noted that in Tab.~\ref{tab:sizes_interferometric_decomposition}, lower indices representing \emph{e}-MERLIN images are omitted, as we cannot probe the nuclear-extended emission in pure \emph{e}-MERLIN images using the method discussed in Sec.~\ref{sec:interferometric_decomposition}.

A notable observation from the 50\% regions is that the sizes increase considerably in maps where the contribution of core-compact emission has been removed, compared to the original maps. This can be seen when comparing with the measurements in Tab.~\ref{tab:source_global_properties}. For instance, in VV\,705 N, the intermediate radial size that encloses half of the total flux is approximately 100 pc. Nonetheless, this still incorporates significant information pertaining to the core-compact structure, rather than the diffuse one. {After removing its contribution, the intermediate effective convolved size ranges from 329 to $\sim$ 372 pc. This suggests that if this emission is due to pure star formation, the majority of  the activity is confined in a region within $\sim$ 400 pc, and not within 100 pc. Across all sources, the average effective size $R_{50}$ of diffuse regions ranges from about $\sim$ 300 to $\sim$ 450 pc, this more indicative of the size of the region dominated by the diffuse component alone. This can be contrasted with the average global effective sizes of $\sim$ 250 pc, as computed in Tab.~\ref{tab:source_global_properties}.}

\subsection{Star formation from the Radio Continuum Emission}
\label{sec:SFR}
To estimate the star formation rate of our sources, we can assume that all the flux density is non-thermal. However, that is it not completely true since we expect a fraction of thermal contribution at 6 GHz. Adopting the calibration for the SFR from \citep{Murphy2011,Murphy2012,Tabatabaei2017}, where it is assumed that the non-thermal emission is connected to the star formation via supernovae rates, the following expression for the SFR combines both thermal and non-thermal contribution from the radio continuum emission (RC):
\begin{align}
\label{eq:sfr}
\left( 
\frac{\text{SFR}^{\mathrm{RC}}_\nu}{\mathrm{M}_{\odot} {\rm yr}^{-1}}
\right)
= 10^{-27} 
&\left[ 
2.18 \left(
\frac{T_e}{10^{4}{\rm K}}
\right)^{0.45}
\left(
\frac{\nu}{{\rm GHz}}
\right)^{-0.1}
+ 15.1
\right.\\ \nonumber
&\quad\left. 
\left(
\frac{\nu}{\rm GHz}^{-\alpha^{\rm NT}}
\right)
\right]^{-1}
\left( 
\frac{L_\nu}{{\rm erg s}^{-1} {\rm Hz}^{-1}}
\right).
\end{align}
Above,  $L_\nu^{\mathrm{}}$ is the spectral luminosity at frequency $\nu$,
\begin{align}
L_\nu^{\rm } = 4 \pi D_{L_\nu}^2 S_\nu^{\rm }, 
\end{align}
with $S_\nu^{\rm }$ being the total flux density computed from the multiscale diffuse emission, and $D_{L_\nu}$, is the luminosity distance of the source in Mpc, given by \citep[e.g.][]{Condon2018}
\begin{align}
D_{L_\nu} = D_L (1+z)^{-(\alpha +1)/2}, \quad D_L = (1+z)D_C.
\end{align}
with $D_C$ the commoving distance in Mpc. 

In Eq.~\ref{eq:sfr}, both thermal and non-thermal radio emission is accounted for \citep{Murphy2012}.
For our purposes to estimate the SFR, We adopt a non-thermal spectral index of $\alpha^{\rm NT} = -0.85$, an electron temperature of $T_{e} = 10^{4}$ K and a frequency of  $\sim $ 6 GHz.

Two particular star formation estimates of interest are: the SFR in nuclear diffuse structures (nuclear star formation), {$S_{\nu}^{\text{ext-nuc}}$},  and the total multiscale SFR in the lower resolution maps after removing the flux contribution from core-compact components, $S_{\nu}^{\text{ext-tot}}$. The nuclear diffuse star-formation is computed in the \emph{e}-MERLIN maps without the contribution from core-compact components. Hence, we have that ${\rm SFR}^{\text{ext-nuc}} = {\rm SFR}_{\rm 6GHz}^{\rm RC}(S_{\nu}^{\text{ext-nuc}})$ (see Eq.~\ref{eq:s_ext_tot} and Tab.~\ref{tab:image_decomposition}) and consequently the nuclear surface density star-formation rate is given by $\Sigma_{\text{ext-nuc}}^{95} = {\rm SFR}^{\text{ext-nuc}}/A_{95}^{\text{ext-nuc}}$. We use the 95\% area instead of 50\% in order to capture the full morphology of the nuclear diffuse emission. {The total multiscale extended emission SFR is given by}
$
{\rm SFR}^{\text{ext-tot}} = {\rm SFR}^{\rm RC}_{6{\rm GHz}} \left( S_{\nu}^{\text{ext-tot}}\right)
$
(see Eq.~\ref{eq:s_ext_tot} and Tab.~\ref{tab:image_decomposition}). 
To compute the total surface density of star formation, we use as reference the maximum deconvolved half-light area of the VLA diffuse emission, $A_{50,\rm d}^{\text{ext}}$. Then, the 6 GHz radio surface density star formation rate at the half-light is $\Sigma_{\text{ext-tot}}^{50} = {\rm SFR}^{\text{ext-tot}}/\max(A_{50,\rm d}^{\text{ext}})$.

To compute these values, the following assumptions are made. In \emph{e}-MERLIN images, the core-compact component does not change significantly in relation to the nuclear diffuse emission when different weighting schemes are used during cleaning (i.e. a \texttt{robust > 0.5}). {We assume that the total flux density of any nuclear diffuse emission will change more than the core-compact flux density. Hence, we consider the size of the nuclear diffuse emission to be the one related to the more natural image, i.e. the one resulting from the larger restoring beam}. Therefore, a good representative area for this nuclear diffuse emission can be expressed in terms of the area of this lower-resolution \emph{e}-MERLIN image, which is a map with the lowest angular resolution possible. In this context, the area will be maximum. This is a safe way to ensure that all the emission will be enclosed in that area. For completeness, the mean value of these areas and variances in terms of the standard deviation are shown in Tab.~\ref{tab:star_formation_rates} (under $A_{95,\rm d}^{\text{ext-nuc}}$). 

{The same assumption is used when using VLA images, where we consider that the total area of the large-scale continuum emission is represented by the image restored with the larger beam.} Additionally, for both cases, we use the variance of the flux measurements in the multiscale diffuse emission as an estimate of uncertainties for the SFR as well as the $\Sigma_{\rm SFR}$, which are also presented in Tab.~\ref{tab:star_formation_rates}.

\section{Discussion}
\label{sec:discussion}

\subsection{Leveraging Limited Sensitivity to separate the nuclear and large scale diffuse emission}
Interferometers with different resolutions and sensitivities pinpoint unique properties of radio sources. With \emph{e}-MERLIN we lack the necessary large-scale sensitivity to characterise the diffuse emission resulting from star formation on scales $\gtrsim$ 100 - 200 pc. Nevertheless, we are able to recover the radio flux density from primarily nuclear (mostly compact) regions. In contrast, the VLA is able to only partially resolve inner regions but performs well at mapping the radio emission of the source, encompassing both the diffuse synchrotron radiation from star formation and the flux originating from the compact/unresolved parts. Hence, when considering the multiscale paradigm, limited sensitivity is not an issue at all, but an advantage. If \emph{e}-MERLIN is detecting the flux mainly from AGN or SB, then it might be possible to subtract that flux density (and structure) from VLA images. Thereby, we can disentangle the radio power arising from different regions, such as circumnuclear star formation, AGN and SB.

Still, further considerations need to be addressed. For example, in UGC\,8696, a considerable portion of nuclear diffuse emission exists at scales of about $\sim$ 200 pc. {For single frequency observations,} to separate the core-compact flux from the nuclear diffuse flux, alternative methodologies must be employed. One could leverage other higher-resolution data, such as VLBI, and implement the technique outlined in Sec.~\ref{sec:interferometric_decomposition}, with \emph{e}-EMERLIN serving as the lower-resolution observation (similar to our use of VLA) or utilise the image fitting on images available at multiple spatial resolutions. {With our current data, in App.~\ref{sec:application} and  Fig.~\ref{fig:mcmc_example_fitting_example_UGC8696}-\ref{fig:mcmc_example_fitting_example_UGC8696_2}, we provide a comprehensive example of how to decompose the nuclear region of UGC\,8696.}

\subsection{Comparison between Methods}
The interferometric decomposition method in Sec.~\ref{sec:interferometric_decomposition} does not disentangle sub-structures in the higher resolution maps, as both core-compact and nuclear diffuse emission are used in the optimisation process. However, we can use the S\'ersic fitting from Sec.~\ref{sec:image_fitting} \mchh{to
separate fluxes from the core-compact region from that of the rest of the emission.}

The residuals provided by the interferometric decomposition approach (Sec.~\ref{sec:interferometric_decomposition}) are significant, $\sim 2$ mJy (Tab.~\ref{tab:image_decomposition_a}). This is an indication of relevant diffuse emission in the outskirts of the source, as can be seen in the lower-right plot of Fig.~\ref{fig:example_sub} (panel showing $\mathcal{R}_T$). {The size results for the dominant nuclear regions, which on average have sizes of $\sim$ 200 - 250 pc, agree with previous studies \citep[e.g.][]{Colina1992}.}  \cite{Song2021} estimated similar sizes for nuclear ring structures in normal SF galaxies and LIRGs. The extent of the nuclear regions is also compatible with the sizes of nuclear disks studied in \cite{Medling2014}.

The fractional fluxes $S_{\nu}^{\text{core-comp}}/{S_{\nu}^{\rm VLA}}$ are compared individually with literature values for each source in the App.~\ref{sec:imaging_results_sources}. 
We observe a significant contribution from emission originating in nuclear diffuse structures ($S_{\nu}^{\text{ext-tot}}/S_{\nu}^{\text{core-comp}}$), \mchh{especially} pronounced in sources with lower luminosities. This observation aligns with the \mchh{concept} that, within local galaxies, the most luminous U/LIRGs have a higher proportion of their total radio power coming from core-compact components {(AGN or SB) rather than diffuse components \citep[e.g][]{Rujopakarn2011}}. However, a substantial portion of this nuclear flux is still situated in dense nuclear regions, which can be strongly influenced by AGN and SB activity. As of now, it remains unclear what fraction of this flux should be attributed to star-formation processes, including \mchh{an} SB, or solely linked to the AGN. \mchh{Hence, we require deeper investigations of these radio structures, which can be robustly analysed using our multiscale approach, in combination with multiple observational data.}

\subsection{Limitations}
\label{sec:limitations}
\paragraph*{Size Estimates:}
The size estimates, converted from pixel area to radii via $A = \pi R^2$, are a good discriminant for spherical components, albeit not so appropriate for asymmetrical or elongated ones. For example, in the case of the VLA image of VV\,250 NW, $R_{95}$ is about 1.3 kpc. \mchh{However, by examining the scale bar on the radio map in Fig.~\ref{fig:results_cont_1}, we see that the semi-major axis and semi-minor axis are about $\sim$ 2.0 kpc and $\sim$ 0.5 kpc, respectively.} 
Hence, one dimension of the source is underestimated while the other is overestimated. In a future work, we will adopt a more robust approach to tackle such asymmetries.

\paragraph*{Deconvolved Sizes:} Deconvolved sizes are obtained using image fitting based on the parameter $R_n$. Furthermore, we have computed values such as $R_{50,\rm d}$ and $R_{95,\rm d}$ from individual model images within the deconvolved image space. In an ideal scenario characterized by infinite signal-to-noise ratio, both $R_n$ and $R_{50,\rm d}$ would exhibit similarity. However, due to the presence of noise in real data, coupled with statistical errors and convolution effects, a brief examination reveals that $R_n$ is not necessarily equal to $R_{50,\rm d}$ for a given fitted component. 

Additionally, we note that for complex emission, the fit can only provide an overall representation of the surface brightness distribution. That is also true for poor quality images. For example, in the case of VV\,250 (SE and NW), the emission recovered by \emph{e}-MERLIN contains low signal-to-noise, with unrecovered structures due to the amount of data flagged, {incomplete $uv$ coverage, and} possible calibration and imaging errors. The model describing those structures is a smooth profile that attempts to fit the overall emission of the source. Hence, the deconvolved size is not accurate and only represents the overall size of the radio structure.

\paragraph*{Data Limitation:} Our data is not able to resolve completely core-compact structures ($\lesssim$ 20 pc) and also not able to map very extended structures ($\gtrsim $ 5 kpc) since we are using the VLA-A configuration. When \emph{e}-MERLIN cannot resolve the most compact structures,  no information can be gathered regarding the origin of the radio {power for a single frequency. In that case,} the core-compact flux $S_{\nu}^{\text{core-comp}}$ will be overestimated. In the VLA end, the A-configuration is less sensitive to extended structures than the D-configuration or a single dish telescope, for example. When there is more flux from larger scales (>10 kpc), the total diffuse flux $S_{\nu}^{\text{ext-tot}}$ may be underestimated. One can use complementary data or available literature information to correct this flux density.

{By using only single-band observations (6 GHz), we can not conduct studies using spectral index maps in order to obtain better discriminants between AGN and SB. That would provide more accurate star formation estimates at the nuclear regions. For now, we are statistically limited in making conclusions on a broader picture of the physical processes in local U/LIRGs. Nevertheless, we focused this paper on introducing a methodology and testing it in a small sample with the available data, so that it will be applied in future work to a larger and multi-frequency sample.}

\section{Conclusions}
\label{sec:conclusions}
\subsection{Summary}

We have presented new high-resolution radio observations ($\sim$ 50 mas) using \emph{e}-MERLIN at C band (6.0 GHz) for four merging local U/LIRGs ($\sim$ 150 Mpc). We have combined this data with sensitive archival observations from the VLA at the same frequency using the A-configuration ($\sim$ 0.3" resolution). \mchh{Introducing two novel image-based methodologies, we extracted valuable physical insights from the radio maps at both nuclear regions ($\sim$ 20 $-$ 200 pc) and larger scales ($\gtrsim$ 1 kpc) (see Tabs.~\ref{tab:sizes_image_fitting} and~\ref{tab:image_decomposition}). Through this analysis, we robustly measured the sizes of the radio emission and computed fractional fluxes between core-compact structures and diffuse emission regions, interpreted as being associated with star formation processes. Consequently, we derived multiscale tracers for the extended flux density, $S_{\nu}^{\mathrm{ext-tot}}$ (refer to Tab.~\ref{tab:image_decomposition_b}), and star formation rate, ${\rm SFR}^{\text{ext-tot}}$ (see Tab.~\ref{tab:star_formation_rates}).}


{
We present below the key findings of our paper:
\begin{enumerate}
\item We have introduced two novel approaches to characterise the multiscale structure of the radio emission in local U/LIRGs, maximising the scientific output from interferometric data sensible to different angular scales. \mchh{Within individual limitations, our results showed that both methods agreed with each other. Our methods are adaptable to other radio frequencies and instruments, making them suitable for the analysis of larger datasets, from both existing and upcoming radio surveys. In particular, our image-fitting approach has potential applications beyond radio astronomy. For instance, in optical studies where S\'ersic modelling is extensively used to morphologically quantify the structural properties of elliptical, spiral, and lenticular galaxies.}
\item The nuclear emission associated with a nuclear disk ($\sim$ 50 - 200 pc), and excluding the most compact ($\lesssim $ 20 pc) emission from potential AGN/SB, is responsible for a significant fraction of the total radio emission at 6 GHz (Tab.~\ref{tab:image_decomposition_b}). Despite not separating the radio emission into AGN and SB components, we robustly determined the sizes and flux densities of the radio emission, providing upper and lower limits through two different ways:
\begin{enumerate}
\item Interferometric decomposition (Sec.~\ref{sec:interferometric_decomposition} and Tab.~\ref{tab:sizes_interferometric_decomposition}): For the nuclear regions across all sources, the estimated averaged effective size is $R_{50} \sim$ 113 pc for radio maps with intermediate resolutions (0.10" $\sim$ 0.25"). With \emph{e}-MERLIN, we found a lower-upper limit of $\sim$ 46 pc, while for VLA, an upper-lower limit of $\sim$ 190 pc. 
After removing the \emph{e}-MERLIN emission from VLA maps (e.g. Fig.~\ref{fig:example_sub}), the extended emission in pure VLA maps provided an averaged effective radius across all sources of $\langle R_{50}\rangle \sim$ 438 pc, with a total extent of $\langle R_{95} \rangle \sim $ 1326 pc. 
\item Image decomposition (Sec.~\ref{sec:image_fitting} and Tabs.~\ref{tab:sizes_image_fitting} and ~\ref{tab:image_decomposition}):
\begin{itemize}
    \item Core-compact components: in maps with intermediate resolutions, we obtained deconvolved effective sizes of $\langle R_{50,\rm d}\rangle \sim$ 67 pc for the core-compact components (averaged across all sources). The fitting on pure \emph{e}-MERLIN maps resulted in lower limits of $\langle R_{50,\rm d}\rangle \sim$ 20 pc, and the fitting in pure VLA images gave upper limits of $\langle R_{50,\rm d}\rangle \sim$ 72 pc. In respect to the full extent of these components, we obtained $\langle R_{95,\rm d}\rangle \sim$ 143 pc on intermediate maps, with  $\langle R_{95,\rm d}\rangle \sim$ 41 pc in \emph{e}-MERLIN and $\langle R_{95,\rm d}\rangle \sim$ 162 pc in VLA maps.
    \item Extended emission: With \emph{e}-MERLIN maps, after subtracting the contribution of the core-compact components (e.g. Tab.~\ref{tab:sizes_image_fitting} and Figs.~\ref{fig:mcmc_example_fitting_example_UGC8696}-\ref{fig:mcmc_example_fitting_example_UGC8696_2}), the nuclear extended emission resulted in an averaged effective radius of $\langle R_{50}\rangle \sim $ 104 pc, having a full extent of $\langle R_{95} \rangle \sim$ 192 pc. Alternatively, the large-scale diffuse emission probed by VLA resulted  $R_{50} \sim $ 447 pc and $R_{95} \sim $ 1.1 kpc, obtained by subtracting the unresolved components (e.g. Fig.~\ref{fig:example_fitting_example_VV705N}). 
\end{itemize}
\end{enumerate}
\mchh{
\item In each of the previously characterized regions, we were able to recover fractions of flux densities compatible with previous studies (see App.~\ref{sec:individual_source_analysis}). Additionally, our method enabled us to investigate how AGN fractions for the same source can exhibit both higher and lower values, biased by limited data. It is also worth mentioning that our results were derived from two interferometers, providing a robust characterisation. This serves as preliminary evidence of the new insights our approach can offer when utilising two or more interferometers.
}
\item From the calculated fractions of flux densities, we found that there is an interplay between the core-compact emission coming from AGN/SB regions in relation to the nuclear diffuse emission. The latter have a significant contribution to the total radio power, which resulted in larger estimates for the total multiscale extended flux density $S_{\nu}^{\mathrm{ext-tot}}$. It also brings down the fraction between the core-compact emission in relation to the total radio emission, $S_{\nu}^{\mathrm{core-comp}}/S_{\nu}^{\rm VLA}$. Hence, if not performing the multiscale structural decomposition from high-resolution to low resolution maps altogether, the contribution from core-compact components can be overestimated. It is noted also that, if we look at the total unresolved flux densities from VLA ($S_{\nu}^{\mathrm{core-unre}}$, Tab.~\ref{tab:image_decomposition}), they are higher almost up to a factor two in relation to the flux density from core-compact components probed by \emph{e}-MERLIN. \mchh{This means that almost half of the total flux density of a unresolved VLA component (at scales of $\sim$ 200$-$400 pc) is resolved on smaller scales, thus not being morphologically a core-compact structure. The other half, then, is core-compact at scales of $\lesssim$ 50 pc. However, we would require other instruments with higher angular resolutions to further dissect these fractions until we reach the sub-parsec physical limit regime.}
\item \mchh{For the broader context of the physics of U/LIRGs, we need to obtain accurate determination of the afformentioned flux densities fractions and associated sizes, understandand their connection with radio morphologies (at the local universe) and related physical mechanisms through spectral analysis. After that, we can study other objects in the distant Universe, where most of their structures will be unresolved, and then infer their properties from multiscale calibrated studies of local systems. Such high-redshift environments may contain sources with structures that harbour a substantial unresolved portion of their nuclear-diffuse {emission}}. We provided a detailed example of this for the local system UGC\,8696 (Figs.~\ref{fig:mcmc_example_fitting_example_UGC8696}-\ref{fig:mcmc_example_fitting_example_UGC8696_2}). 
\item We conclude that it is required to apply the aforementioned methodologies to the full galaxy sample with a wide range of frequencies in order to properly evaluate the fractions of core-compact fluxes, the multiscale diffuse emission and the fractions of thermal and non-thermal processes across the radio spectrum. We are particularly interested to uncover the origin of the nuclear diffuse emission and how it coexists with the central energy source. Understanding these factors will be essential on the study of high-redshift objects. 
\item We have shown that by using combined data with matching baselines and similar sensitivity, we can map the radio emission at different structural scales, therefore obtaining high-resolution images with improved sensitivity. This is true since, with VLA, we considerably reduce the short-spacing issue present in \emph{e}-MERLIN (see Fig.~\ref{fig:uv_coverage}). Hence, imaging combined observations with weighting towards \emph{e}-MERLIN will produce images that are not possible to construct when imaging \emph{e}-MERLIN data alone. The same applies to VLA, in respect to resolution. 
\end{enumerate}
}

\subsection{Future Work}
{
Our approach to quantify the sizes of the radio emission, fractions of flux densities and associated morphologies is independent of the instrument and also on the frequency of observation. This strategy can be used to disentangle more radio components, such as SB from AGN, and jets from diffuse emission. 
We take this opportunity to achieve improvements and turn our methodology completely automated in the near future. This will be essential for the 
next-generation instruments such as the Square Kilometre Array (SKA) and the Next Generation VLA (ngVLA), which will provide observations with high-resolution and simultaneously high sensitivity to all compact and diffuse scales.}

{
Follow-up work will consist in the same analysis conducted in this work, for the complete LIRGI sample and in a multi-frequency basis. Multi-frequency data from VLA from 1.4 GHz to 33 GHz will be combined with existing \emph{e}-MERLIN observations at L band (1.4 GHz) and C band (6 GHz). This will provide a more complete high-resolution sampling of the $uv$ coverage, as well better resolved spectral information for a multi-component and multiscale \mchh{study of the spectral energy distribution}. Thus, establishing a comprehensive understanding of the physical processes over the radio spectrum that drives the evolution of local U/LIRGs, and enabling a new framework to explore distant mergers.}

\section*{Acknowledgements}

We acknowledge financial support from Grant LINKB20064 (Spanish National Research Council Program of Scientific Cooperation for Development i-LINK+2020).
G.L. acknowledge financial support for a PhD studentship from STFC (grant 2627016).
J.M., M.P.-T. and A.A. acknowledge financial support from the grant CEX2021-001131-S funded by MCIN/AEI/ 10.13039/501100011033. J.M acknowledges financial support from the grant PID2021-123930OB-C21 funded by MCIN/AEI/ 10.13039/501100011033, by ``ERDF A way of making Europe'' and by the ``European Union'' and by the grant TED2021-130231B-I00 funded by MCIN/AEI/ 10.13039/501100011033 and by the ``European Union NextGenerationEU/PRTR''. M.P.T. and A.A. acknowledges financial support through grant PID2020-117404GB-C21, funded by MCIN/AEI/ 10.13039/501100011033.
J.M. acknowledges the Spanish Prototype of an SRC (SPSRC) service and support funded by the Spanish Ministry of Science, Innovation and Universities, by the Regional Government of Andalusia, by the European Regional Development Funds and by the European Union NextGenerationEU/PRTR. The SPSRC acknowledges financial support from the State Agency for Research of the Spanish MCIU through the ``Center of Excellence Severo Ochoa'' award to the Instituto de Astrofísica de Andalucía (SEV-2017-0709) and from the grant CEX2021-001131-S funded by MCIN/AEI/ 10.13039/501100011033.
This project has received funding from the European Union’s Horizon 2020 research and innovation programme under grant agreement No 101004719 (ORP).
We would like to acknowledge the support of the \emph{e}-MERLIN Legacy project ``LIRGI'', upon which this study is based. \emph{e}-MERLIN and, formerly, MERLIN, is a National Facility operated by the University of Manchester at Jodrell Bank Observatory on behalf of the STFC.
{Open access and page charges to this article will be paid by the University of Manchester. We would like to thank the anonymous reviewer for many useful questions and suggestions that helped improve the manuscript.}
The National Radio Astronomy Observatory is a facility of the National Science Foundation operated under cooperative agreement by Associated Universities, Inc.
This research has made use of the NASA/IPAC Extragalactic Database (NED), which is operated by the Jet Propulsion Laboratory, California Institute of Technology, under contract with the National Aeronautics and Space Administration. 


\section*{Data Availability}
Data used in this work is available under request to the corresponding author, in multiple formats, {such as phase-calibrated and self-calibrated visibilities (single and combined) and images.} Code documentation, development notes and Jupyter Notebooks to conduct the analysis are publicly available at the following GitHub repository: \url{https://github.com/lucatelli/morphen}.




\bibliographystyle{mnras}
\bibliography{references} 




\clearpage
\appendix
\section{Comments on Individual Sources}
\label{sec:imaging_results_sources}
\label{sec:individual_source_analysis}
The detailed analysis in the next paragraphs was derived with our multiscale methods using mulit-resolution radios maps, where some examples are shown in Fig.~\ref{fig:pre_results} and Fig.~\ref{fig:results_cont_1}. For reference on derived quantities, see Tabs.~\ref{tab:source_global_properties},~\ref{tab:sizes_interferometric_decomposition},~\ref{tab:sizes_image_fitting},~\ref{tab:image_decomposition} and~\ref{tab:star_formation_rates}.

\subsection*{VV\,705}
{\bf \emph{System Description} -- }
This is a binary merger system (classification M3 \citep{rupke2013,Larson2016}) with a nuclear separation of $\sim 6-7$kpc,  featuring long tidal tails ($\sim $40") revealed by HST data. VV\,705 N is brighter than VV\,705 S in the infrared. Over the past few years, uncertainties about the existence of an AGN remained. As well, the AGN contribution fraction is unclear: using the [N \textsc{II}]/H$\alpha$ diagram, \citep{Yuan2010} estimated AGN fractions of 0.5 and 0.3 for the north and south components, respectively, while no concrete evidence for an AGN was established by \citep{rupke2013}, giving a fraction of $10\%$ or less. But more recent studies suggest that this system is facing high star-formation and AGN activity, multiphase outflows of ionised gas and good evidence of feedback \citep[e.g.][]{Yuan2018,Perna2019}.

\paragraph*{VV\,705 N}
{\bf \emph{Global Properties} -- }
The higher resolution maps indicate a very core-compact component ($\lesssim$ 27pc) with some signs of nuclear diffuse emission ($\sim$ 76 pc). More weight towards VLA reveals a prominent and smoothly distributed emission around the unresolved/core-compact component, extending to a radial scale over  $\sim $ 2kpc. From our images, it is also possible to see a faint blob structure at the north region of VV\,705 N, located at a distance of $\sim$ 1.7 kpc from the core region (component 3 in Fig.~\ref{fig:example_fitting_example_VV705N}), which was not mentioned/identified in previous works (\citep{Iwasawa2011} in X-rays, \citep{loreto2015} at 33GHz and \citep{Vardoulaki2015} at 8GHz). For this system (N and S), the 6 GHz images in this work are the highest resolution presented until now in literature.  The total recovered flux for this source in \emph{e}-MERLIN is $\sim$ 7 mJy, while in VLA is $\sim$ 12.7 mJy. The half-light global radii are in average $\sim$ 109 - 235 pc while the maximum extent is $\sim$ 1.8 kpc.

{\bf \emph{Decomposition Results} -- }

Half-light sizes for the core-compact component are $R_{50}\sim$ 30pc ($R_{50,\rm d}\sim$ 23pc) while the full sizes are $R_{95}\sim$ 69pc ($R_{95,\rm d}\sim$ 52pc). This region contains a total integrated flux density of $\sim$ 3.7 mJy. Using $R_{50,\rm d}$ and Eq.~\ref{eq:Tb}, the estimated brightness temperature is 3.3 $\times$ 10 $^{5}$ K. This is a good indication of an AGN (> 10$^{5}$ K) which was identified by recent studies \citep{Yuan2018,Perna2019}. Regarding the nuclear extended component, the computed sizes are $R_{50} \sim $ 48 pc and $R_{95} \sim $ 77 pc, with a total integrated flux of $\sim$ 3.4 mJy.
For VLA images, the unresolved component encloses a total flux of $\sim$ 6.1 mJy, with estimated sizes of $R_{50,\rm d} \sim $ 56 pc and $R_{50,\rm d} \sim $ 140 pc. 

By using total nuclear extended emission with the VLA diffuse $S_{\nu}^{\text{ext}}$ the estimated total multiscale extended emission provides a total flux of $\sim$ 9.3mJy, so the fraction between the core-compact flux to the total radio flux at 6 GHz is $\sim$ 0.28, while if using the unresolved VLA component, it gives a fraction of 0.48, a factor of almost two. 

{\bf \emph{SFR Results} -- }
In the nuclear region we obtained SFR$^{\text{ext-nuc}} \sim 17.6$ \Moyr with a high nuclear surface density SFR of $\Sigma_{\text{ext-nuc}}^{95}\sim$ 454\Moyr kpc$^{-2}$. The total SFR for this source is $\sim$ 63 \Moyr (which already includes the nuclear contribution). This value agree with previous studies, 70-80 \Moyr \citep{Hickox2018,Paspaliaris2021,esposito2022}, but we still have to account for the S component below. 

\paragraph*{VV\, 705 S}
{\bf \emph{Global Properties} -- }
In \emph{e}-MERLIN the main core-compact component is visible with an emission size of $R_{50}\sim $ 19 pc and $R_{95}\sim $ 39 pc. There are no signs of significant nuclear diffuse emission.  Going \mchh{through} the lower resolution maps another structure is visible, at about 0.65 kpc NE of the nuclear region, which significantly contributes to the total flux (e.g. the half-light region encompasses this structure), and this component is not mentioned in previous studies. The total recovered flux for the S source is about 1.0 mJy in \emph{e}-MERLIN and 6.2 mJy in VLA. Half of the total VLA flux is enclosed in a radius of $R_{50}\sim$ 148-384 pc within a total maximum radial size of about 1.7 kpc. 

{\bf \emph{Decomposition Results} -- }
In \emph{e}-MERLIN, the core-compact component has an average half-light of $R_{50}\sim$ 21pc ($R_{50,\rm d}\sim$ 15 pc) and a full extent of $R_{95}\sim$ 42 pc ($R_{95,\rm d}\sim$ 29 pc) containing $\sim$ 1.0 mJy in flux. For the brightness temperature, this  yields $T_B\sim $ 0.4 $\times$ 10$^{5}$K. Additionally, nuclear extended emission is almost absent (or negligible, $<$ 0.2 mJy, within errors). 

With combined data, the nuclear region has an average size of $R_{95}\sim $ 258 pc ($R_{95, \rm d} \sim $ 138 pc). On pure VLA images, a significant amount of diffuse emission appears as sensitivity is gained, with sizes of $R_{50} \sim $ 583 pc ($R_{50,\rm d} \sim $ 513) up to a full size of $R_{95} \sim $ 1.3 kpc ($R_{95,\rm d} \sim $ 1.1 kpc).

\mchh{
The fraction of core-compact flux density in relation to the multiscale diffuse flux density is 0.14 for VV\,705 S. For comparison, when using the total unresolved VLA component, that fraction would be 0.36. Combining the fractions between VV\,705 N \& S, the total ratio between the flux density of core-compact components in relation to the total radio flux density is 0.24. This value agrees well with the measured fraction of 0.25 by \cite{Dietrich2018} using the infrared bolometric luminosity. This result is larger than the 0.1 measured by \cite{rupke2013}. Additionally, if we use
}
the unresolved VLA component as the core-compact structure, the fractions 0.48 and 0.36 agrees with \cite{Yuan2010} using [N II]/H$_{\alpha}$ diagrams, where 0.5 and 0.3 were obtained, respectively for N and S sources.

{\bf \emph{SFR Results} -- }
The nuclear diffuse flux density is almost negligible, so no attempt was made to compute SFR and $\Sigma_{\rm SFR}$. The total SFR resulted in SFR$^{\text{ext-tot}}\sim$ 37\Moyr with a density of $\Sigma_{\text{ext-tot}}^{50,\rm d} \sim $ 15 \Moyrk. The SFR is higher than the value of $\sim$ 11\Moyr determined by \cite{Yuan2018}. Hence, combining the SFR from sources N \& S  sums to an estimate of $\sim$ 100\Moyr which is close to values obtained by \cite{rupke2013,DeLooze2014,Paspaliaris2021}.



\subsection*{UGC\,5101}
{\bf \emph{System Description} -- }
UGC\,5101 is a ULIRG of merger class 5 \citep{haan2011,U_2019} and classified as Sy 1.5, LINER. 
Previous studies pointed out that this source has a compact unresolved core smaller than $\sim$ 200 pc \citep{Soifer2000}, and VLBI observations show that the nuclear region is composed of 3 main core-compact components \citep{Lonsdale2003}, with diameters smaller than $\sim$ 4 pc. There are multiple indications that UGC\,5101 hosts an AGN \citep{Imanishi_2001,Lonsdale2003,Grimes_2005,Iwasawa2011,U_2019,Dietrich2018}. 

{\bf \emph{Global Properties} -- }
Our \emph{e}-MERLIN images feature two clear components. The main one contains a large fraction of the flux density, while the faint south-east component contains about $\sim$ 5\% of the total \emph{e}-MERLIN integrated flux density. The main structure is not completely resolved, as can be seen by its asymmetric shape. The reason is that, from the VLBI study, its nature is represented by the two core-compact sources which turns to split into five smaller components. The half-light averaged size of the nuclear region probed by \emph{e}-MERLIN is $\sim$ 41pc.  The faint diffuse blob in the southeast region of the main component seen in our maps is revealed to be a diffuse component by the VLBI study. Additionally, there are some signs of diffuse emission when using a more natural weighting scheme in the higher resolution maps, giving an extent of $R_{95} \sim $ 175 pc. The total integrated flux density recovered in the nuclear region with \emph{e}-MERLIN is $\sim$ 37 mJy.

In the lower resolution maps (VLA), one feature of the emission is that the core region has a different orientation in relation to the diffuse emission, and this orientation is also distinct from the overall orientation of the nuclear region probed by \emph{e}-MERLIN. It is possible also to see a second component in the north direction of the core region, with a total integrated flux density of about $\sim$ 1 mJy. At larger scales, the half-light radius of UGC\,5101 is $R_{50} \sim$ 132-214 pc, reaching a full extent of $R_{95} \sim$ 1.7 kpc. The total integrated flux density recovered by VLA is $\sim$ 60 mJy. We additionally report that the full extended emission is only recoverable when using a natural weight during deconvolution (restoring beam of 0.60" $\times$ 0.49", lower rightmost plot in Fig.~\ref{fig:pre_results}), and that structure represents a fraction of 0.1 of the total flux density.

{\bf \emph{Decomposition Results} -- }
From the higher resolution maps, the estimated total size of the core-compact region is $R_{95}\sim $ 82 pc ($R_{95,\rm d}\sim $ 60 pc deconvolved) with a half-light radius of $R_{50} \sim $ 29 pc ($R_{50,\rm d} \sim $ 14 pc deconvolved). This component encloses a total flux density of $S_{\nu}^{\text{core-comp}} \sim $ 24 mJy. Using the deconvolved size for the core-compact component, we obtain a value of $\sim$ 19 $\times$ 10 $^{5}$ K for the brightness temperature, representing a good indication for an AGN. With respect to the nuclear diffuse emission, the estimated extent is $R_{95}\sim $ 240 pc ($R_{95,\rm d}\sim $ 177 pc deconvolved) with a half-light radius of $R_{50} \sim $ 130 pc ($R_{50,\rm d} \sim $ 101 pc deconvolved). The integrated flux density of the nuclear diffuse emission is $S_{\nu}^{\text{ext-nuc}} \sim$ 18 mJy.  

The unresolved component by VLA yields a total flux density of $\sim$ 33 mJy and the VLA extended flux density is around $\sim$ 26 mJy. As before, by removing the core-compact flux, we estimate that the total diffuse flux density (small scales and large scales) is $S_{\nu}^{\text{ext-tot}} \sim$ 37.5 mJy and the fraction $S_{\nu}^{\text{core-comp}}\ /S_{\nu}^{\rm VLA}$ is 0.38. For comparison, by using the unresolved VLA component, that fraction would be 0.53. Results from literature for this fraction span a wide range of measurements, using different approaches. We just report some of those values: \citep{Dietrich2018} found a value of  0.76 while \cite{DiazSantos_2017} provided 0.25 using combined diagnostics for the bolometric AGN fraction and a value of 0.41 using combined MIR diagnostics, which is close to our estimate.

{\bf \emph{SFR Results} -- }
The star formation rates for this source is significantly high, in total $\sim$ 254 \Moyr and at the nuclear region, the estimated value is also significant, $\sim$ 95 \Moyr. The total SFR is considerably higher than previous studies, which pointed to: $\sim$ 100 \Moyr \cite{esposito2022,Yamada2023} using X-ray and radio wavelengths; and $\sim$ 25 \Moyr \citep{Dietrich2018} using IR luminosity. However, the value found in this work is similar to \citep{DeLooze2014}, which used a far-infrared tracer, getting an estimate of $\sim$ 223 \Moyr. 

The surface densities of SFR are 227 \Moyrk for the nuclear diffuse emission and $\sim$ 420 \Moyrk within the half-light area after minimising the contribution from the core-compact flux. By observations made by \cite{Lonsdale_2003}, which point out the possible location of an AGN on smaller scales, we conclude that these high star formation rates using radio emission as a tracer are plausible. We have not considered the effects of jets in this computation, as our data is limited to detect any evidence of their presence.

\subsection*{UGC\,8696 N - NE - SW}
{\bf \emph{System Description} -- }
Also known as Mrk\,273, it is a late-stage merger and dual-AGN ULIRG and classified as a Seyfer 1 \citep{U_2013,Liu_2019,U2022}. It is also classified as AGN by X-ray observations \citep{Yamada2021}. This is a source that shows significant changes in radio emission across multiple scales. {High-resolution observations by \cite{Klockner2004} using EVN showed OH maser emission within a circumstellar disk at the N component. Also, \cite{Bondi2005} provided high resolution observations of a dense nuclear starburst at the nuclear region of the N component.
Additionally, recent works provided} good indications of multiphase outflows \citep{Zaur2014,Lutz2020,Zubovas2022} that can extend up to 5 kpc \citep{Tadhunter2018} as also pinpointed by LOFAR observations \citep{Kukreti2022}. These latter work argues about the possibility that this source harbours three AGNs instead of two.



{\bf \emph{Global Properties} -- }
The \emph{e}-MERLIN imaging results reveal with enough sensitivity the diffuse emission around the N component as well resolution to partially resolve the inner region into three main components (see Fig.~\ref{fig:mcmc_example_fitting_example_UGC8696}) where we see the main component at the centre, one spherical structure in the SW direction at a distance of $\sim$ 70 pc and one elongated structure in the NE direction. The sizes of this region in \emph{e}-MERLIN is $R_{50} \sim $ 81 pc up to $R_{95} \sim $ 192pc, with a total integrated flux density of $\sim $ 40 mJy.

In the VLA image, on kpc scales, the source becomes more complex, where sensitivity is enough to map more components: NW, NE and SW (SW1 and SW2). We can clearly see these structures in \citep{Vardoulaki2015,Kukreti2022}. We report that for this source in particular, it was necessary to perform nine steps of self-calibration (i.e. 3$\times$ (\texttt{phase+phase+ampphase})) to be able to recover the extended structures and remove deconvolution artefacts in the NW component. 

Since the structure of this source is complex, the sizes are just an estimation of the overall morphology of the radio emission, which in VLA provided an extent of $\sim$ 0.9kpc. For the total integrated flux density, we were not able to recover from the maps the already reported value of 60 mJy (including the SE component, see below) \citep{loreto2017}. However, inspecting the amplitude versus $uv$ distance, the amplitude at the shortest baselines appears to converge to $\sim$ 60 mJy.

\paragraph*{UGC\, 8696 N}
{\bf \emph{Global Properties} -- }
By using \emph{e}-MERLIN, we are able to resolve the nuclear region into two main features: a noticeable circumstellar disk of diffuse emission around a small region, which may be the location of at least three core-compact components. These were observed with EVN/VLBI in a much higher resolution (5 mas) in \citep{Bondi2005} where evidence is presented supporting the idea of a high supernova rate. These authors also found that the dominating region size where extreme star formation is taking place is of about $\sim$ 30 pc. The sizes computed in \emph{e}-MERLIN maps were $R_{50} \sim $ 87 pc and $R_{95} \sim $ 206 pc, with a total flux density of $\sim$ 31.6 mJy. To determine $T_b$ for this source, we have used only the deconvolved fitted size of the main component, i.e. $\mathrm{ID_1}$ in Fig.~\ref{fig:mcmc_example_fitting_example_UGC8696}, which provided a value of $T_b \sim$ 6.7 $\times$ 10$^{5}$ K.

Moving to the large scale structure of this source, we notice a very clear relation between the alignment of the nuclear region components with the emission on scales > 1 kpc, where new components are visible -- here, we use the same labels as \cite{Kukreti2022} --- NE, NW and SW. The SW region consists of two blobs (SW1 and SW2) with a distance of $\sim$ 0.85 kpc and $\sim$ 1.6 kpc from the N component. On the north-west side, $\sim$ 1.2 kpc apart from N, there is another blob of emission. Each one of these have, roughly, a total integrated flux of $\sim$ 1 mJy. On the NE side of the source, located at $\sim$ 1.6 kpc from N, we see a diffuse component ($\sim$ 1 mJy) that is almost aligned with the N and SW components. We consider components SW2 and NW as part of blob/core-compact emission, while the NE and SW1 as diffuse emission.

{\bf \emph{Decomposition Results} -- }
For the core-compact region, we obtained half-light radii of $R_{50} \sim$ 27 pc ($R_{50,\rm d} \sim$ 22 pc) and having a full extent of $R_{95} \sim$ 53 pc ($R_{95,\rm d} \sim$ 44 pc).  The integrated flux density within these components is about 10 mJy. Strinkingly, the component that most contributes to the total flux density is from the diffuse emission (see Sec. \ref{sec:application}), in total, $\sim$ 24 mJy. Our size estimates gives $R_{50}\sim $ 140 pc and $R_{95} \sim $ 281 pc, with similar deconvolved counterparts.

Taking the total corrected diffuse emission of $\sim$ 40 mJy against the core-compact flux, 14.5 mJy (the sum of both two core-compact regions, see discussion for UGC\,8696 SE below), the calculated fraction is 0.36. That is larger than the 0.08 using X-rays studies from \cite{Yamada2023} and similar to 0.34 by \cite{Veilleux2013} in relation to the bolometric luminosity combining different tracers (see their paper for details). Our value is less than half of the one provided by \cite{Dietrich2018}, 0.66.

{\bf \emph{SFR Results} -- }
The complexity of the N component is clear. Taking only the diffuse emission of the circumstellar disk, we estimate a nuclear SFR$^{\text{ext-nuc}}$ of $\sim$ 157 \Moyr while the surface density is the highest between our sources, $\Sigma_{\text{ext-nuc}}^{95} \sim $ 1000 \Moyrk. The surface density for the VLA images considering the half-light area resulted in an estimate of $\Sigma_{\text{ext-tot}}^{50,\rm d} \sim $ 542 \Moyrk.   


\paragraph*{UGC\, 8696 SE}
{\bf \emph{Global Properties} -- }
The SE component of UGC\,8696 is a core-compact source. With our \emph{e}-MERLIN observations, we were not able to probe extended emission. In \cite{Bondi2005}, with higher resolution observations, there is evidence of diffuse emission around this core-compact component, at scales smaller than 10 pc. This can explain the fact that even unresolved, the SE component is not completely symmetric. In VLA images, we can see a faint diffuse structure (< 1 mJy) towards the SW direction. We also see a connection between the N and SE components, but in this case, it may be due to the fact that both (N and SE) are unresolved and almost blended.


{\bf \emph{Decomposition Results} -- }
For this source, no decomposition was needed because a single model component was enough to describe the source structure. 
 We obtained a deconvolved half-light radii of $\sim$ 21 pc with a full deconvolved extent of $R_{95,\rm d} \sim$ 42 pc.
 After removing the core-compact component, which yielded an integrated flux of density of $\sim$ 5 mJy, no significant sign of diffuse emission is present. Using its core-compact flux density and the deconvolved sizes, we obtain a brightness temperature of $T_b$ 5.2 $\times$ 10 $^{5}$ K, which is enough to support that UGC\,8696 SE host an AGN. Also, since that both VLA and \emph{e}-MERLIN can not resolve this component, no star-formation estimates were attempted.  


\subsection*{Accurately measuring the core-compact nuclear regions of UGC\,8696: Is it a dual AGN?}
\label{sec:application}
When the radio emission from a nuclear region appears to be compact, additional data should be used to verify \mchh{whether} the emission is physically compact \citep[e.g.][]{Biggs2010} -- AGN  -- or a dense region dominated by pure star formation and/or a compact nuclear star-cluster and/or a \mchh{starburst}. \mchh{If an AGN is confirmed, it becomes necessary to separate its flux density from the total source flux density to accurately determine star formation rates.}

We take as an example the case of UGC\,8696. Using \emph{e}-MERLIN observations, it is clear that the core region of this source features diffuse emission embedded with the other three components. When using interferometric decomposition for a multi-feature source it is almost impracticable to perform the selection of an optimised threshold as both structures coexist and a significant fraction of the extended emission is added to $I_{1}^{\rm mask}$ (Eq.~\ref{eq:I1_mask_opt}, see also Fig.~\ref{fig:int_decomposition}) to be removed from the VLA image. In this sense, we apply a detailed analysis in the main nuclear region of this source using the image fitting approach in order to disentangle the diffuse flux from these other compact components. 

One of the combined images of UGC\,8696 N was selected, where the three components are partially resolved (using a robust parameter of -1.0). We then performed a minimisation in order to recover their deconvolved physical sizes and fluxe densities. After that, the nuclear diffuse emission was separated and the total integrated flux density obtained for the three compact components was $\sim$ 10 mJy (plus the $\sim$ 4.5 mJy from the SE component), while the flux density on the nuclear extended component resulted in $\sim$ 30 mJy. In Fig.~\ref{fig:mcmc_example_fitting_example_UGC8696} we provide the minimisation results alongside a Monte-Carlo simulation for each one of the effective radii (e.g. N source only). We also show the radio maps for the deconvolved and convolved images, as well the resulting nuclear diffuse emission.

From this analysis, we may infer that the contribution to the total flux in the nuclear region is 10 (N) + 4.5 (SE) mJy for the core-compact components, i.e. 14.5 mJy (see Fig.~\ref{fig:mcmc_example_fitting_example_UGC8696}) and the contribution to nuclear diffuse emission becomes $\sim$ 27 mJy, totalling $\sim$ 40 mJy for the full diffuse emission, small scales, and larger scales. Therefore, the fraction between the compact to full diffuse flux density results in $S_{\nu}^{\text{core-comp}}/S_{\nu}^{\text{ext-tot}}\sim$ 0.35 (see Tab.~\ref{tab:image_decomposition_b}) which agrees with \cite{Veilleux2013,Lutz2020,DiazSantos_2017}. Also,the fraction between the core-compact with the nuclear diffuse is 0.48.

There are some debates about UGC\,8696 hosting a dual AGN \citep{U_2013,Iwasawa2011,Liu_2019} (N and SW) and even a triplet adding the SE component \citep{Vardoulaki2015,Kukreti2022}. 
Considering the compactness of the SE component, its brightness temperature and observations from \cite{Carilli_2000,Bondi2005} one can possibly point out the existence of an AGN. For the SW component, using X-ray \cite{Iwasawa2011,Liu_2019} demonstrated evidence of a dusty, buried and absorbed AGN. However, at 6 GHz, this component is only detected in VLA and is not visible in \emph{e}-MERLIN, which leaves unclear its location. We require using multi-wavelength radio data in a future work to pinpoint this.

\subsection*{VV\,250}
{\bf \emph{System Description} -- }
VV250 is a composite early-stage merging system (predecessor-merger [PM] \citep{Jin2019a}; M2 \citep{Larson2016} ) with two morphologically similar sources, having a nuclear separation of $\sim 22$ kpc \citep{Larson2016}. The origin of the radio emission is yet unclear: H \textsc{II} emission was detected 
\citep{Leech2010}; \cite{Vardoulaki2015} suggest that the SE source is SB in nature due to its flat spectrum while that characterisation is uncertain for the NW source; \cite{Hattori_2004} suggests that most of the H$_{\alpha}$ flux originates by a compact source in the SE source and that the H$_{\alpha}$ flux in the NW source is extended in nature.

\paragraph*{VV\,250 SE}
{\bf \emph{Global Properties} -- }
On our higher-resolution maps, this source is faint ($\sim$ 1 mJy) \mchh{in the shape of a smooth blob}. However, flux \mchh{density} estimates are uncertain because of the low signal-to-noise ratio of the resulting map. The overall shape of the emission with a more natural weighting reveals a dense nuclear region with asymmetric structure featuring multiple blobs of emission. For VV\,250, deconvolution was performed by using a sky taper function of 0.025 and 0.05 arcsec in order to improve the signal-to-noise of the image in the balance of resolution, which helped to deal with PSF effects caused by \emph{e}-MERLIN \citep[e.g.][]{Muxlow2020,Harrison2020}. 
\mchh{
Since the nuclear region features some diffuse structures, the \emph{e}-MERLIN half-light radius is larger than the sizes of the other sources, $\sim$ 80 pc, extending up to 158 pc. The same quantity derived from higher resolution maps provided $\sim$ 166$-$241 pc, having a maximum extent $R_{95} \sim$ 1.3 kpc on VLA maps. The total flux density recovered by \emph{e}-MERLIN is $\sim$ 8.6 mJy and $\sim$ 18 mJy with VLA. 
}

{\bf \emph{Decomposition Results} -- }
From the \emph{e-}MERLIN observation, we can infer that the nuclear region is composed by a smooth distribution of flux, but characterized by the presence of two blobs at the centre. We report that the following decomposition results are approximations, since we have poor signal-to-noise in our images. We also had to use tapering to recover the emission of this source, therefore size estimates represent upper limits. 

The estimated sizes are $R_{50} \sim $ 61 pc ($R_{50,\rm d} \sim$ 64 pc) and $R_{95} \sim $ 115 pc ($R_{95,\rm d} \sim $ 96 pc). 
The nuclear diffuse emission has a circular radius of $R_{95} \sim$ 113 pc ($R_{95,\rm d}\sim $ 89 pc deconvolved). The decomposed nuclear emission is highly uncertain. If the two mentioned blobs are considered as core-compact, the total integrated flux density is below $\sim$ 6 mJy, hence the nuclear diffuse flux density is $\gtrsim$ 2 mJy. \mchh{For this source, the value for $T_b$ is estimated to be of the order of $0.2 \times 10^{5}$ K. }

In VLA maps, the extent of the diffuse emission is about 0.8 to 1.3 kpc, with a total integrated flux density above 11.7 mJy. It was not possible to measure exactly the total extended emission in nuclear regions in combination with the diffuse VLA scales due to uncertainties in quantifying the flux density in \emph{e}-MERLIN. Considering that the core-compact structure is represented by the two main components at the centre, we estimate that  its contribution to the flux density is below 0.34. A prediction made by \citep{diaz_santos_2017} points to a fraction below 0.05.

{\bf \emph{SFR Results} -- }
The surface density star formation rate in the nuclear region of this source is lower than the others, $\Sigma_{\text{ext-tot}}^{95} \sim $ 154 \Moyrk and the value for the nuclear SFR is $\sim$ 7 \Moyr. We see, however, that in both these measurements, they contain a large fractional error. 

If we consider all the nuclear emission as SF, the density of SF is not that high.  Hence, we require additional data to investigate this further, to probe if the most compact emission is actually core-compact or pure SF in origin. We note that the morphology of this source is similar to that of NGC\,1614 \citep[e.g.][]{Olsson2010,Herrero-Illana_2014}, i.e. the nuclear region is driven by star-formation processes. Our total SFR estimated for this source is $\sim$ 57 \Moyr. This value is plausible with previous measurements of 75 \Moyr \citep{DeLooze2014}, 64 \Moyr \citep{Vivian2012} and 55 \Moyr \citep{Cao_2016}, in contrast to an estimate of $\sim$ 110 \Moyr from \citep{Paspaliaris2021,Howell2010}.





\paragraph*{VV\,250 NW}
{\bf \emph{Global Properties} -- }
Similarly as the SE component, NW is faint with no clear signs of a core-compact region, but in the form of a blob, it shows some evidence of a dense nuclear region. In general, for this source, our \emph{e}-MERLIN observations are not sensitive enough to probe the origin of radio emission at the nuclear region. When using a natural weighting imaging, no emission with reasonable signal-to-noise is visible in \emph{e}-MERLIN. 
We made images including all the VV\,250 components. However, the best images of the NW component were obtained with a combination of $uv$-tapering at 0.025", and a Brigg's weighting tending to natural-like values (using a robustness parameter between 1.0 and 2.0). In the case of combined data, we obtained images with \emph{e}-MERLIN features also using the same sky-taper and a \texttt{robust} of $\sim$ -1.0, which yielded a slight better image in relation to the previous case. In both cases, we obtained an emission with a smooth distribution of surface brightness, of about $R_{50} \sim$ 50 pc and $R_{95} \sim$ 79 pc, with an approximated integrated flux density of 1.5 mJy. Nevertheless, to unravel the properties of the nuclear region of VV250 NW on parsec scales, we require further observations with better sensitivity. 

On larger scales, diffuse emission is clear, extending over an average radial linear distance of 1.3 kpc from the centre (with a semi-major axis of $\sim$ 3.0kpc). The total flux density recovered by VLA is $\sim$ 3.5.

{\bf \emph{Decomposition Results} -- }
Using the images with a restoring beam modified by a $uv$ taper of 0.025", we were able to fit a two-component model to the \emph{e}-MERLIN image, recovering the sizes of the blob at the nuclear region of $R_{50} \sim $ 45 pc and $R_{50, d}\sim$ 26 pc. Due to the poor signal-to-noise of our images, we only obtained a crude estimate for the most compact region, ranging from $\sim$ 0.5 to $\sim$ 1.0 mJy. The full radial sizes of the nuclear diffuse component resulted in $R_{95} \sim $ 98 pc and $R_{95,\rm d} \sim $ 73 pc, with a flux density of $\sim$ 0.9 mJy. For $T_b$, using the previously calculated sizes and using the flux density from the most compact component, we obtained a value smaller than $\sim$ 0.1 $\times$ 10$^{5}$ K. 

Using VLA images, we recovered the upper-limits of the unresolved component to be $R_{50,\rm d} \sim $ 86 pc and $R_{95,\rm d} \sim $ 187 pc, which is a bit larger than the nuclear-diffuse size mapped by \emph{e}-MERLIN.  The diffuse component in VLA was modelled with one component, giving a half-light radii of $\sim$ 407 pc and with a full averaged radial extent of $\sim $ 697 pc.

{\bf \emph{SFR Results} -- }

The nuclear star formation rate is about $\sim $ 1.7 \Moyr, but highly uncertain. However, considering the same nature of this source as VV\,250 SE and NGC\,1614, we can compute the nuclear star formation rate considering the total radio emission, providing $\sim$ 4 \Moyr, with a nuclear surface star formation density of $\sim $ 51 \Moyrk (again, likely uncertain). These values are reasonable if we consider that the nuclear region is purely dominated by star formation. For the total star formation rate, considering all scales,  we obtain an estimate of $\sim$ 11 \Moyr. Previous calculations of SFR for VV\,250 NW resulted in $\sim$ 11 \Moyr \citep{DeLooze2014} and $\sim$ 8 \Moyr \citep{Cao_2016,Dutta2018a}.

\section{Multi S\'ersic Decomposition}\label{app:multi_sersic}
{
To conduct the analysis of the current work, we started the development of data processing tools combining functionalities with other existing astrophysical packages (e.g. \textsc{Morfometryka}, PetroFit, CASA, PyBDSF). These will facilitate the analysis of radio astronomical images, image structural decomposition, common tasks for radio interferometry (e.g. self-calibration and imaging) and upcoming features that will be implemented as we progress on this research. Code availability, usage instructions and additional information can be found at the GitHub repository called \textsc{Morphen}\footnote{\url{https://github.com/lucatelli/morphen}.}.}

\subsection{Subcomponent detection of the radio emission}
\label{sec:sub_region_analysis}

To identify multiple radio components, being core-compact, blobs or extended, we approach the problem by performing a source extraction analysis from the radio emission maps. However, carrying out the identification of multiple components in radio images, especially in wide-field images from radio surveys, still presents challenges, due to the complexity of radio structures \citep[e.g.][]{Song_2022}. Additionally, fitting model components to complex or even simple image structures can be problematic if the initial parameters of the model are not well set up \citep[see][for examples in optical studies]{Haussler_2007,Andrae2011}, or at least, not constrained to an interval of values that represent the actual physical structure of the source. 

In this work, we attempt to disentangle the radio emission into different components: core-compact/unresolved; nuclear diffuse and large-scale diffuse emission. Below, we list the general steps behind this, to construct suitable initial conditions for the minimisation, which has proved to be very effective:
\begin{enumerate}
\item Firstly, cutouts are made for all images (\emph{e}-MERLIN, combined and VLA) around the radio emission and centred in reference to the position of the main \emph{e}-MERLIN component. The image sizes are large enough to enclose all the diffuse VLA emission. 
\item Secondly, a source finding is applied to a combined image ($I_2$) in order to identify the locations of relevant radio emission. These regions are labelled and sorted by total flux density (see an example in the lower left panel of Fig.~\ref{fig:example_fitting_example_VV705N}). In this routine, the minimum subcomponent detection size (or its total number of pixels) is delimited to have the same size as the number of pixels within the restoring beam of that particular image -- this helps with issues related to our over-sampled VLA images. 
\item Then, for each one of these regions, relevant structural properties can be extracted via basic photometry, such as the local peak of flux and its position, the local half-light radius \mchh{$R_{50}$} and the contour at $R_{50}$, which is interpreted as a possible initial guess for $I_{50}$. Other critical quantities are also recorded, such as the major and minor semi-axis of the emission, $R_a$ and $R_b$ respectively, and the position angle $PA$. To perform these initial measurements, the Python packages \textsc{PetroFit} \citep{Geda_2022} and \texttt{SEP} \citep{Bertin1996,barbary2016} are used. To maximise run-time efficiency and minimise fitting issues due to the complexity of the mathematical approach and also sources structures, the coordinates $(x_0,y_0)$ of the detected structures (IDs) are fixed to their detection positions, with a free interval of $+/-$5 pixels. Providing these initial conditions to the model-fitting minimisation proved critical in the code's performance because we are relying on constraining the parameter space \mchh{with} intrinsic properties computed on the actual emission. This can be generalised and used for any particular astronomical image where the S\'ersic model can be used as a proxy (e.g. optical studies). The only parameter that is not initially constrained in this study is the S\'ersic index $n$ \citep[for that, see][]{Andrae2011,Breda2019}, but we keep it to $n=0.5$ or $n=1$ in some exceptions when modelling diffuse emission.
\item The general case of a core-compact region is that it is surrounded by nuclear diffuse emission, e.g. circumstellar regions of star formation and accretion processes. In this case, a single mathematical model may not be able to describe the emission completely. For each detected structure, a single-component fitting is performed, and then we analyse the residuals. Any significant residual flux can be linked to (nuclear) extended emission, therefore careful inspections are conducted to check if another model component must be added to the fitting, with the same coordinates as the parent ID object (see the bottom panel of Fig.~\ref{fig:example_fitting_example_VV705N}). If that is deemed true, the minimisation is repeated. Since large-scale structures are more asymmetric, the parameter range for the central coordinates of these additional components can be larger than the detected ones, usually $\pm$ 30$-$50 pixels. Currently, this is done manually by specifying the parent ID where a new component should be added so that the emission can be successfully modelled with two or three components\footnote{In future versions, this will be done automatically by computing a first-order residual map on the nuclear emission as well others regions. If significant residuals are present, then that is an indication that a diffuse emission component is required to be modelled altogether.}.
\item Before the fitting, additional options can be changed, such as keeping the S\'ersic index fixed or not; as well as the positions of the components and providing the maximum ranges. 
\end{enumerate}

After the above steps, we proceed with the minimisation. The fitting implementation in this work is performed with the python package \textsc{LMFIT}\footnote{See at \url{https://lmfit.github.io/lmfit-py/}.} \citep{imfit_software2014} using a non-linear least-squares minimisation, based on the Levenberg-Marquardt and the Trust Region Reflective \citep{Branch1999} methods. Each optimisation is run twice for better convergence of parameters. We use this to compute the standard errors from the covariance matrix. The parameters that can be minimised are those from equations~\ref{eq:sersic_law} and~\ref{eq:general_ell}: $I_n$, $R_n$, $x_0$, $y_0$, $q$, PA and, when requested, $n$ and $C$.

Due to the complexity of radio emission, this implementation was made available online with examples. For optimisation purposes, it is designed to run using standard Python libraries in combination with \textsc{JaX} \citep{jax2018github}, which features auto-multiprocessing and/or CUDA GPU acceleration (if present). 
Comprehensive examples are provided in the online documentation\footnote{\label{foot:code_usage}Python notebooks, packages, and documentation are made available online at \url{https://github.com/lucatelli/morphen} and are expected to receive updates and be maintained in the future.}.

\subsection{Fitting the deconvolved image plane}
Each interferometric data has a different restoring beam, and a combined observation will have one which is intermediate to the two interferometers alone \citep{Muxlow2005}. So, for each image cleaned with different weights  will have distinct restoring beam shapes. Their shapes are defined by the averaged distribution of baselines and weights. In the context of image synthesis, the beam size can be thought of as the size of the Gaussian PSF that blurs the real image of the source brightness distribution to produce what we actually see on a radio map -- which means, convolution \citep{thompson2017}. {Therefore, accessing information from the deconvolved data would provide closer measurements to the true structure of the source, such as true sizes.}

In the context of image fitting, it would be ideal to get the true physical sizes of components, instead of the convolved ones. Consider that $\mathfrak{D}$ is the observed data (radio map) and $\mathfrak{M}$ is the model fitted to it. In the image decomposition approach, one can work the minimisation problem $\min (\mathfrak{D} - \mathfrak{M})$ in two different ways:
\begin{enumerate}
\item minimize the parameters of a {convolved model} such as 
\begin{align}
    \label{eq:convolved_mini}
    \min(\mathfrak{D} - \mathfrak{M}) \sim 0 ; \quad {\rm or}
\end{align}
\item minimise the parameters of a {deconvolved model} such as 
\begin{align}
    \label{eq:deconvolved_mini}
    \min(\mathfrak{D} - \texttt{PSFBEAM}*\mathfrak{M})\sim 0.
\end{align}
\end{enumerate}
where \texttt{PSFBEAM} is the elliptical PSF Gaussian representing the restoring beam of an image. In this work, we use the second approach and for that, we use the CASA function \texttt{componentlist} to generate a 2D PSF image with the same size/shape as the restoring beam and this PSF will be used subsequently during the image decomposition implementation.

\subsection{Modelling the Radio Emission}
{
It is known that there are degeneracy issues when fitting multiple S\'ersic functions to a common region of an image \citep[e.g][]{DeJong2004,Andrae2011}. However, multiple functions can be fitted on distinct regions. We model the radio emission of the galaxies in our sample through a combination of S\'erscic functions (Eq.\ref{eq:sersic_law}). In general, we have 
\begin{align}
\mathfrak{S}^T(R) = \sum_i^{N} \mathfrak{S}^i(R). 
\end{align}
where $N$ is the total number of model components fitted to the data. The exact $N$ is determined by performing component extraction (labelled by IDs, lower-left panel of Fig.~\ref{fig:example_fitting_example_VV705N}) and by inspecting the continuum radio images. Typically, $N$ is not larger than $N=3$, and in Sec.~\ref{sec:sersic_caveats} we further discuss a way to minimise problems associated with multiple components. After that, for each detected structure ID, we have looked into \emph{e}-MERLIN and VLA images to spot how many structures can be characterised with such modelling. This inspection is relevant to conclude when a region can not be modelled by a single S\'ersic or Gaussian function (a model component ``COMP\_'').}

When the radio emission is complex, asymmetric and extended (i.e. not following shapes described by elementary mathematical functions) we do not attempt to achieve an accurate model by fitting multiple components. Instead, we choose to adjust a component that captures the overall shape of the emission. This  leaves significant residuals, however, in this work, we analyse and quantify residual properties to understand the total output coming from faint/diffuse residual emission.

The minimisation strategy is based on $\chi^2$  provided by \textsc{LMFIT} in order to match the model $\mathfrak{M}$ against the data $\mathfrak{D}$. To simulate a more robust model, we include in the minimisation the residual map $\mathfrak{Re}$ that was generated in the interferometric deconvolution. 
{\color{black}
However, we do not add the actual residual image because it is already a convolved image, and would require to obtain a deconvolved version of the residual map. Hence, we apply a random pixel shuffling transformation to the residual map. This transformation will remove the effect of convolution and therefore simulate a deconvolved background noise image, denoted by $\mathfrak{N}_{\rm bkg}$,
\begin{align}
\mathfrak{N}_{\rm bkg} = S_{\oplus} \mathfrak{Re}.
\end{align}
{where $S_{\oplus}$ is a random shuffling operator.}
The critical point here is that the signal level of this mock residual is exact the same as the original residual image, so that its total flux density is conserved. We also add a flat sky component $\mathfrak{F}$ to take into account any additional offset.} Therefore, the final model image is written as 
\begin{align}
\mathfrak{M} = \mathfrak{S}_{\nu}^{T} + \mathfrak{N} + \mathfrak{F}
\end{align}
The final metric for the optimisation consists of $\min (\mathfrak{D} - \mathfrak{M})$. Note that the data image $\mathfrak{D}$ consisting of radio emission is actually the \textsc{wsclean} (or \textsc{CASA}) model convolved with the restoring beam added to the interferometric residual map, that is $$\mathfrak{D} = \texttt{BEAM}*\mathfrak{M}_{\rm clean} + \mathfrak{Re}.$$ 
Therefore, this ensures that there are no residual flux offsets when $\mathfrak{D}$ is subtracted from $\mathfrak{M}$.

\subsection{Solving Caveats on Multi-S\'ersic Image Decomposition}
\label{sec:sersic_caveats}

Multi-S\'ersic decomposition is known for presenting challenges during minimisation and by resulting in the degeneracy of parameters \citep[e.g.][]{Andrae2011}. Also, small perturbations of the input of initial conditions can cause large variations in the outcome optimisation. 
To work around this issue, we have found a very suitable way to help with this minimisation problem and try to minimise the effects of initial conditions on the outcome. We use basic a \emph{priori} photometric information computed before the minimisation, for example, via a simple Petrosian analysis (Sec.~\ref{sec:sub_region_analysis} and Fig.~\ref{fig:example_fitting_example_VV705N}). In this strategy, the positions of subcomponents and their $R_{50}$, $I_{50}$, $q=R_b/R_a$ and $PA$ are calculated. This provides good starting points and constraints for the minimisation, which is naturally based on intrinsic properties of the source structure. This prevents the algorithm for searching a non-physical parameter space. This is an approach that was never been done before, even in optical image decomposition. We suggest that this approach can be generalized and easily incorporated within other imaging data. We are planning a future work to conduct a deeper analysis on simulated data in order to evaluate the reliability of the method.

\section{Size Relations}\label{app:sizes_complement}
\subsection{Relation between the S\'ersic $R_n$ and the Gaussian FWHM}
In order to relate the effective radius of a S\'ersic function when $n=0.5$ (Gaussian equivalent), namely $R_{n=0.5}$, and the FWHM of a Gaussian, namely $\theta$, we can equal the S\'ersic profile $\mathfrak{S}(n,R)$ with the Gaussian distribution $G(R)$,
\begin{align}
G(R) = a \exp\left\{ -\frac{R^2}{2c^2}\right\}, \quad {\rm FWHM} \equiv \theta = 2\sqrt{2 \log 2}c
\end{align}
then, setting $n=0.5$ in Eq. \ref{eq:sersic_law} we have
\begin{align}
\mathfrak{S}(n=0.5,R) &= G(R)\nonumber\\ 
a \exp\left\{ -\frac{R^2}{2c^2}\right\}
&=
I_{n=0.5} \exp\left\{ 
-b_{0.5} \left[ 
\left( 
\frac{R}{R_{n=0.5}}
\right)^2 - 1
\right]
\right\}
\end{align}
Since this holds for any $R>0$, we can set by convenience $R = R_{n=0.5}$, which simplifies to 
\begin{align}
\label{eq:R05_ext}
\frac{a}{I_{n=0.5}} = \exp\left\{ 
\frac{R^2_{n=0.5}}{2c^2}\right\} \quad \Rightarrow \quad
R^2_{n=0.5} = 2 c^2 \log\left( 
\frac{a}{I_{n=0.5}}
\right).
\end{align}
The effective intensity $I_n$ relates to the scaled intensity (or amplitude) of the Gaussian as 
\begin{align}
I_n = I_0 e^{-b_n} \quad {\rm with} \quad  a\equiv I_0 \qquad [{\rm because}\ G(0) = \mathfrak{S}(n=0.5,0)]
\end{align}
therefore, for $n=0.5$ we have $I_{n=0.5} = a e^{-2/3}$. Using this and $c = {\rm FWHM}/(2\sqrt{2\log 2})$ in Eq.~\ref{eq:R05_ext} we obtain 
\begin{align}
\label{eq:Rn05_fwhm}
R_{n=0.5} = \frac{{\rm \theta}}{\sqrt{6 \log 2}} \approx \frac{{\rm \theta}}{2}.
\end{align}
So, the effective radius of a S\'ersic profile with a S\'ersic index equal to $n=0.5$ is half of the FWHM of an equivalent Gaussian function.

\subsection{Deconvolved Sizes}
\label{sec:deconvolved_sizes_app}
We may also express the relation between the convolved and deconvolved sizes for both FWHM and $R_n(n=0.5)$. For a Gaussian-shaped radio source, when observed by a beam having semi-major and minor axis $\theta_{\rm maj}$ and $\theta_{\rm min}$ (mean beam width of $\theta_{1/2}$), the relation between the true size of its major axis $\phi_{\rm maj}$ with its convolved semi-major size $\varphi_{\rm maj}$ {is \citep[e.g.][]{Condon_1998,Murphy_2017}}
\begin{align}
\label{eq:phi_maj}
\phi_{\rm maj} = \sqrt{\varphi_{\rm maj}^2 - \theta_{\rm maj}^2}
\end{align}
and similarly for the minor axis, 
\begin{align}
\label{eq:phi_min}
\phi_{\rm min} = \sqrt{\varphi_{\rm min}^2 - \theta_{\rm min}^2}.
\end{align}

The link between $R_n$ when $n=0.5$, namely $R_{n=0.5}$, and {$\phi$} -- the associated Gaussian deconvolved FWHM -- is (see Eq. \ref{eq:Rn05_fwhm})
\begin{align}
R_{n=0.5} \approx \frac{\phi}{2}
\end{align}
Consider now that the convolved effective radius is ${R}_{n=0.5}^{*}$ and its deconvolved counterpart is ${R}_{n=0.5}$ (as before), hence
\begin{align}
{R}_{n=0.5} = \frac{\sqrt{4 ({R}_{n=0.5}^{*})^2 - (\theta_{1/2})^2}}{2}. 
\end{align}
This means that ${R}_{n=0.5}^{*}\approx \texttt{PSFBEAM}*{R}_{n=0.5}$. Note that these relations only hold for Gaussian functions and $n=0.5$. In this sense, when the brightness distribution of the radio  emission is not Gaussian-shaped, we face challenges in recovering the true physical sizes. We, therefore, propose to optimise a minimisation problem considering the convolution with the \texttt{PSFBEAM} in Eq. \ref{eq:deconvolved_mini} so that the best parameters of the model represents deconvolved quantities.

{Usually,} single core-compact radio sources follow a Gaussian distribution. Hence, if the half-light radii $R_{50}$ can be measured in the convolved image plane via simple statistics, a good approximation is to set $R_{50} \sim {R}_{n=0.5}^{*}$ and therefore one can obtain the approximated deconvolved half-light radii $R_{50,\rm d}$ via 
\begin{align}
{R}_{50,\rm d} \approx \frac{\sqrt{4 {R}_{50}^2 - \theta_{1/2}^2}}{2}. 
\end{align}
This is particularly interesting since we do not require fitting model components to the image, hence obtaining the deconvolved effective radius is straightforward.

\subsection{Limits of Resolution}
\label{app:lim_res}

In order to check our limits in estimating sizes, we can make some analysis using the over-resolution power that an interferometer has to resolve compact sources \citep{Kovalev_2005,Popkov2021}
\begin{align}
\label{eq:overres}
\theta_{\rm lim} = \theta_{\rm maj} 
\sqrt{
\frac{4 \ln 2}{\pi} \ln
\left( 
\frac{{\rm SNR}}{{\rm SNR} -1 }
\right) 
}
,\quad 
{\rm SNR} \approx \frac{S_{\rm p}}{\sigma_{\rm image} \sqrt{N_A-3}}
\end{align}
where $N_A$ is the number of antennas (i.e. 6 for our \emph{e}-MERLIN data\footnote{The Lovell Telescope was not used in our observations.}) and $\sigma_{\rm image}$ is the local standard deviation of the image where the radio component is.

For a particular source, UGC\,5101, it features a compact and faint component south-east from the nuclei, see \emph{e}-MERLIN map in Fig.~\ref{fig:results_cont_1}. This component has a size equal to the restoring beam size and is detectable with enough signal-to-noise. The beam FWHM for this image is 0.03" (or $\sim$ 24 pc) but using Eq. \ref{eq:overres}, we obtain that $\theta_{\rm lim} \approx $ 0.013" or $\approx$ 10 pc, which is particularly the diameter of the component located at the south-east position ($\sim$ 30 pc) from the central core. This tells us that the size of that component is smaller than 10 pc. This is confirmed by \citep{Lonsdale2003} and, in fact, this structure is completely resolved using VLBI observations.

\section{Extra material}
\begin{figure*}
\centering
\includegraphics[width=0.54\linewidth]{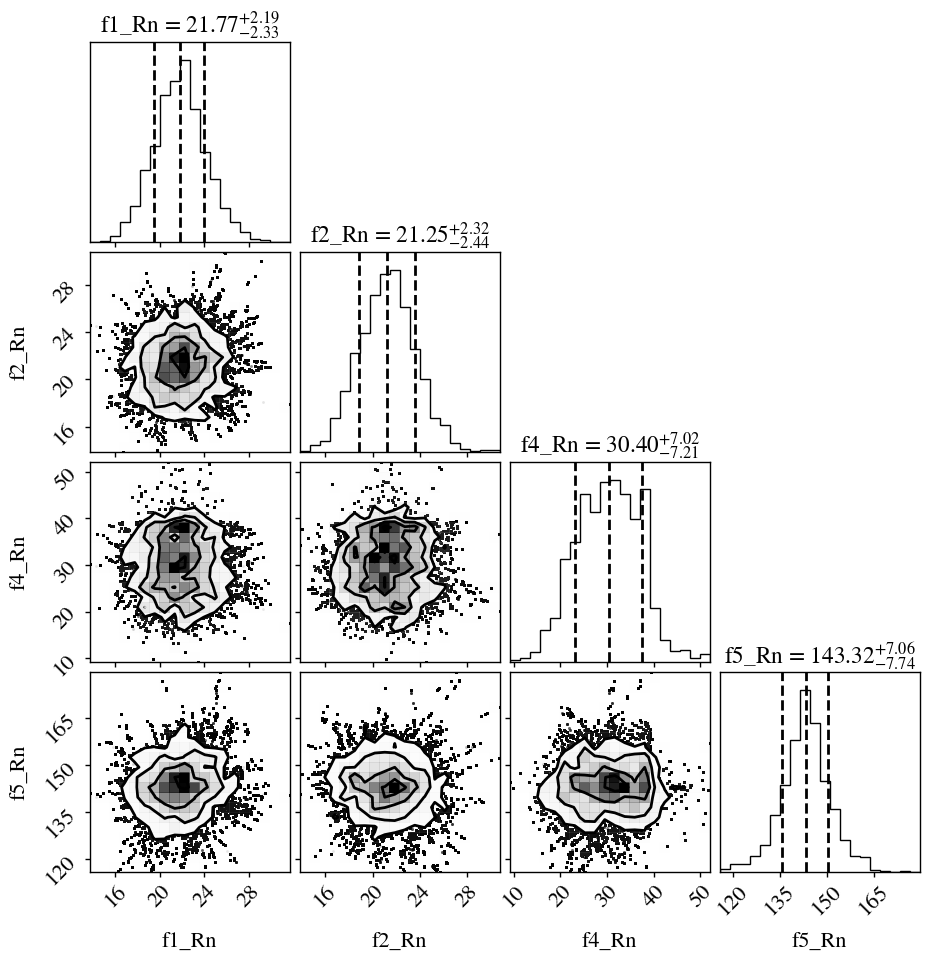}
\includegraphics[width=0.44\linewidth]{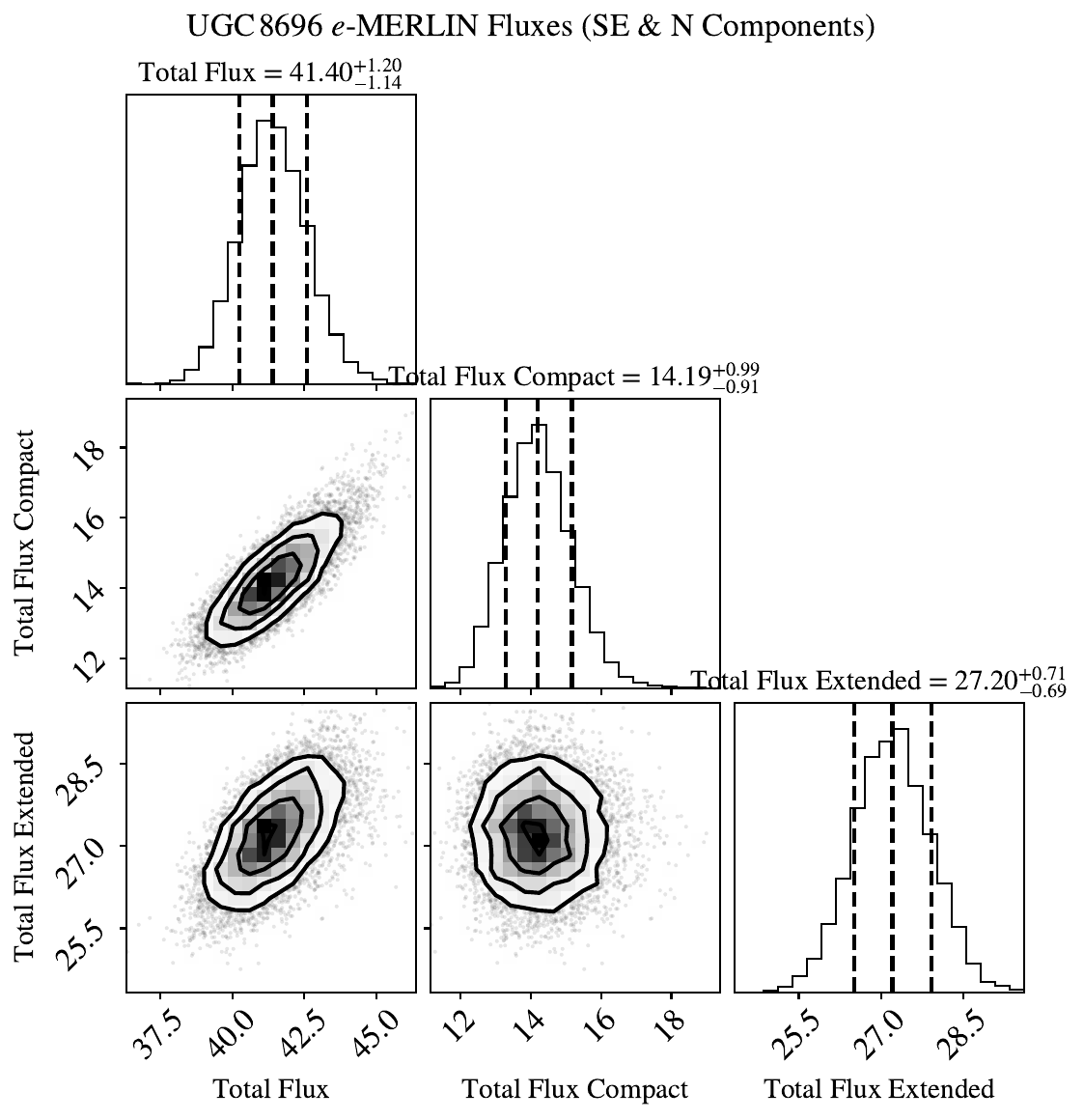}
\caption{MCMC distribution for the size model parameters (left) in terms of $R_n$ (given in parsec) and disentangled flux densities (right) of the nuclear region of UGC\,8696, featuring three core-compact components. Including the SE component of UGC\,8696 (not shown in Fig.~\ref{fig:mcmc_example_fitting_example_UGC8696_2}), the total core-compact flux density is $\sim$ 14 mJy. The nuclear diffuse component, containing a major fraction of the total flux density, is $\sim$ 27 mJy in the \emph{e}-MERLIN image.  The total multiscale extended flux density sums up to 39.9 mJy when the VLA is taken into account.}
\label{fig:mcmc_example_fitting_example_UGC8696}
\end{figure*}

\begin{figure*}
\centering
\includegraphics[width=0.26\linewidth]{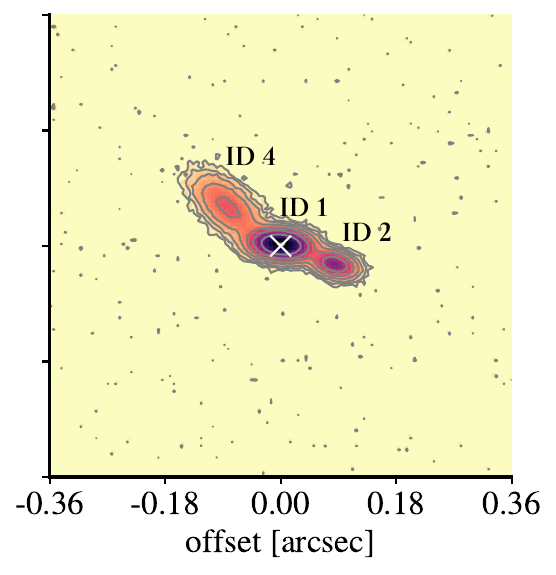}
\includegraphics[width=0.26\linewidth]{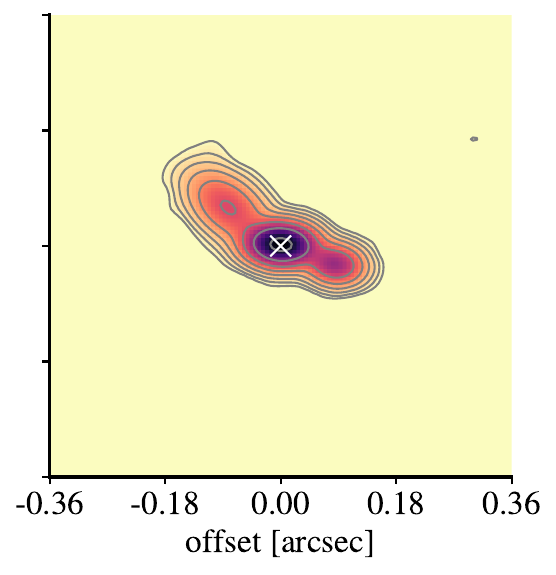}
\includegraphics[width=0.26\linewidth]{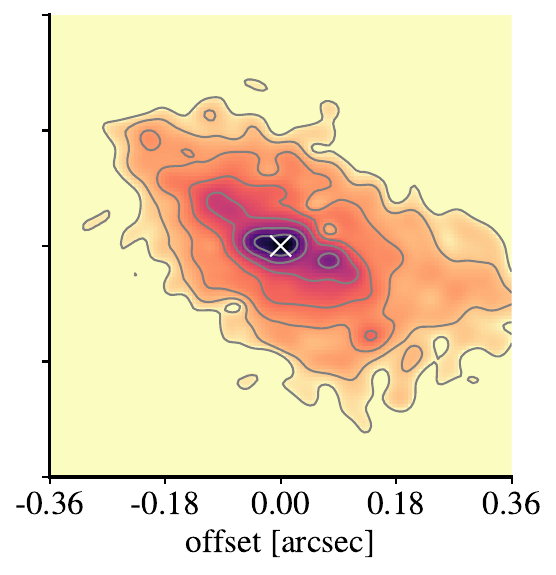}
\includegraphics[width=0.26\linewidth]{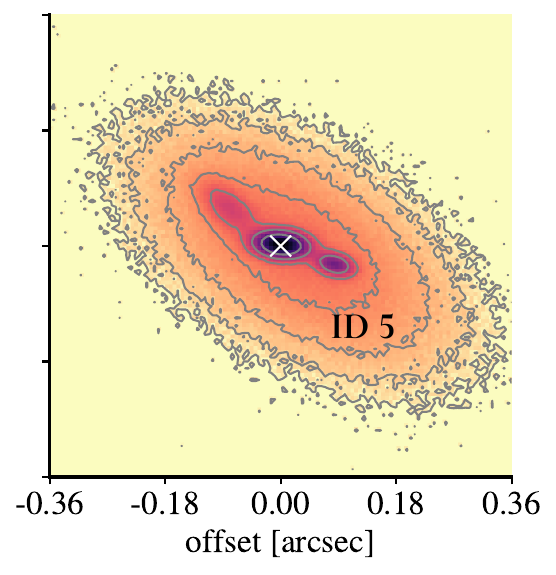}
\includegraphics[width=0.26\linewidth]{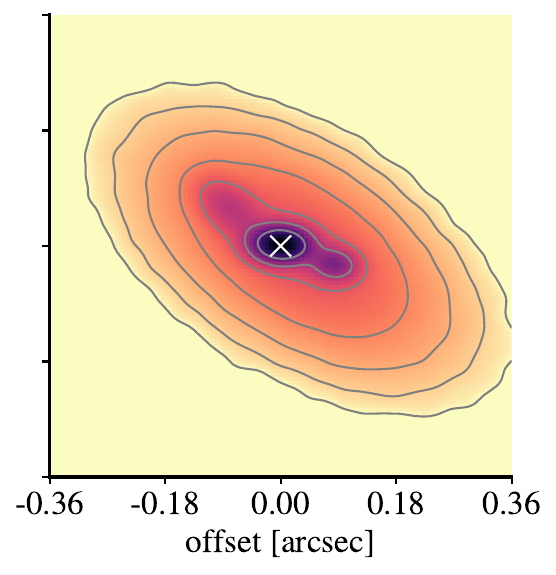}
\includegraphics[width=0.26\linewidth]{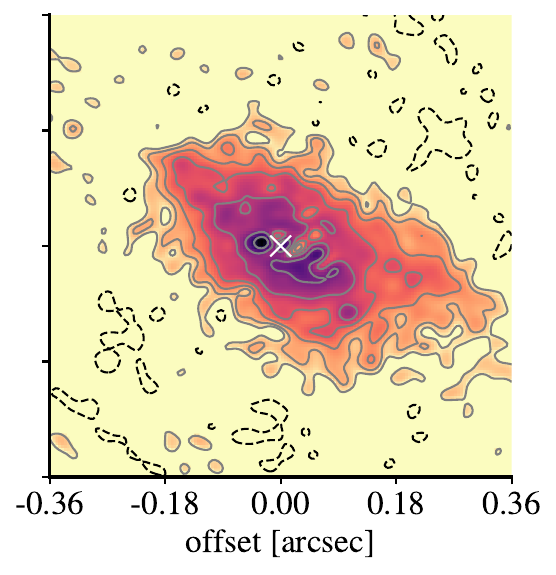}
\caption{
{
Top row, from left to right: Deconvolved core-compact regions, convolved core-compact regions and the original radio map. Bottom row, from left to right: total deconvolved modelled emission, total convolved model emission and the recovered nuclear extended emission. Total flux densities for this emission are shown in Fig.~\ref{fig:mcmc_example_fitting_example_UGC8696}.}}
\label{fig:mcmc_example_fitting_example_UGC8696_2}
\end{figure*}

{
We present here extra figures complementary to the main text discussion.
Notes to some figures: 
\begin{itemize}
    \item In Fig.~\ref{fig:mcmc_example_fitting_example_UGC8696}, the flux densities for the diffuse and core-compact slightly differ from that of Tab.~\ref{tab:image_decomposition_b} because those values were computed in multiple images, and in Fig.~\ref{fig:mcmc_example_fitting_example_UGC8696}, in just one image.
    \item In Fig.~\ref{fig:mcmc_example_fitting_example_UGC8696_2}, the south component of UGC\,8696 (would be \texttt{ID\_3}) is absent in the plots. We just wanted to highlight the N component.
\end{itemize}
}

\begin{figure}
\centering
\includegraphics[width=0.95\linewidth]{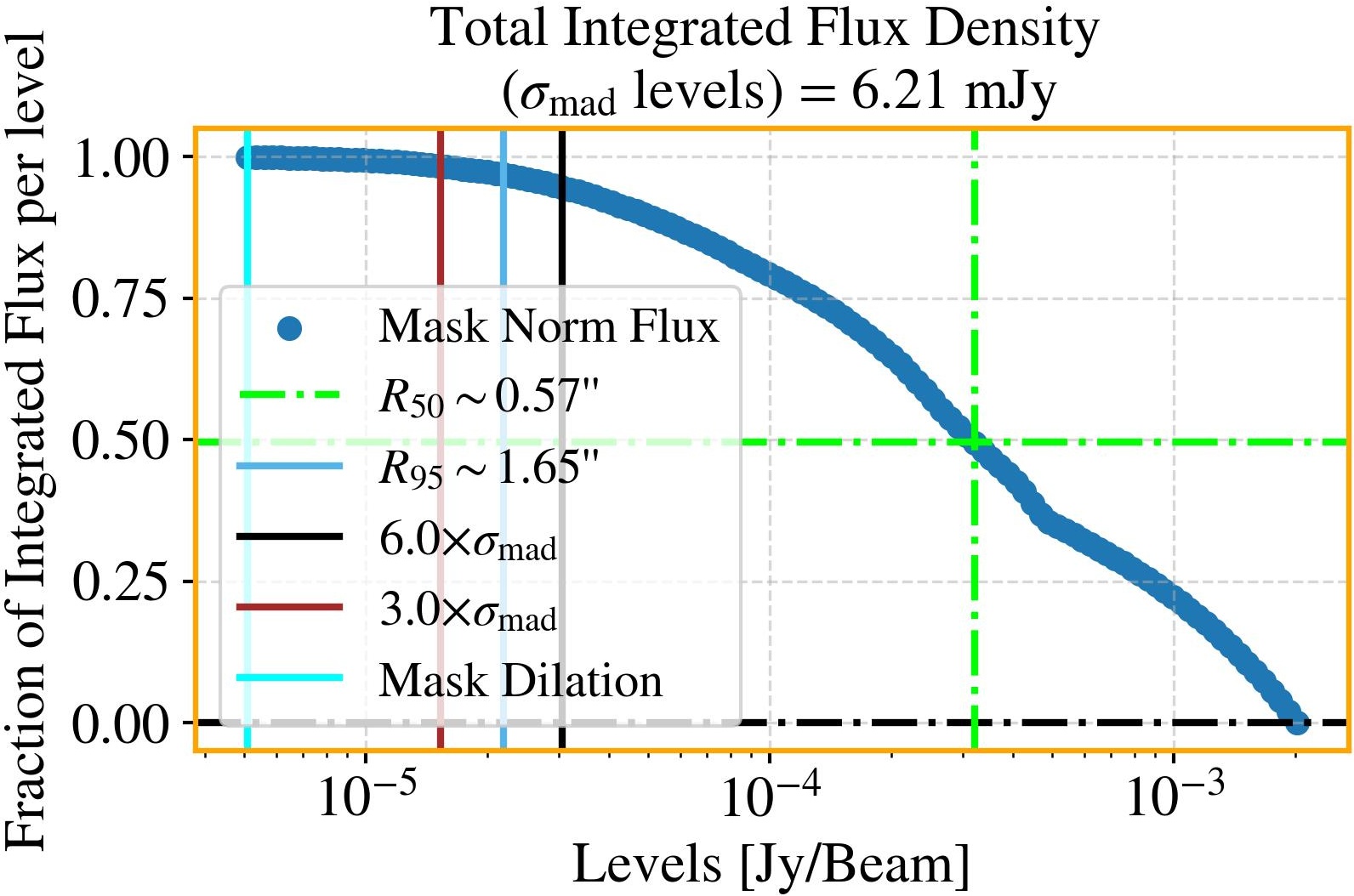}
\includegraphics[width=0.60\linewidth]{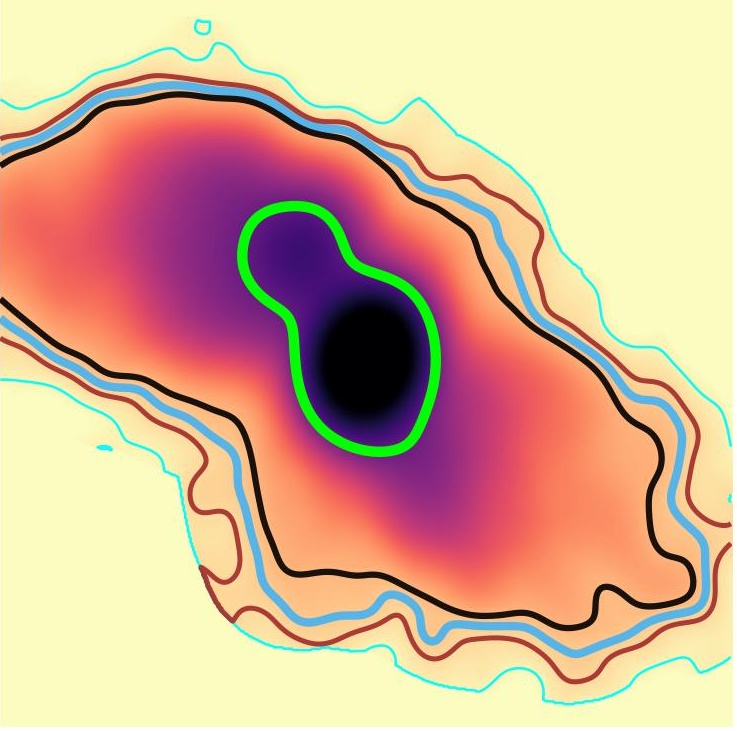}
\caption{Flux density estimation using the mask dilation approach (see Sec.\ref{sec:mask_dilation}). Top: Total fraction of integrated flux density per level from the image. Bottom: Radio emission with relevant contour levels (see below). First, a mask at $6\sigma_{\rm mad}$ is created (black solid contour) and then dilated (cyan solid contour). The value for sigma is taken to be the $\sigma_{\rm mad}$ of the cleaned residual image from deconvolution with \textsc{wsclean} (or CASA). The lime green solid line represents the region enclosing half of the total flux density, which is converted to a representative circular radius using $A = \pi R^{2}$. The light-blue solid line represents the region of $95\%$ of enclosed flux density (for these sizes, see Sec.~\ref{sec:image_shape_analysis}). The brown line indicates the $3\sigma_{\mathrm{mad}}$ region.}
\label{fig:segmentation_decomposition}
\end{figure}

\begin{figure*}
\centering
\includegraphics[width=0.80\linewidth]{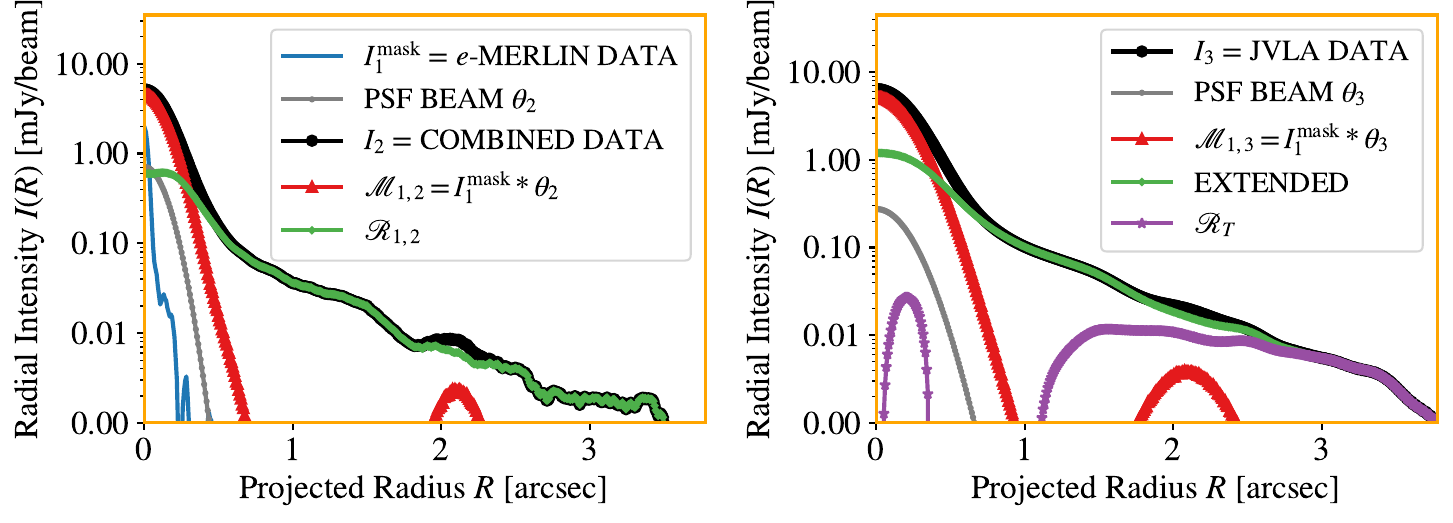}
\caption{Radial profile intensities highlighting the results from Eqs.~\ref{eq:R12},~\ref{eq:M12},~\ref{eq:M13_M23} and \ref{eq:RT}, and Fig.~\ref{fig:example_sub}. Each label corresponds to the same labels showed in Fig.~\ref{fig:example_sub}.}
\label{fig:int_decomposition_radial_profiles}
\end{figure*}

\begin{figure*}
\centering
\includegraphics[width=0.33\linewidth]{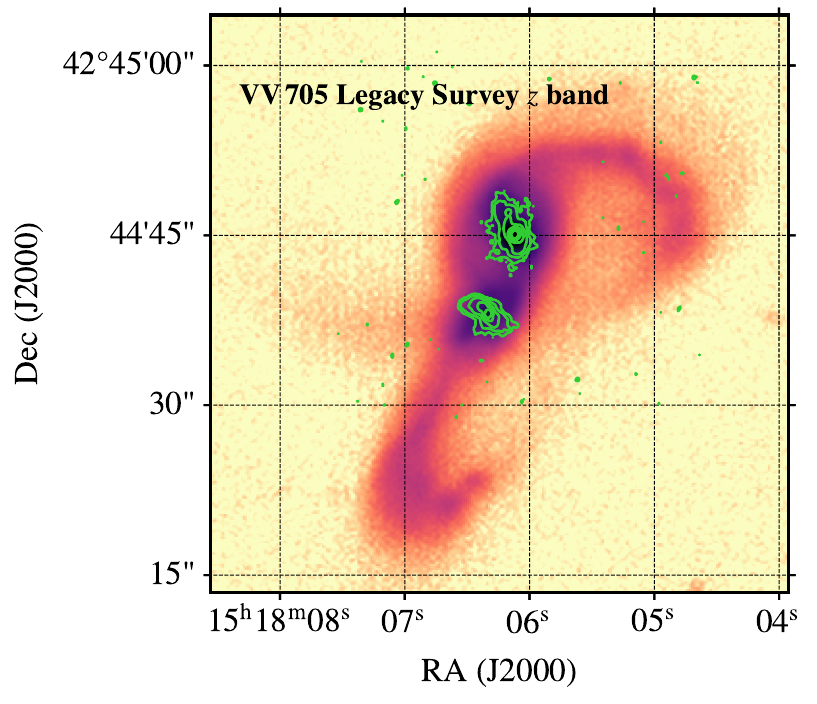}
\includegraphics[width=0.33\linewidth]{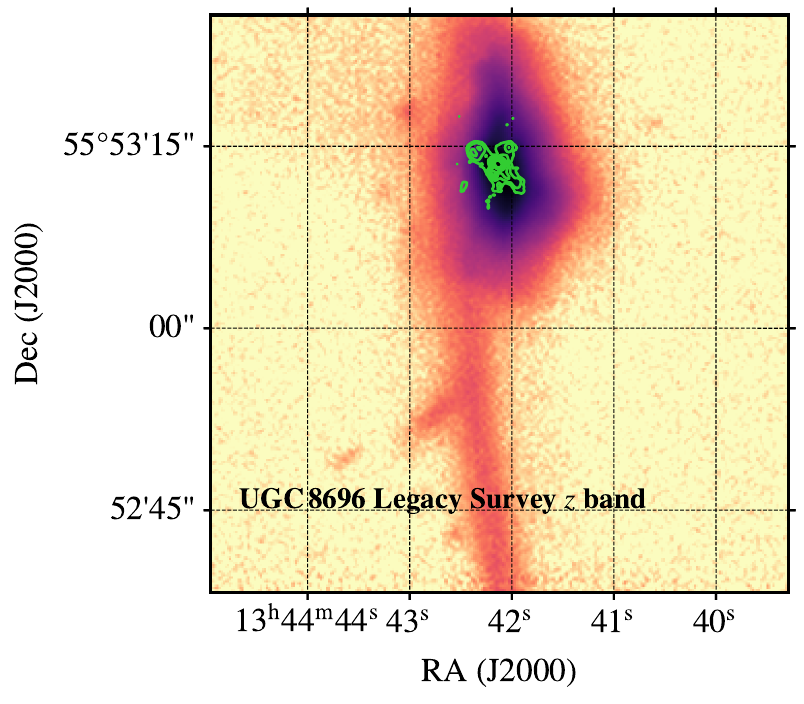}
\includegraphics[width=0.33\linewidth]{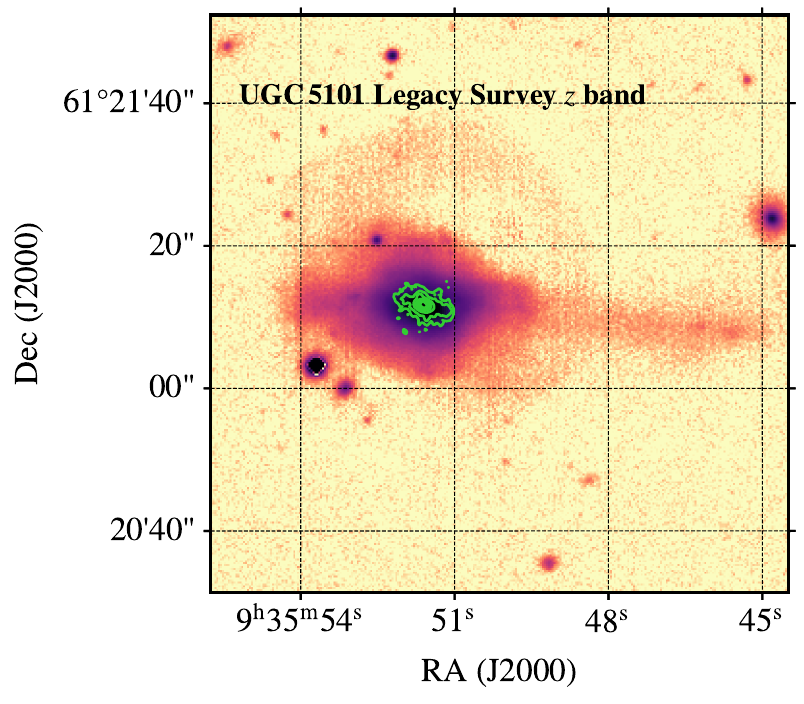}
\includegraphics[width=0.75\linewidth]{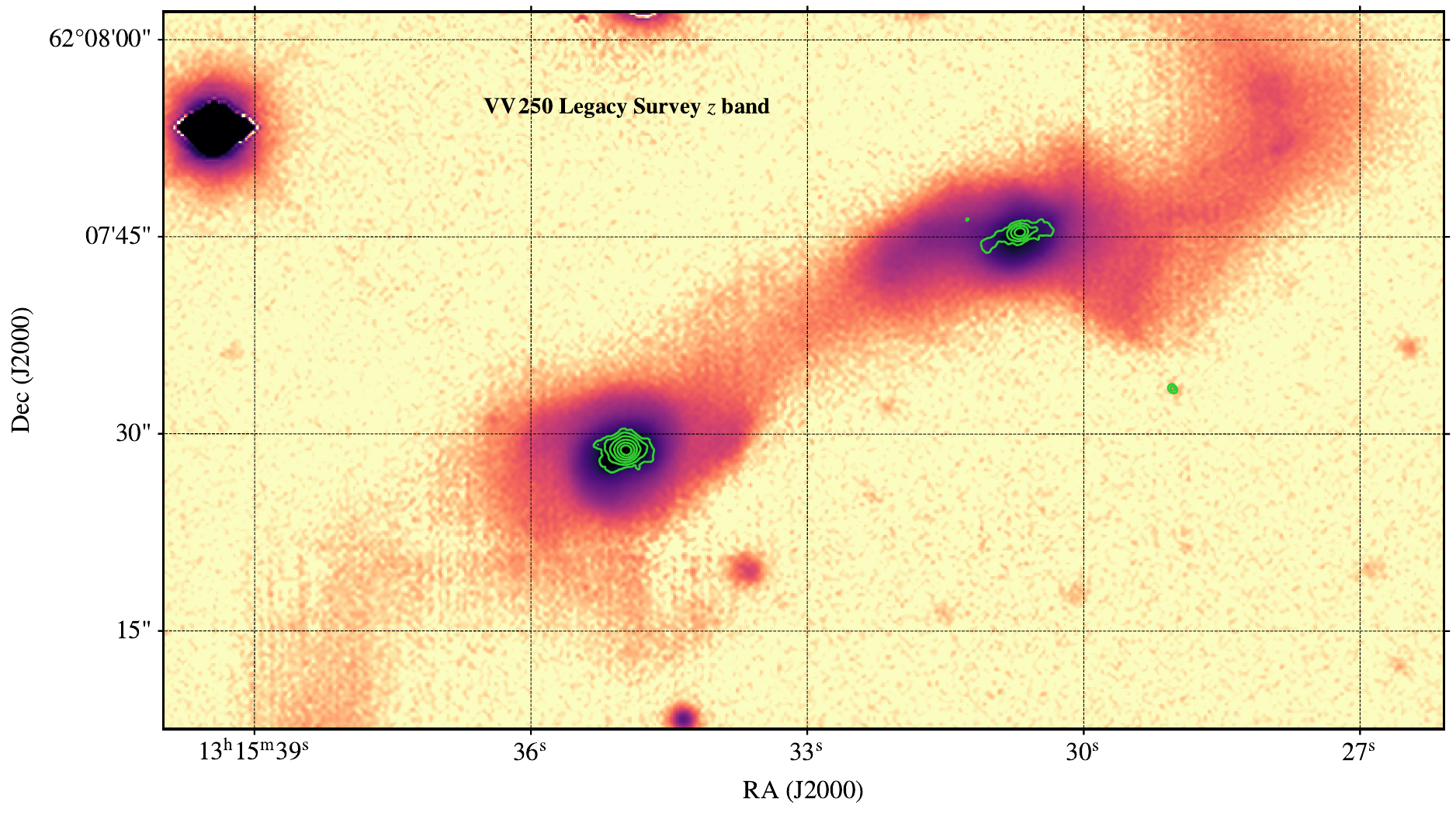}
\caption{Images from the DESI Legacy Imaging Survey\protect\footnotemark at $z$ band (of VV\,705, UGC\,8696, UGC\,5101 and VV\,250) with VLA radio contours at 6 GHz.}
\label{fig:overlay_vv705}
\end{figure*}
\footnotetext{The Legacy Surveys consist of three individual and complementary projects: the Dark Energy Camera Legacy Survey (DECaLS; Proposal ID \#2014B-0404; PIs: David Schlegel and Arjun Dey), the Beijing-Arizona Sky Survey (BASS; NOAO Prop. ID \#2015A-0801; PIs: Zhou Xu and Xiaohui Fan), and the Mayall z-band Legacy Survey (MzLS; Prop. ID \#2016A-0453; PI: Arjun Dey). DECaLS, BASS and MzLS together include data obtained, respectively, at the Blanco telescope, Cerro Tololo Inter-American Observatory, NSF’s NOIRLab; the Bok telescope, Steward Observatory, University of Arizona; and the Mayall telescope, Kitt Peak National Observatory, NOIRLab. Pipeline processing and analyses of the data were supported by NOIRLab and the Lawrence Berkeley National Laboratory (LBNL). The Legacy Surveys project is honored to be permitted to conduct astronomical research on Iolkam Du’ag (Kitt Peak), a mountain with particular significance to the Tohono O’odham Nation.}


\bsp	
\label{lastpage}
\end{document}